\documentclass[
aps,prl,twocolumn,preprintnumbers,
nofootinbib,floatfix,
amsmath,amssymb,
longbibliography,superscriptaddress
]{revtex4-2}

\usepackage[utf8]{inputenc}
\usepackage{siunitx}
\usepackage{amsfonts}
\usepackage{braket}
\usepackage{mathtools}
\usepackage{cancel}
\usepackage{slashed}
\usepackage{pifont}
\usepackage{soul}
\usepackage{comment}

\usepackage{yhmath} 

\usepackage{booktabs,array}
\usepackage{hhline}
\usepackage{siunitx}
\usepackage{graphicx}
\usepackage[caption=false]{subfig}
\usepackage{multirow}

\usepackage{dcolumn}
\usepackage{mathtools}
\usepackage{upgreek}
\newcolumntype{T}{D{.}{.}{10}}
\newcolumntype{E}{D{.}{.}{11}}
\newcolumntype{F}{D{.}{.}{5}}

\usepackage[normalem]{ulem}

\usepackage[dvipsnames]{xcolor}
\usepackage{hyperref}
\hypersetup{
    colorlinks=true,     
    linkcolor=blue,      
    citecolor=blue,      
    filecolor=blue,      
    urlcolor=blue        
}

\usepackage{bm}

\newcommand{\vev}[1]{\ensuremath{\langle #1 \rangle}}
\renewcommand{\vec}[1]{{\mathbf{#1}}}

\makeatletter 
\renewcommand\onecolumngrid{
\do@columngrid{one}{\@ne}%
\def\set@footnotewidth{\onecolumngrid}
\def\footnoterule{\kern-6pt\hrule width 1.5in\kern6pt}%
}
\renewcommand\twocolumngrid{
        \def\footnoterule{
        \dimen@\skip\footins\divide\dimen@\thr@@
        \kern-\dimen@\hrule width.5in\kern\dimen@}
        \do@columngrid{mlt}{\tw@}
}%
\makeatother

\begin{document}
\title{Gravitational form factors of the proton from lattice QCD}
\author{Daniel C. Hackett}
\affiliation{Fermi National Accelerator Laboratory, Batavia, IL 60510, U.S.A.}
\affiliation{Center for Theoretical Physics, Massachusetts Institute of Technology, Cambridge, MA 02139, U.S.A.}
\author{Dimitra A. Pefkou}
\affiliation{Department of Physics, University of California, Berkeley, CA 94720, U.S.A}
\affiliation{Center for Theoretical Physics, Massachusetts Institute of Technology, Cambridge, MA 02139, U.S.A.}
\affiliation{Nuclear Science Division, Lawrence Berkeley National Laboratory, Berkeley, CA 94720, USA}
\author{Phiala E. Shanahan}
\affiliation{Center for Theoretical Physics, Massachusetts Institute of Technology, Cambridge, MA 02139, U.S.A.}

\begin{abstract}

The gravitational form factors (GFFs) of a hadron encode fundamental aspects of its structure, including its shape and size as defined from e.g., its energy density. This work presents a determination of the flavor decomposition of the GFFs of the proton from lattice QCD, 
in the kinematic region $0\leq -t\leq 2~\text{GeV}^2$. The decomposition into up-, down-, strange-quark, and gluon contributions provides first-principles constraints on the role of each constituent in generating key proton structure observables, such as its mechanical radius, mass radius, and $D$-term.

\end{abstract}

\preprint{MIT-CTP/5630,FERMILAB-PUB-23-592-T}

\maketitle

Achieving a quantitative description of the structure of the proton and other hadrons in terms of their quark and gluon constituents is a defining challenge for hadronic physics research. 
The gravitational structure of the proton, encoded in its gravitational form factors (GFFs), has come under particular investigation~\cite{Neubelt:2019sou,Polyakov:2018exb,Azizi:2019ytx,Fujita:2022jus,Amor-Quiroz:2023rke,Chakrabarti:2020kdc,Choudhary:2022den,Lorce:2022cle,Owa:2021hnj,Anikin:2019kwi,Won:2022cyy,Fiore:2021wuj,Mamo:2019mka,Mamo:2022eui,Mamo:2021krl,deTeramond:2021lxc,Alexandrou:2019ali,Shanahan:2018nnv,Shanahan:2018pib,Pefkou:2021fni,Bali:2018zgl,Won:2023zmf,Won:2023zmf,Guo:2023qgu} since the first extraction of one of its GFFs from experimental measurements in 2018~\cite{Burkert:2018bqq}. 
Defined from the matrix elements of the energy-momentum tensor (EMT) in a hadron state, GFFs describe fundamental properties such as the mass and spin of a state, the less well-known but equally fundamental $D$-term (or ``Druck” term), and information that can be interpreted in terms of the distributions of energy, angular momentum, and various mechanical properties of the system~\cite{Lorce:2018egm,Polyakov:2002yz,Polyakov:2018zvc,Burkert:2023wzr}.

The proton GFFs $A(t)$, $J(t)$, and $D(t)$ are defined as
\begin{equation}\label{eq:protonME}
\begin{gathered}
\bra{N(\vec{p}',s')} \hat{T}^{\mu\nu} \ket{N(\vec{p},s)} =\frac{1}{m} \bar{u}(\vec{p}',s')  \bigg[
P^{\mu}P^{\nu} A(t) + \\
iP^{\{\mu}\sigma^{\nu\}\rho}\Delta_{\rho}
J(t)  
 + \frac{1}{4}(\Delta^{\mu}
\Delta^{\nu} - g^{\mu\nu}\Delta^2) D(t) \bigg] u(\vec{p},s),
\end{gathered}
\end{equation} 
where $a^{\{\mu}b^{\nu\}}=(a^{\mu}b^{\nu}+a^{\nu}b^{\mu})/2$,
$N(\vec{p},s)$ is a proton state with three-momentum $\vec{p}$ and spin eigenvalue $s=\pm\frac{1}{2}$, $u(\vec{p},s)$ is the Dirac spinor, $P=(p+p')/2$, $\Delta=p'-p$, $t=\Delta^2$, and $\sigma_{\mu\nu}=\frac{i}{2}[\gamma_{\mu},\gamma_{\nu}]$ where $\gamma_\mu$ are the Dirac matrices. $\hat{T}^{\mu\nu}$ is the renormalization-scale independent~\cite{Nielsen:1977sy} symmetric EMT of QCD~\cite{Belinfante}. It can be decomposed into quark and gluon contributions as $\hat{T}^{\mu\nu} = \sum_{i\in\{q,g\}}\hat{T}_{i}^{\mu\nu}$, where
\begin{align} \label{eq:belifante}
\begin{split}\;
\hat{T}_{g}^{\mu\nu} &= 2\; \mathrm{Tr}\left[- F^{\mu\alpha}F^{\nu}_{\;\alpha} + \frac{1}{4}
g^{\mu\nu}F^{\alpha\beta}F_{\alpha\beta}\right],\\
\hat{T}_{q}^{\mu\nu} &=\sum_{f}\left[ i\bar{\psi}_{f}D^
{\{\mu}\gamma^{\nu\}} \psi_{f}\right],
\end{split}
\end{align}
$F^{\mu\alpha}$ is the gluon field strength tensor, $\psi_f$ is a quark field of flavor $f$, and $D^{\mu}=\partial^{\mu}+igA^{\mu}$ is the covariant derivative.
The matrix elements of $\hat{T}_{i}^{\mu\nu}$ define the renormalization scheme- and scale-dependent partonic contributions to the GFFs, which can also be interpreted as moments of generalized parton distributions (GPDs)~\cite{Muller:1994ses,Ji:1996ek,Radyushkin:1996nd}. The corresponding forward limits $A_i(0)$, $J_i(0)$, and $D_i(0)$ describe the partonic decomposition of the proton’s momentum, spin, and $D$-term, respectively. Poincar\'{e} symmetry imposes the sum rules~\cite{Bakker:2004ib,Ji:1996ek,Kobzarev:1962wt,Pagels:1966zza,Lorce:2022cle} $A(0)=1$ and $J(0)=1/2$, while the value of $D(0)$ is conserved but not constrained from spacetime symmetries~\cite{Polyakov:1999gs}. 
The $t$-dependence of the GFFs encodes additional information about the quark and gluon contributions to densities in the proton~\cite{Lorce:2018egm,Polyakov:2002yz,Polyakov:2018zvc}. While determination of the flavor decomposition of the proton's momentum and spin have a long history, reviewed in Refs.~\cite{Fan:2022qve,Ethier:2020way,Detmold:2019ghl,Liu:2021lke}, 
constraints on the $D$-term~\cite{Duran:2022xag,Burkert:2018bqq,LHPC:2007blg,Alexandrou:2019ali,Shanahan:2018pib,Shanahan:2018nnv,Pefkou:2021fni} and the $t$-dependence~\cite{LHPC:2007blg,Alexandrou:2019ali,Shanahan:2018pib,Shanahan:2018nnv,Pefkou:2021fni,Lin:2020rxa,Bali:2018zgl,Alexandrou:2022dtc} of the proton's GFFs have been comparatively recent.

This work presents a flavor-decomposition of the total GFFs of the proton, $A(t)$, $J(t)$, and $D(t)$, into gluon, up-, down-, and strange-quark contributions, achieved through a lattice QCD calculation with quark masses yielding a close-to-physical value of the pion mass. 
The $D_{u+d}(t)$ and $D_g(t)$ GFFs are found to be consistent with the recent experimental extractions of these quantities~\cite{Burkert:2018bqq,Duran:2022xag}, while the $t$-dependence of $A_g(t)$ is consistent with one of the two analyses of experimental data presented in Ref.~\cite{Duran:2022xag} and in tension with the other (including the updated analysis of Ref.~\cite{Guo:2023pqw}). The individual up-, down-, and strange-quark GFFs are quantified for the first time from experiment or first-principles theory, albeit on a single lattice QCD ensemble, but fully accounting for mixing with the gluon contribution. From the GFFs, the energy and radial force densities of the proton, and the associated mass and mechanical radii, are computed, allowing a quantitative comparison of these different measures of the proton's size.

\textit{Lattice QCD calculation}: The lattice QCD calculation is performed using a single ensemble of gauge field configurations generated by the JLab/LANL/MIT/WM groups~\cite{ensembles}, using the L\"uscher-Weisz gauge action~\cite{Luscher:1984xn} and $N_f=2+1$ flavors of clover-improved Wilson quarks~\cite{Sheikholeslami:1985ij} with clover coefficient set to the tree-level tadpole-improved value and constructed using stout-smeared links~\cite{Morningstar:2003gk}. The light-quark masses are tuned to yield a pion mass of $m_{\pi}\approx170~\text{MeV}$, and the lattice spacing and volume are ${a \approx 0.091~\text{fm}}$~\cite{Park:2021ypf,BMW:2012hcm} and $L^3 \times T = 48^3 \times 96$. 
The technical details of the lattice QCD calculation are as for the determination of the pion GFFs in Ref.~\cite{Hackett:2023nkr} and are summarized below. Additional details, including analysis hyperparameter choices and figures illustrating intermediate results, are included in the supplementary material.

First, the bare matrix elements of  $\hat{T}_g^{\mu\nu}$, and of the singlet and non-singlet quark flavor combinations of the EMT, i.e.,
\begin{align}
\text{singlet:}\quad &\hat{T}_q^{\mu\nu} = \hat{T}_u^{\mu\nu}+\hat{T}_d^{\mu\nu}+\hat{T}_s^{\mu\nu} \\
\text{non-singlet:}\quad &\hat{T}_{v_1}^{\mu\nu} = \hat{T}_u^{\mu\nu}-\hat{T}_d^{\mu\nu}, \\ &\hat{T}_{v_2}^{\mu\nu} = \hat{T}_u^{\mu\nu}+\hat{T}_d^{\mu\nu}-2\hat{T}_s^{\mu\nu},
\end{align}
are constrained from ratios of three-point and two-point  functions that are proportional to the bare matrix elements of the EMT, Eq.~\eqref{eq:protonME}, at large Euclidean times. 
The three-point function of the gluon EMT is measured on 2511 configurations, averaged over 1024 source positions per configuration, with the gluon EMT measured on gauge fields that have been Wilson flowed~\cite{Luscher:2010iy,Narayanan:2006rf,Lohmayer:2012hs} to $t_{\text{flow}}/a^2=2$, for all sink and operator momenta with $|\vec{p}'|^2\leq 10(2\pi/L)^2$ and $|\vec{\Delta}|^2\leq25(2\pi/L)^2$, and all four spin channels, $s,s'\in\{\pm 1/2\}$. The connected part of the quark three-point function is measured on $1381$ configurations using the sequential source method, inverting through the sink for 11 choices of source-sink separation in the range $[6a,18a]$, with the number of sources varying between $9$ and $32$ for the different source-sink separations. The momenta measured are $\vec{p}' \in 2\pi/L\{(1,0,-1), (-2,-1,0),(-1,-1,-1)\}$ and all $\vec{\Delta}$ with $|\vec{\Delta}|^2\leq25(2\pi/L)^2$, for a single spin channel with $s=s'=1/2$. 
The disconnected parts are stochastically estimated on the same $1381$ configurations as the connected parts, using $2$ samples of $Z_4$ noise~\cite{doi:10.1080/03610919008812866}, diluting in spacetime using hierarchical probing~\cite{Stathopoulos:2013aci,Gambhir:2017the} with $512$ Hadamard vectors, and computing the spin-color trace exactly.
Measurements are made for all $|\vec{p}'|^2\leq 10(2\pi/L)^2$, $|\vec{\Delta}|^2\leq25(2\pi/L)^2$, and all four spin channels, 

\begin{figure}[!]
\centering
\includegraphics[width=0.48\textwidth]{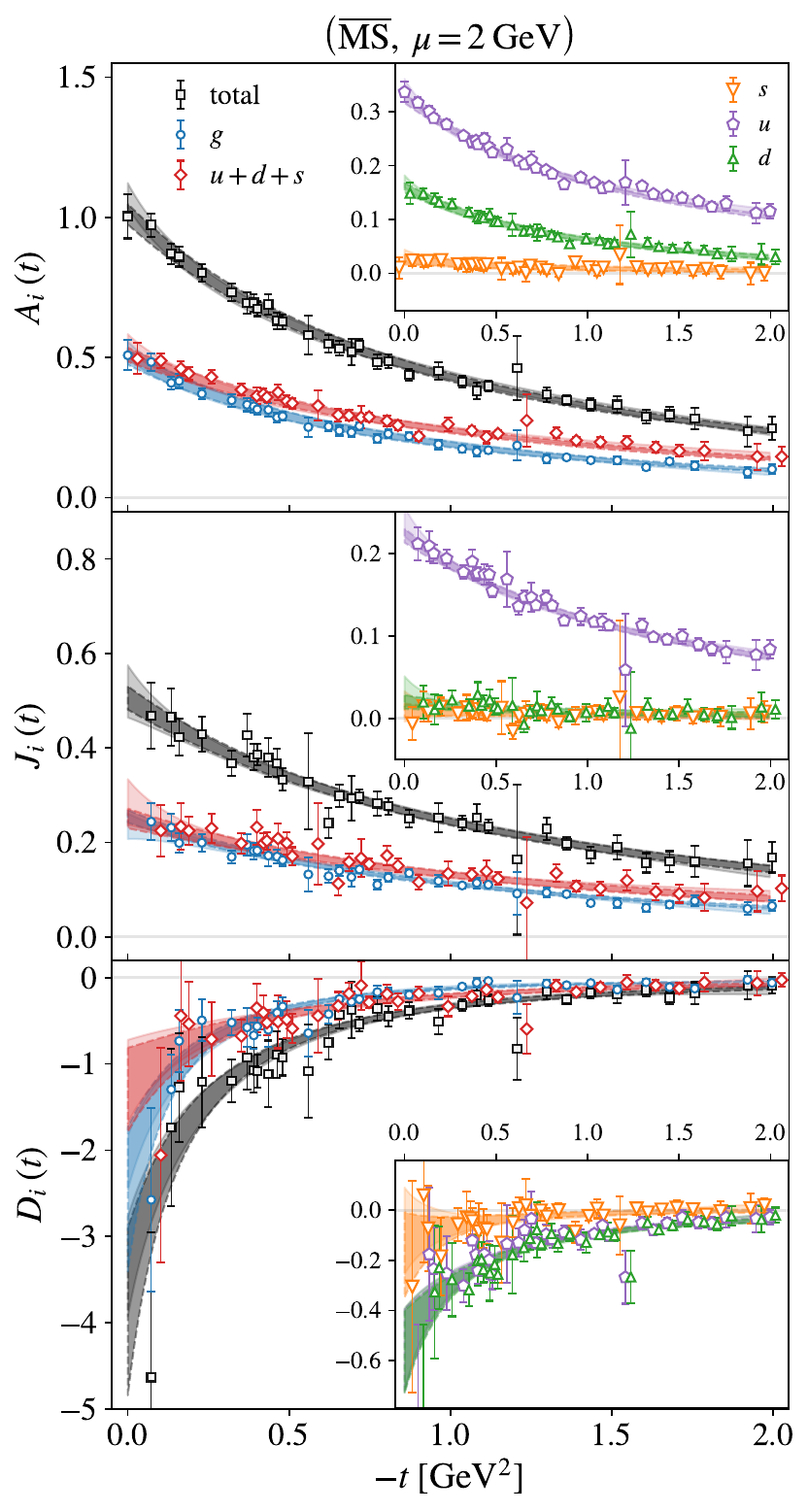}
\caption{ 
The three GFFs of the proton, computed on the lattice QCD ensemble of this work, and their decomposition into gluon and total quark contributions, are shown as functions of $t$. Inset figures show the isosinglet quark GFFs further decomposed into up-, down-, and strange-quark contributions. The total GFFs are renormalization scheme- and scale-independent, while all other GFFs are shown in the $\overline{\text{MS}}$ scheme at $\mu=2~\text{GeV}$. The dark bands represent dipole fits to the data in the case of $g$ and $q=u+d+s$, and linear combinations of the dipole fits to $q$, $v_1$, and $v_2$ in all other cases. The lighter bands show analogous fits using the $z$-expansion.}
\label{fig:GFFflavors}
\end{figure}

Second, ratios of three- and two-point functions that correspond to the same linear combination of GFFs---as defined in Eq.~\eqref{eq:protonME}, and up to an overall sign---are averaged. The summation method~\cite{Capitani:2012gj,Maiani:1987by,Dong:1997xr,Djukanovic:2021cgp} is used to fit the Euclidean time-dependence of the averaged ratios and extract the bare matrix elements. In all cases, $1000$ bootstrap ensembles are used to estimate statistical uncertainties, and systematic uncertainties in fits are propagated using model averaging with weights dictated by the Akaike information criterion~\cite{Akaike:1998zah} (AIC)~\cite{Jay:2020jkz,Rinaldi:2018osy,Beane:2015yha}. Since connected measurements exist for only a subset of the matrix elements, the disconnected contributions to the bare GFFs of $T_q$ and $T_{v_2}$\footnote{$T_{v_1}$ is purely connected, as the disconnected contributions cancel in the difference.} are fit separately using all available data, with the results used to obtain better constraints for the subset for which connected parts are available and thus to obtain the full matrix elements of $T_q$ and $T_{v_2}$.
Finally, the matrix elements are divided into 34 $t$-bins using $k$-means clustering~\cite{kmeans1d}, and the GFFs are extracted by solving the resulting linear systems of equations, with the renormalization performed non-perturbatively using the results and procedure presented in Ref.~\cite{Hackett:2023nkr}.

\begin{figure}
\centering
\includegraphics[width=0.48\textwidth]{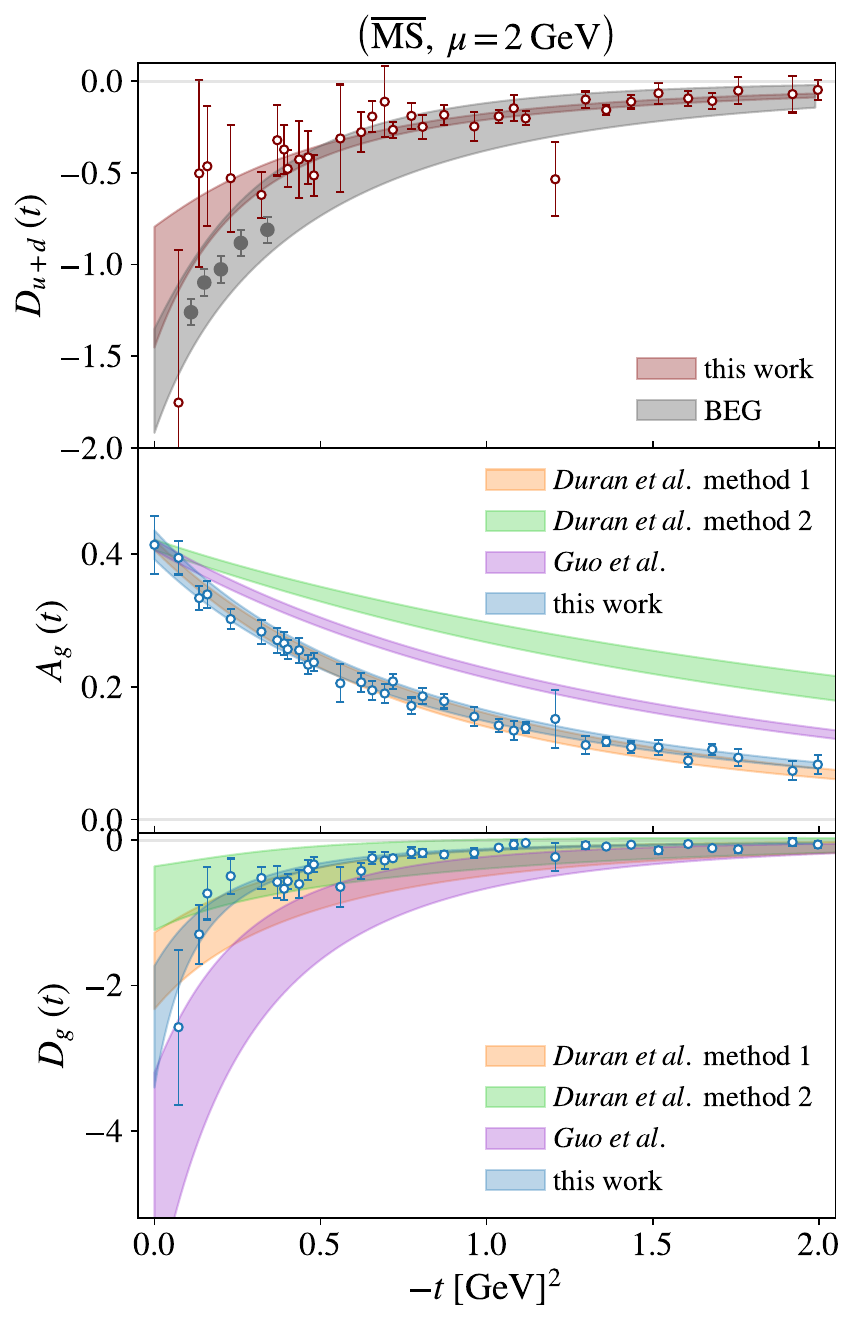}
\caption{The proton GFFs $A_g(t)$, $D_g(t)$, and $D_{u+d}(t)$ and corresponding dipole fits, 
computed on the lattice QCD ensemble of this work, are compared with the experimentally constrained multipole parametrizations of Refs.~\cite{Burkert:2018bqq} (BEG),~\cite{Duran:2022xag} (Duran et al.), and~\cite{Guo:2023pqw} (Guo et al.). For this comparison, the results for $A_g(t)$ are re-scaled such that the gluon momentum fraction is $A_g(0)=0.414(8)$~\cite{Hou:2019efy}, which is the value used as an input in the extraction of Refs.~\cite{Duran:2022xag} and~\cite{Guo:2023pqw}. This does not affect the $t$-dependence of the GFF.}
\label{fig:expcomparison}
\end{figure}

\begin{table*} 
\begin{center}
\begin{ruledtabular}
\begin{tabular}{SSSSSSS}
& \multicolumn{3}{c}{Dipole} & \multicolumn{3}{c}{$z$-expansion}  \\ \midrule
 & {$A_i(0)$} & {$J_i(0)$} & {$D_i(0)$} & {$A_i(0)$} & {$J_i(0)$} & {$D_i(0)$} \\ \midrule
{$u$} & 0.3255(92) & 0.2213(85) & -0.56(17) & 0.349(11) & 0.238(18) & -0.56(17)  \\[2pt]
{$d$} & 0.1590(92) & 0.0197(85) & -0.57(17) & 0.171(11) & 0.033(18) & -0.56(17) \\[2pt]
{$s$} & 0.0257(95) & 0.0097(82) & -0.18(17) & 0.032(12) & 0.014(19) & -0.08(17) \\[2pt]
{$u+d+s$} & 0.510(25) & 0.251(21) & -1.30(49) & 0.552(31) & 0.286(48) & -1.20(48) \\[2pt]
{$g$} & 0.501(27) & 0.255(13) & -2.57(84) & 0.526(31) & 0.234(27) & -2.15(32) \\[2pt]
{Total} & 1.011(37) & 0.506(25) & -3.87(97) & 1.079(44) & 0.520(55) & -3.35(58) \\
\end{tabular}
\end{ruledtabular}
\end{center}
\caption{\label{tab:forwardlimit}The flavor decomposition of the momentum fraction, spin, and $D$-term of the proton, computed on the lattice QCD ensemble of this work, obtained from dipole and $z$-expansion fits to the proton GFFs, renormalized at $\mu = 2\;\text{GeV}$ in the $\overline{\text{MS}}$ scheme. The fit parameters of the two models are included in the Supplementary Information.
}
\end{table*}

\textit{Results:} The flavor decomposition of the renormalized GFFs, computed on the lattice QCD ensemble of this work, is presented in Fig.~\ref{fig:GFFflavors}. To guide the eye, the 
GFFs of currents $g$ and $q$ are fit using both a multipole ansatz with $n=2$ (dipole), chosen as the integer yielding the lowest $\chi^2$ per degree of freedom for the majority of the fits, as well as the more expressive $z$-expansion~\cite{Hill:2010yb}. 
The GFFs are further decomposed to yield the individual quark flavor contributions $\vec{G}(t)=(A(t),J(t),D(t))$ from the data and fits for currents $q$, $v_1$, and $v_2$, using
\begin{align}
\vec{G}_{u}(t) = \frac{1}{3}\vec{G}_q(t)+\frac{1}{6}\vec{G}_{v_2}(t)+\frac{1}{2}\vec{G}_{v_1}(t) \;,
\end{align}
\begin{align}
\vec{G}_{d}(t) &= \frac{1}{3}\vec{G}_q(t)+\frac{1}{6}\vec{G}_{v_2}(t)-\frac{1}{2}\vec{G}_{v_1}(t) \;,\\
\vec{G}_{s}(t) &= \frac{1}{3}\vec{G}_q(t)-\frac{1}{3}\vec{G}_{v_2}(t) \;.
\end{align}
The functional forms of the fit models, along with the resulting fit parameters, are given in the supplementary material. 

The flavor decomposition of the forward limits $A(0)$, $J(0)$, and $D(0)$ is summarized in Table~\ref{tab:forwardlimit}, and can be compared with other recent lattice QCD calculations of the decomposition of the momentum and spin fractions of the proton. 
Specifically, recent studies of the forward-limit quantities at or extrapolated to the physical pion mass report $A_q(0)=0.491(20)(23), A_g(0)=0.509(20)(23), J_q(0)=0.270(11)(22), J_g(0)=0.231(11)(22)$~\cite{Wang:2021vqy}, $A_q(0)=0.618(60), A_g(0)=0.427(92), J_q(0)=0.285(45), J_g(0)=0.187(46)$~\cite{Alexandrou:2020sml}, and $A_g(0)=0.492(52)(49)$~\cite{Fan:2022qve}.
In the present calculation, the sum rules for the total momentum fraction and spin are satisfied, and the total quark and gluon contributions to these quantities are approximately equal. The calculated gluon momentum fraction is, however, several standard deviations larger than the global fit result $A_g(0)=0.414(8)$~\cite{Hou:2019efy}, which can likely be attributed to remaining systematic uncertainties that could not be estimated from this calculation using a single ensemble of lattice QCD gauge fields. In particular, the continuum limit has not be taken, and renormalization coefficients were computed on an ensemble with larger lattice spacing and quark masses~\cite{Hackett:2023nkr}. 
Moreover, the $m_{\pi} L\approx 3.8$ of the ensemble used in this work is less than the typical rule-of-thumb bound $m_{\pi} L > 4$ targeted to limit finite volume effects to the percent level. 
The expected magnitudes of these various additional sources of systematic uncertainty are discussed in the Supplementary Information.
The calculated result for the total $D$-term satisfies the chiral perturbation theory prediction for its upper bound~\cite{Gegelia:2021wnj}, $D(0)/m\leq-1.1(1)~\text{GeV}^{-1}$, and is in agreement with chiral models~\cite{Goeke:2007fp,Schweitzer:2002nm,Ossmann:2004bp,Goeke:2007fq,Wakamatsu:2007uc,Cebulla:2007ei,Jung:2013bya}.

Figure~\ref{fig:expcomparison} presents a comparison of the dipole fit results with the available experimentally constrained multipole parametrization results for $D_{u+d}$, $A_g$, and $D_g$, presented in Refs.~\cite{Burkert:2018bqq,Duran:2022xag,Guo:2023pqw}. The fits to $D_{u+d}$ are found to be consistent, but the uncertainties of the lattice data are comparatively large at the small values of $|t|$ for which experimental data is available. Extractions from new~\cite{CLAS:2021lky} and future experimental data over a larger $|t|$-range, as well as better control of the uncertainties of the lattice QCD result at low $|t|$, will be necessary for a robust comparison.
For the gluon GFFs, the lattice QCD results are found to be consistent with the `holographic QCD' inspired approach~\cite{Mamo:2022eui,Mamo:2021krl} (method 1) to the analysis of experimental data in Ref.~\cite{Duran:2022xag} and disfavor the `generalized parton distribution' (GPD) inspired approach~\cite{Guo:2021ibg} (method 2). A more recent analysis~\cite{Guo:2023pqw} including an update to the GPD inspired analysis method, as well as additional experimental data~\cite{GlueX:2023pev}, is in less tension with the lattice QCD results presented here.
These comparisons illustrate the continued synergy and complementarity between lattice and experimental results for these quantities. 

\begin{figure}
\includegraphics[width=0.48\textwidth]{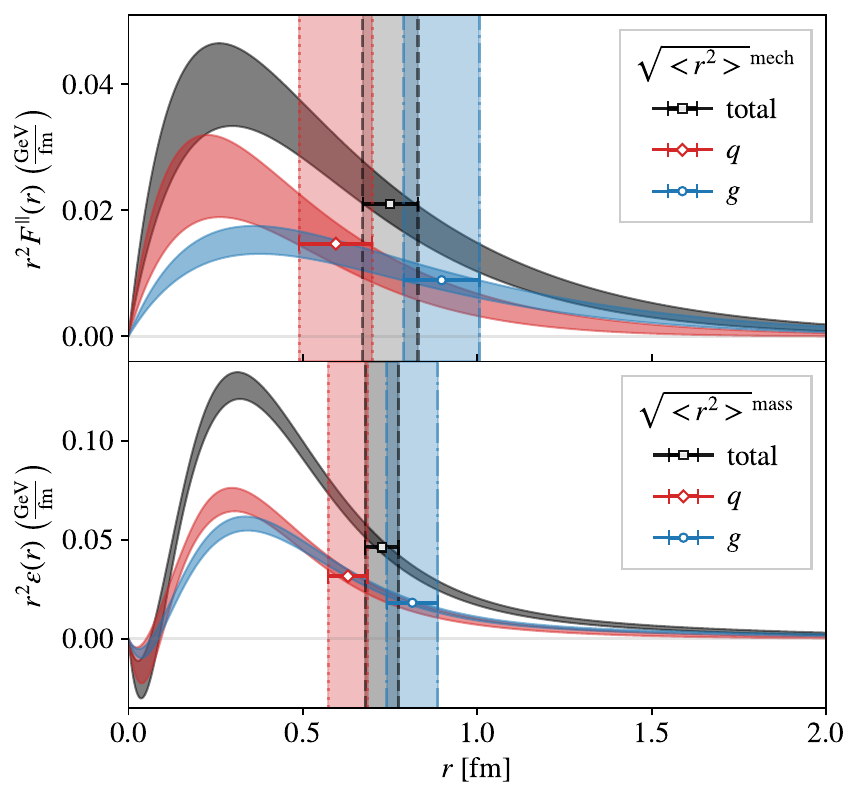}
\caption{The quark, gluon, and total contributions to the longitudinal force (upper) and energy (lower)
densities in the Breit frame, computed on the lattice QCD ensemble of this work, are shown as functions of the radial distance from the center of the proton. The corresponding quark, gluon, and total mechanical and mass radii are marked as data points on the corresponding curves.}
\label{fig:densities_radii}
\end{figure}

\begin{figure}[t]
\includegraphics[width=\linewidth]{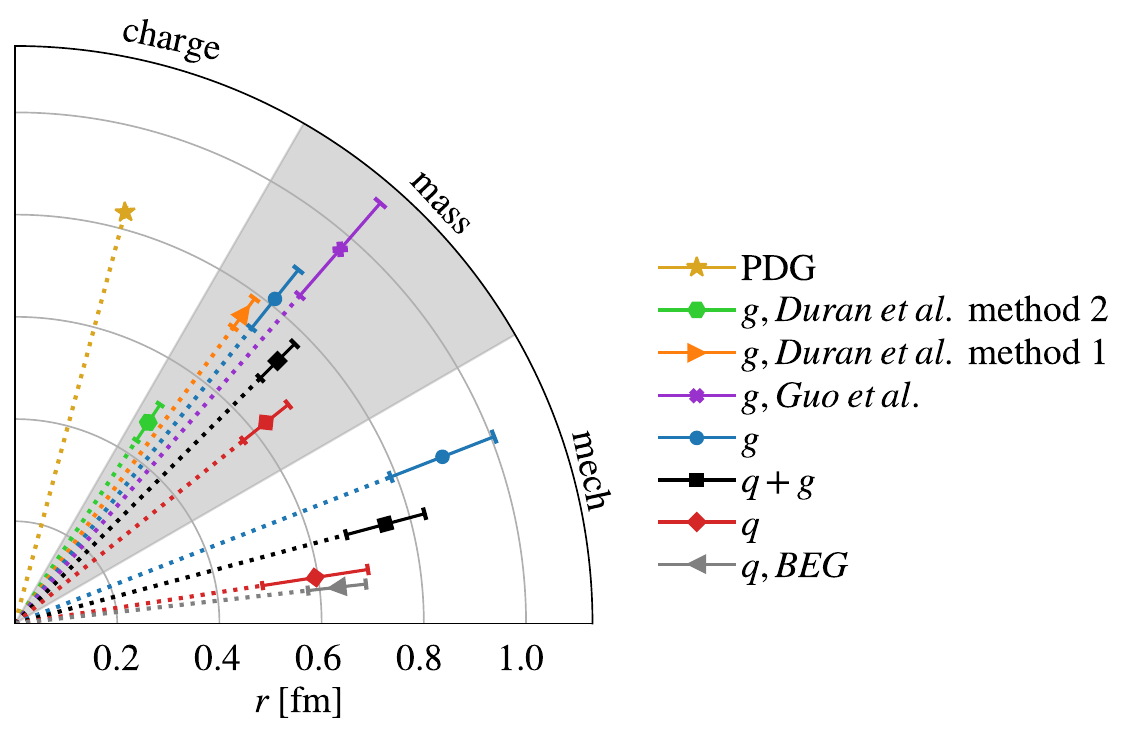}
\caption{Comparison of different proton radii. In addition to the results obtained on the lattice QCD ensemble of this work, the charge radius from Ref.~\cite{ParticleDataGroup:2022pth} (PDG), the gluonic mass radii from Ref.~\cite{Duran:2022xag} (Duran et al.) and Ref.~\cite{Guo:2023pqw} (Guo et al.), and the quark mechanical radii from Ref.~\cite{BEG:2023mech} (BEG) are shown. While the latter only includes the light quark contributions, they are compared directly as the strange quark contribution is found to be negligible for this quantity. The uncertainty of the charge radius is too small to be visible.}
\label{fig:testrad}
\end{figure}

\textit{Densities}: Through the definition in terms of the EMT, and by analogy\footnote{The physical significance of these analogies is debated~\cite{Ji:2021mfb,Jaffe:2020ebz,Panteleeva:2022uii,Freese:2021mzg}.} to mechanical systems, the $t$-dependence of the GFFs also gives insight into various densities in the proton.
Specifically, the Breit-frame distributions $\varepsilon_{i}(r)$, $p_{i}(r)$, and $s_{i}(r)$, defined as\footnote{The  quark and gluon contributions to the pressure and energy densities additionally depend on the GFF $\bar{c}_i(t)$, which appears in the decomposition of the matrix elements of $\hat{T}^{\mu\nu}_i$ due to the quark and gluon EMT terms not being individually conserved. This contribution, which is not constrained in this work, vanishes for the total densities since $\bar{c}_q(t)+\bar{c}_g(t)$=0.}
\begin{align} 
\varepsilon_{i}(r) &= 
m \left[A_{i}(t) \small{-} \frac{t(D_{i}(t) \small{+} A_{i}(t)\small{-}2J_i(t))}{4m^2}\right]_{\text{FT}}\;,
\end{align}
\begin{align}
p_{i}(r) &=\frac{1}{6 m}\frac{1}{r^2}\frac{d}{dr}r^2\frac{d}{dr}[D_{i}(t)]_{\text{FT}}\;,
\end{align}
\begin{align}
s_{i}(r) &= -\frac{1}{4m}r\frac{d}{dr}\frac{1}{r}\frac{d}{dr}
[D_{i}(t)]_{\text{FT}}\;,
\end{align}
where $r=|\vec{r}|$,
\begin{equation}
[f(t)]_{\text{FT}} = \int \frac{d^3\vec{\Delta}}{(2\pi)^3}e^{-i\vec{\Delta}\cdot \vec{r}}f(t) \;,
\end{equation}
and $i\in\{q,g,q\small{+}g\}$, can be interpreted as energy, pressure, and shear force distributions respectively~\cite{Lorce:2018egm,Polyakov:2002yz,Polyakov:2018zvc}.
The root-mean-square radii of the energy density and the longitudinal force density 
\begin{equation}
  F^{||}_{i}(r)=p_{i}(r)+2s_{i}(r)/3  
\end{equation} 
yield the mass and mechanical radii of the proton~\cite{Polyakov:2018zvc}, 
\begin{align} \label{eq:nucradii}
\braket{r^2_{i}}^{\mathrm{mass}} &= \frac{\int d^3\vec{r}\, r^2 \varepsilon_i(r)}{\int d^3\vec{r}\, \varepsilon_i(r)}\;, \nonumber \\ 
\braket{r^2_{i}}^{\mathrm{mech}} &= \frac{\int d^3\vec{r}\, r^2 F^{||}_i(r)}{\int d^3\vec{r}\,F^{||}_i(r)}.
\end{align}
Figure~\ref{fig:densities_radii} shows the quark, gluon, and total densities and corresponding radii obtained analytically from the dipole fits to the GFFs.
For both densities, the gluonic radius is found to be larger than the quark radius. For the case of the gluon mass radius, the result is consistent with that predicted using the `holographic QCD' inspired model in the phenomenological extraction of Ref.~\cite{Duran:2022xag} and the updated analysis of Ref.~\cite{Guo:2023pqw}, as shown in Fig.~\ref{fig:testrad}. The results for the quark mechanical radius are consistent with a recent extraction from deeply virtual Compton scattering cross sections data~\cite{BEG:2023mech}. They are also consistent with the soliton model prediction~\cite{Cebulla:2007ei,Goeke:2007fp} that the proton mechanical radius is slightly smaller than the charge radius~\cite{ParticleDataGroup:2022pth}, and with the equality of the two radii in the non-relativistic limit shown in the bag model~\cite{Lorce:2022cle,Neubelt:2019sou}.

\textit{Summary}: The flavor decomposition of the proton's $A(t)$, $J(t)$, and $D(t)$ GFFs into their up-, down-, strange-quark, and gluon contributions is determined for the first time for a kinematic range $0\leq -t\leq 2~\text{GeV}^2$, using a first-principles lattice QCD calculation. The results reveal that, while the contributions of quarks and gluons to the proton's momentum, spin, and $D$-term are approximately equal, the gluon contributions act to extend the radial size of the proton over that defined by the quark contributions as quantified through the mass and mechanical radii encoded in the $t$-dependence of the GFFs. To improve upon these first results, it is crucial that this study be repeated using ensembles with different lattice volumes and lattice spacings in order for these systematic uncertainties to be fully accounted for. Moreover, it will be important to improve the renormalization procedure, e.g., by exploring the use of gauge-invariant renormalization schemes~\cite{Costa:2021iyv}. Nevertheless, these first-principles results permit first comparisons between theory and experiment for several aspects of these fundamental measures of proton structure.

The lattice QCD results for the $D_{u+d}(t)$ and $D_g(t)$ GFFs are consistent with the recent experimental results of Refs.~\cite{Burkert:2018bqq,Duran:2022xag}, but $D_{u+d}(t)$ is constrained over a greater kinematic range. For $A_g(t)$, however, the comparison of first-principles theory with the experimental results of Ref.~\cite{Duran:2022xag} provides important additional constraints that distinguish between different analyses of the experimental data. Moreover, the results for the separate up-, down-, and strange-quark GFFs presented here are the first constraints on these quantities from first principles theory or from experiment. This work thus sets important benchmarks on these fundamental aspects of proton structure for future measurements at Thomas Jefferson National Accelerator Facility~\cite{JeffersonLabSoLID:2022iod,doi:10.1146/annurev-nucl-101917-021129,Arrington:2021alx}  and at a future Electron-Ion Collider~\cite{AbdulKhalek:2021gbh}.

\begin{acknowledgements}
\textit{Acknowledgements}: The authors thank Will Detmold, Zein-Eddine Meziani, and Ross Young for useful feedback and suggestions. This work is supported in part by the U.S.~Department of Energy, Office of Science, Office of Nuclear Physics, under grant Contract Number DE-SC0011090 and by Early Career Award DE-SC0021006, and has benefited from the QGT Topical Collaboration DE-SC0023646. PES is supported in part by Simons Foundation grant 994314 (Simons Collaboration on Confinement and QCD Strings) and by the U.S. Department of Energy SciDAC5 award DE-SC0023116. DAP is supported from the Office of Nuclear Physics, Department of Energy, under contract DE-SC0004658. This manuscript has been authored by Fermi Research Alliance, LLC under Contract No.~DE-AC02-07CH11359 with the U.S. Department of Energy, Office of Science, Office of High Energy Physics. This research used resources of the National Energy Research Scientific Computing Center (NERSC), a U.S. Department of Energy Office of Science User Facility operated under Contract No.~DE-AC02-05CH11231, as well as resources of the Argonne Leadership Computing Facility, which is a DOE Office of Science User Facility supported under Contract DE-AC02-06CH11357, and the Extreme Science and Engineering Discovery Environment (XSEDE), which is supported by National Science Foundation grant number ACI-1548562. Computations were carried out in part on facilities of the USQCD Collaboration, which are funded by the Office of Science of the U.S. Department of Energy. The authors thank Robert Edwards, Rajan Gupta, Balint Jo{\'o}, Kostas Orginos, and the NPLQCD collaboration for generating the ensemble used in this study, and Eloy Romero for assistance with software deployment. 
The Chroma~\cite{Edwards:2004sx}, QLua~\cite{qlua},  QUDA~\cite{Clark:2009wm,Babich:2011np,Clark:2016rdz}, QDP-JIT~\cite{6877336}, QPhiX~\cite{10.1007/978-3-319-46079-6_30} , and Chromaform~\cite{chromaform} software libraries were used in this work. Code for disconnected diagrams was adapted from LALIBE~\cite{lalibe}, including the hierarchical probing implementation by Andreas Stathopoulos~\cite{Stathopoulos:2013aci}.
Data analysis used NumPy~\cite{harris2020array}, SciPy~\cite{2020SciPy-NMeth}, pandas~\cite{jeff_reback_2020_3715232,mckinney-proc-scipy-2010}, lsqfit~\cite{peter_lepage_2020_4037174}, and gvar~\cite{peter_lepage_2020_4290884}.
Figures were produced using matplotlib~\cite{Hunter:2007}.
\end{acknowledgements}

\bibliography{main}

\begin{thebibliography}{118}%
\makeatletter
\providecommand \@ifxundefined [1]{%
 \@ifx{#1\undefined}
}%
\providecommand \@ifnum [1]{%
 \ifnum #1\expandafter \@firstoftwo
 \else \expandafter \@secondoftwo
 \fi
}%
\providecommand \@ifx [1]{%
 \ifx #1\expandafter \@firstoftwo
 \else \expandafter \@secondoftwo
 \fi
}%
\providecommand \natexlab [1]{#1}%
\providecommand \enquote  [1]{``#1''}%
\providecommand \bibnamefont  [1]{#1}%
\providecommand \bibfnamefont [1]{#1}%
\providecommand \citenamefont [1]{#1}%
\providecommand \href@noop [0]{\@secondoftwo}%
\providecommand \href [0]{\begingroup \@sanitize@url \@href}%
\providecommand \@href[1]{\@@startlink{#1}\@@href}%
\providecommand \@@href[1]{\endgroup#1\@@endlink}%
\providecommand \@sanitize@url [0]{\catcode `\\12\catcode `\$12\catcode
  `\&12\catcode `\#12\catcode `\^12\catcode `\_12\catcode `\%12\relax}%
\providecommand \@@startlink[1]{}%
\providecommand \@@endlink[0]{}%
\providecommand \url  [0]{\begingroup\@sanitize@url \@url }%
\providecommand \@url [1]{\endgroup\@href {#1}{\urlprefix }}%
\providecommand \urlprefix  [0]{URL }%
\providecommand \Eprint [0]{\href }%
\providecommand \doibase [0]{https://doi.org/}%
\providecommand \selectlanguage [0]{\@gobble}%
\providecommand \bibinfo  [0]{\@secondoftwo}%
\providecommand \bibfield  [0]{\@secondoftwo}%
\providecommand \translation [1]{[#1]}%
\providecommand \BibitemOpen [0]{}%
\providecommand \bibitemStop [0]{}%
\providecommand \bibitemNoStop [0]{.\EOS\space}%
\providecommand \EOS [0]{\spacefactor3000\relax}%
\providecommand \BibitemShut  [1]{\csname bibitem#1\endcsname}%
\let\auto@bib@innerbib\@empty
\bibitem [{\citenamefont {Neubelt}\ \emph {et~al.}(2020)\citenamefont
  {Neubelt}, \citenamefont {Sampino}, \citenamefont {Hudson}, \citenamefont
  {Tezgin},\ and\ \citenamefont {Schweitzer}}]{Neubelt:2019sou}%
  \BibitemOpen
  \bibfield  {author} {\bibinfo {author} {\bibfnamefont {M.~J.}\ \bibnamefont
  {Neubelt}}, \bibinfo {author} {\bibfnamefont {A.}~\bibnamefont {Sampino}},
  \bibinfo {author} {\bibfnamefont {J.}~\bibnamefont {Hudson}}, \bibinfo
  {author} {\bibfnamefont {K.}~\bibnamefont {Tezgin}},\ and\ \bibinfo {author}
  {\bibfnamefont {P.}~\bibnamefont {Schweitzer}},\ }\bibfield  {title}
  {\bibinfo {title} {{Energy momentum tensor and the D-term in the bag
  model}},\ }\href {https://doi.org/10.1103/PhysRevD.101.034013} {\bibfield
  {journal} {\bibinfo  {journal} {Phys. Rev. D}\ }\textbf {\bibinfo {volume}
  {101}},\ \bibinfo {pages} {034013} (\bibinfo {year} {2020})},\ \Eprint
  {https://arxiv.org/abs/1911.08906} {arXiv:1911.08906 [hep-ph]} \BibitemShut
  {NoStop}%
\bibitem [{\citenamefont {Polyakov}\ and\ \citenamefont
  {Son}(2018)}]{Polyakov:2018exb}%
  \BibitemOpen
  \bibfield  {author} {\bibinfo {author} {\bibfnamefont {M.~V.}\ \bibnamefont
  {Polyakov}}\ and\ \bibinfo {author} {\bibfnamefont {H.-D.}\ \bibnamefont
  {Son}},\ }\bibfield  {title} {\bibinfo {title} {{Nucleon gravitational form
  factors from instantons: forces between quark and gluon subsystems}},\ }\href
  {https://doi.org/10.1007/JHEP09(2018)156} {\bibfield  {journal} {\bibinfo
  {journal} {JHEP}\ }\textbf {\bibinfo {volume} {09}},\ \bibinfo {pages}
  {156}},\ \Eprint {https://arxiv.org/abs/1808.00155} {arXiv:1808.00155
  [hep-ph]} \BibitemShut {NoStop}%
\bibitem [{\citenamefont {Azizi}\ and\ \citenamefont
  {\"Ozdem}(2020)}]{Azizi:2019ytx}%
  \BibitemOpen
  \bibfield  {author} {\bibinfo {author} {\bibfnamefont {K.}~\bibnamefont
  {Azizi}}\ and\ \bibinfo {author} {\bibfnamefont {U.}~\bibnamefont
  {\"Ozdem}},\ }\bibfield  {title} {\bibinfo {title}
  {{Nucleon\textquoteright{}s energy\textendash{}momentum tensor form factors
  in light-cone QCD}},\ }\href {https://doi.org/10.1140/epjc/s10052-020-7676-5}
  {\bibfield  {journal} {\bibinfo  {journal} {Eur. Phys. J. C}\ }\textbf
  {\bibinfo {volume} {80}},\ \bibinfo {pages} {104} (\bibinfo {year} {2020})},\
  \Eprint {https://arxiv.org/abs/1908.06143} {arXiv:1908.06143 [hep-ph]}
  \BibitemShut {NoStop}%
\bibitem [{\citenamefont {Fujita}\ \emph {et~al.}(2022)\citenamefont {Fujita},
  \citenamefont {Hatta}, \citenamefont {Sugimoto},\ and\ \citenamefont
  {Ueda}}]{Fujita:2022jus}%
  \BibitemOpen
  \bibfield  {author} {\bibinfo {author} {\bibfnamefont {M.}~\bibnamefont
  {Fujita}}, \bibinfo {author} {\bibfnamefont {Y.}~\bibnamefont {Hatta}},
  \bibinfo {author} {\bibfnamefont {S.}~\bibnamefont {Sugimoto}},\ and\
  \bibinfo {author} {\bibfnamefont {T.}~\bibnamefont {Ueda}},\ }\bibfield
  {title} {\bibinfo {title} {{Nucleon D-term in holographic quantum
  chromodynamics}},\ }\href {https://doi.org/10.1093/ptep/ptac110} {\bibfield
  {journal} {\bibinfo  {journal} {PTEP}\ }\textbf {\bibinfo {volume} {2022}},\
  \bibinfo {pages} {093B06} (\bibinfo {year} {2022})},\ \Eprint
  {https://arxiv.org/abs/2206.06578} {arXiv:2206.06578 [hep-th]} \BibitemShut
  {NoStop}%
\bibitem [{\citenamefont {Amor-Quiroz}\ \emph {et~al.}(2023)\citenamefont
  {Amor-Quiroz}, \citenamefont {Focillon}, \citenamefont {Lorc\'e},\ and\
  \citenamefont {Rodini}}]{Amor-Quiroz:2023rke}%
  \BibitemOpen
  \bibfield  {author} {\bibinfo {author} {\bibfnamefont {A.}~\bibnamefont
  {Amor-Quiroz}}, \bibinfo {author} {\bibfnamefont {W.}~\bibnamefont
  {Focillon}}, \bibinfo {author} {\bibfnamefont {C.}~\bibnamefont {Lorc\'e}},\
  and\ \bibinfo {author} {\bibfnamefont {S.}~\bibnamefont {Rodini}},\
  }\bibfield  {title} {\bibinfo {title} {{Energy-momentum tensor in the scalar
  diquark model}},\ }\href@noop {} {\  (\bibinfo {year} {2023})},\ \Eprint
  {https://arxiv.org/abs/2304.10339} {arXiv:2304.10339 [hep-ph]} \BibitemShut
  {NoStop}%
\bibitem [{\citenamefont {Chakrabarti}\ \emph {et~al.}(2020)\citenamefont
  {Chakrabarti}, \citenamefont {Mondal}, \citenamefont {Mukherjee},
  \citenamefont {Nair},\ and\ \citenamefont {Zhao}}]{Chakrabarti:2020kdc}%
  \BibitemOpen
  \bibfield  {author} {\bibinfo {author} {\bibfnamefont {D.}~\bibnamefont
  {Chakrabarti}}, \bibinfo {author} {\bibfnamefont {C.}~\bibnamefont {Mondal}},
  \bibinfo {author} {\bibfnamefont {A.}~\bibnamefont {Mukherjee}}, \bibinfo
  {author} {\bibfnamefont {S.}~\bibnamefont {Nair}},\ and\ \bibinfo {author}
  {\bibfnamefont {X.}~\bibnamefont {Zhao}},\ }\bibfield  {title} {\bibinfo
  {title} {{Gravitational form factors and mechanical properties of proton in a
  light-front quark-diquark model}},\ }\href
  {https://doi.org/10.1103/PhysRevD.102.113011} {\bibfield  {journal} {\bibinfo
   {journal} {Phys. Rev. D}\ }\textbf {\bibinfo {volume} {102}},\ \bibinfo
  {pages} {113011} (\bibinfo {year} {2020})},\ \Eprint
  {https://arxiv.org/abs/2010.04215} {arXiv:2010.04215 [hep-ph]} \BibitemShut
  {NoStop}%
\bibitem [{\citenamefont {Choudhary}\ \emph {et~al.}(2022)\citenamefont
  {Choudhary}, \citenamefont {Gurjar}, \citenamefont {Chakrabarti},\ and\
  \citenamefont {Mukherjee}}]{Choudhary:2022den}%
  \BibitemOpen
  \bibfield  {author} {\bibinfo {author} {\bibfnamefont {P.}~\bibnamefont
  {Choudhary}}, \bibinfo {author} {\bibfnamefont {B.}~\bibnamefont {Gurjar}},
  \bibinfo {author} {\bibfnamefont {D.}~\bibnamefont {Chakrabarti}},\ and\
  \bibinfo {author} {\bibfnamefont {A.}~\bibnamefont {Mukherjee}},\ }\bibfield
  {title} {\bibinfo {title} {{Gravitational form factors and mechanical
  properties of the proton: Connections between distributions in 2D and 3D}},\
  }\href {https://doi.org/10.1103/PhysRevD.106.076004} {\bibfield  {journal}
  {\bibinfo  {journal} {Phys. Rev. D}\ }\textbf {\bibinfo {volume} {106}},\
  \bibinfo {pages} {076004} (\bibinfo {year} {2022})},\ \Eprint
  {https://arxiv.org/abs/2206.12206} {arXiv:2206.12206 [hep-ph]} \BibitemShut
  {NoStop}%
\bibitem [{\citenamefont {Lorc\'e}\ \emph {et~al.}(2022)\citenamefont
  {Lorc\'e}, \citenamefont {Schweitzer},\ and\ \citenamefont
  {Tezgin}}]{Lorce:2022cle}%
  \BibitemOpen
  \bibfield  {author} {\bibinfo {author} {\bibfnamefont {C.}~\bibnamefont
  {Lorc\'e}}, \bibinfo {author} {\bibfnamefont {P.}~\bibnamefont
  {Schweitzer}},\ and\ \bibinfo {author} {\bibfnamefont {K.}~\bibnamefont
  {Tezgin}},\ }\bibfield  {title} {\bibinfo {title} {{2D energy-momentum tensor
  distributions of nucleon in a large-Nc quark model from ultrarelativistic to
  nonrelativistic limit}},\ }\href
  {https://doi.org/10.1103/PhysRevD.106.014012} {\bibfield  {journal} {\bibinfo
   {journal} {Phys. Rev. D}\ }\textbf {\bibinfo {volume} {106}},\ \bibinfo
  {pages} {014012} (\bibinfo {year} {2022})},\ \Eprint
  {https://arxiv.org/abs/2202.01192} {arXiv:2202.01192 [hep-ph]} \BibitemShut
  {NoStop}%
\bibitem [{\citenamefont {Owa}\ \emph {et~al.}(2022)\citenamefont {Owa},
  \citenamefont {Thomas},\ and\ \citenamefont {Wang}}]{Owa:2021hnj}%
  \BibitemOpen
  \bibfield  {author} {\bibinfo {author} {\bibfnamefont {S.}~\bibnamefont
  {Owa}}, \bibinfo {author} {\bibfnamefont {A.~W.}\ \bibnamefont {Thomas}},\
  and\ \bibinfo {author} {\bibfnamefont {X.~G.}\ \bibnamefont {Wang}},\
  }\bibfield  {title} {\bibinfo {title} {{Effect of the pion field on the
  distributions of pressure and shear in the proton}},\ }\href
  {https://doi.org/10.1016/j.physletb.2022.137136} {\bibfield  {journal}
  {\bibinfo  {journal} {Phys. Lett. B}\ }\textbf {\bibinfo {volume} {829}},\
  \bibinfo {pages} {137136} (\bibinfo {year} {2022})},\ \Eprint
  {https://arxiv.org/abs/2106.00929} {arXiv:2106.00929 [hep-ph]} \BibitemShut
  {NoStop}%
\bibitem [{\citenamefont {Anikin}(2019)}]{Anikin:2019kwi}%
  \BibitemOpen
  \bibfield  {author} {\bibinfo {author} {\bibfnamefont {I.~V.}\ \bibnamefont
  {Anikin}},\ }\bibfield  {title} {\bibinfo {title} {{Gravitational form
  factors within light-cone sum rules at leading order}},\ }\href
  {https://doi.org/10.1103/PhysRevD.99.094026} {\bibfield  {journal} {\bibinfo
  {journal} {Phys. Rev. D}\ }\textbf {\bibinfo {volume} {99}},\ \bibinfo
  {pages} {094026} (\bibinfo {year} {2019})},\ \Eprint
  {https://arxiv.org/abs/1902.00094} {arXiv:1902.00094 [hep-ph]} \BibitemShut
  {NoStop}%
\bibitem [{\citenamefont {Won}\ \emph {et~al.}(2022)\citenamefont {Won},
  \citenamefont {Kim},\ and\ \citenamefont {Kim}}]{Won:2022cyy}%
  \BibitemOpen
  \bibfield  {author} {\bibinfo {author} {\bibfnamefont {H.-Y.}\ \bibnamefont
  {Won}}, \bibinfo {author} {\bibfnamefont {J.-Y.}\ \bibnamefont {Kim}},\ and\
  \bibinfo {author} {\bibfnamefont {H.-C.}\ \bibnamefont {Kim}},\ }\bibfield
  {title} {\bibinfo {title} {{Gravitational form factors of the baryon octet
  with flavor SU(3) symmetry breaking}},\ }\href
  {https://doi.org/10.1103/PhysRevD.106.114009} {\bibfield  {journal} {\bibinfo
   {journal} {Phys. Rev. D}\ }\textbf {\bibinfo {volume} {106}},\ \bibinfo
  {pages} {114009} (\bibinfo {year} {2022})},\ \Eprint
  {https://arxiv.org/abs/2210.03320} {arXiv:2210.03320 [hep-ph]} \BibitemShut
  {NoStop}%
\bibitem [{\citenamefont {Fiore}\ \emph {et~al.}(2021)\citenamefont {Fiore},
  \citenamefont {Jenkovszky},\ and\ \citenamefont
  {Oleksiienko}}]{Fiore:2021wuj}%
  \BibitemOpen
  \bibfield  {author} {\bibinfo {author} {\bibfnamefont {R.}~\bibnamefont
  {Fiore}}, \bibinfo {author} {\bibfnamefont {L.}~\bibnamefont {Jenkovszky}},\
  and\ \bibinfo {author} {\bibfnamefont {M.}~\bibnamefont {Oleksiienko}},\
  }\bibfield  {title} {\bibinfo {title} {{On matter and pressure distribution
  in nucleons}},\ }\href@noop {} {\  (\bibinfo {year} {2021})},\ \Eprint
  {https://arxiv.org/abs/2112.00605} {arXiv:2112.00605 [hep-ph]} \BibitemShut
  {NoStop}%
\bibitem [{\citenamefont {Mamo}\ and\ \citenamefont
  {Zahed}(2020)}]{Mamo:2019mka}%
  \BibitemOpen
  \bibfield  {author} {\bibinfo {author} {\bibfnamefont {K.~A.}\ \bibnamefont
  {Mamo}}\ and\ \bibinfo {author} {\bibfnamefont {I.}~\bibnamefont {Zahed}},\
  }\bibfield  {title} {\bibinfo {title} {{Diffractive photoproduction of
  $J/\psi$ and $\Upsilon$ using holographic QCD: gravitational form factors and
  GPD of gluons in the proton}},\ }\href
  {https://doi.org/10.1103/PhysRevD.101.086003} {\bibfield  {journal} {\bibinfo
   {journal} {Phys. Rev. D}\ }\textbf {\bibinfo {volume} {101}},\ \bibinfo
  {pages} {086003} (\bibinfo {year} {2020})},\ \Eprint
  {https://arxiv.org/abs/1910.04707} {arXiv:1910.04707 [hep-ph]} \BibitemShut
  {NoStop}%
\bibitem [{\citenamefont {Mamo}\ and\ \citenamefont
  {Zahed}(2022)}]{Mamo:2022eui}%
  \BibitemOpen
  \bibfield  {author} {\bibinfo {author} {\bibfnamefont {K.~A.}\ \bibnamefont
  {Mamo}}\ and\ \bibinfo {author} {\bibfnamefont {I.}~\bibnamefont {Zahed}},\
  }\bibfield  {title} {\bibinfo {title} {{J/\ensuremath{\psi} near threshold in
  holographic QCD: A and D gravitational form factors}},\ }\href
  {https://doi.org/10.1103/PhysRevD.106.086004} {\bibfield  {journal} {\bibinfo
   {journal} {Phys. Rev. D}\ }\textbf {\bibinfo {volume} {106}},\ \bibinfo
  {pages} {086004} (\bibinfo {year} {2022})},\ \Eprint
  {https://arxiv.org/abs/2204.08857} {arXiv:2204.08857 [hep-ph]} \BibitemShut
  {NoStop}%
\bibitem [{\citenamefont {Mamo}\ and\ \citenamefont
  {Zahed}(2021)}]{Mamo:2021krl}%
  \BibitemOpen
  \bibfield  {author} {\bibinfo {author} {\bibfnamefont {K.~A.}\ \bibnamefont
  {Mamo}}\ and\ \bibinfo {author} {\bibfnamefont {I.}~\bibnamefont {Zahed}},\
  }\bibfield  {title} {\bibinfo {title} {{Nucleon mass radii and distribution:
  Holographic QCD, Lattice QCD and GlueX data}},\ }\href
  {https://doi.org/10.1103/PhysRevD.103.094010} {\bibfield  {journal} {\bibinfo
   {journal} {Phys. Rev. D}\ }\textbf {\bibinfo {volume} {103}},\ \bibinfo
  {pages} {094010} (\bibinfo {year} {2021})},\ \Eprint
  {https://arxiv.org/abs/2103.03186} {arXiv:2103.03186 [hep-ph]} \BibitemShut
  {NoStop}%
\bibitem [{\citenamefont {de~T\'eramond}\ \emph {et~al.}(2021)\citenamefont
  {de~T\'eramond}, \citenamefont {Dosch}, \citenamefont {Liu}, \citenamefont
  {Sufian}, \citenamefont {Brodsky},\ and\ \citenamefont
  {Deur}}]{deTeramond:2021lxc}%
  \BibitemOpen
  \bibfield  {author} {\bibinfo {author} {\bibfnamefont {G.~F.}\ \bibnamefont
  {de~T\'eramond}}, \bibinfo {author} {\bibfnamefont {H.~G.}\ \bibnamefont
  {Dosch}}, \bibinfo {author} {\bibfnamefont {T.}~\bibnamefont {Liu}}, \bibinfo
  {author} {\bibfnamefont {R.~S.}\ \bibnamefont {Sufian}}, \bibinfo {author}
  {\bibfnamefont {S.~J.}\ \bibnamefont {Brodsky}},\ and\ \bibinfo {author}
  {\bibfnamefont {A.}~\bibnamefont {Deur}} (\bibinfo {collaboration} {HLFHS}),\
  }\bibfield  {title} {\bibinfo {title} {{Gluon matter distribution in the
  proton and pion from extended holographic light-front QCD}},\ }\href
  {https://doi.org/10.1103/PhysRevD.104.114005} {\bibfield  {journal} {\bibinfo
   {journal} {Phys. Rev. D}\ }\textbf {\bibinfo {volume} {104}},\ \bibinfo
  {pages} {114005} (\bibinfo {year} {2021})},\ \Eprint
  {https://arxiv.org/abs/2107.01231} {arXiv:2107.01231 [hep-ph]} \BibitemShut
  {NoStop}%
\bibitem [{\citenamefont {Alexandrou}\ \emph
  {et~al.}(2020{\natexlab{a}})\citenamefont {Alexandrou} \emph
  {et~al.}}]{Alexandrou:2019ali}%
  \BibitemOpen
  \bibfield  {author} {\bibinfo {author} {\bibfnamefont {C.}~\bibnamefont
  {Alexandrou}} \emph {et~al.},\ }\bibfield  {title} {\bibinfo {title}
  {{Moments of nucleon generalized parton distributions from lattice QCD
  simulations at physical pion mass}},\ }\href
  {https://doi.org/10.1103/PhysRevD.101.034519} {\bibfield  {journal} {\bibinfo
   {journal} {Phys. Rev. D}\ }\textbf {\bibinfo {volume} {101}},\ \bibinfo
  {pages} {034519} (\bibinfo {year} {2020}{\natexlab{a}})},\ \Eprint
  {https://arxiv.org/abs/1908.10706} {arXiv:1908.10706 [hep-lat]} \BibitemShut
  {NoStop}%
\bibitem [{\citenamefont {Shanahan}\ and\ \citenamefont
  {Detmold}(2019{\natexlab{a}})}]{Shanahan:2018nnv}%
  \BibitemOpen
  \bibfield  {author} {\bibinfo {author} {\bibfnamefont {P.}~\bibnamefont
  {Shanahan}}\ and\ \bibinfo {author} {\bibfnamefont {W.}~\bibnamefont
  {Detmold}},\ }\bibfield  {title} {\bibinfo {title} {{Pressure Distribution
  and Shear Forces inside the Proton}},\ }\href
  {https://doi.org/10.1103/PhysRevLett.122.072003} {\bibfield  {journal}
  {\bibinfo  {journal} {Phys. Rev. Lett.}\ }\textbf {\bibinfo {volume} {122}},\
  \bibinfo {pages} {072003} (\bibinfo {year} {2019}{\natexlab{a}})},\ \Eprint
  {https://arxiv.org/abs/1810.07589} {arXiv:1810.07589 [nucl-th]} \BibitemShut
  {NoStop}%
\bibitem [{\citenamefont {Shanahan}\ and\ \citenamefont
  {Detmold}(2019{\natexlab{b}})}]{Shanahan:2018pib}%
  \BibitemOpen
  \bibfield  {author} {\bibinfo {author} {\bibfnamefont {P.}~\bibnamefont
  {Shanahan}}\ and\ \bibinfo {author} {\bibfnamefont {W.}~\bibnamefont
  {Detmold}},\ }\bibfield  {title} {\bibinfo {title} {{Gluon gravitational form
  factors of the nucleon and the pion from lattice QCD}},\ }\href
  {https://doi.org/10.1103/PhysRevD.99.014511} {\bibfield  {journal} {\bibinfo
  {journal} {Phys. Rev. D}\ }\textbf {\bibinfo {volume} {99}},\ \bibinfo
  {pages} {014511} (\bibinfo {year} {2019}{\natexlab{b}})},\ \Eprint
  {https://arxiv.org/abs/1810.04626} {arXiv:1810.04626 [hep-lat]} \BibitemShut
  {NoStop}%
\bibitem [{\citenamefont {Pefkou}\ \emph {et~al.}(2022)\citenamefont {Pefkou},
  \citenamefont {Hackett},\ and\ \citenamefont {Shanahan}}]{Pefkou:2021fni}%
  \BibitemOpen
  \bibfield  {author} {\bibinfo {author} {\bibfnamefont {D.~A.}\ \bibnamefont
  {Pefkou}}, \bibinfo {author} {\bibfnamefont {D.~C.}\ \bibnamefont
  {Hackett}},\ and\ \bibinfo {author} {\bibfnamefont {P.~E.}\ \bibnamefont
  {Shanahan}},\ }\bibfield  {title} {\bibinfo {title} {{Gluon gravitational
  structure of hadrons of different spin}},\ }\href
  {https://doi.org/10.1103/PhysRevD.105.054509} {\bibfield  {journal} {\bibinfo
   {journal} {Phys. Rev. D}\ }\textbf {\bibinfo {volume} {105}},\ \bibinfo
  {pages} {054509} (\bibinfo {year} {2022})},\ \Eprint
  {https://arxiv.org/abs/2107.10368} {arXiv:2107.10368 [hep-lat]} \BibitemShut
  {NoStop}%
\bibitem [{\citenamefont {Bali}\ \emph {et~al.}(2019)\citenamefont {Bali},
  \citenamefont {Collins}, \citenamefont {G\"ockeler}, \citenamefont {R\"odl},
  \citenamefont {Sch\"afer},\ and\ \citenamefont {Sternbeck}}]{Bali:2018zgl}%
  \BibitemOpen
  \bibfield  {author} {\bibinfo {author} {\bibfnamefont {G.~S.}\ \bibnamefont
  {Bali}}, \bibinfo {author} {\bibfnamefont {S.}~\bibnamefont {Collins}},
  \bibinfo {author} {\bibfnamefont {M.}~\bibnamefont {G\"ockeler}}, \bibinfo
  {author} {\bibfnamefont {R.}~\bibnamefont {R\"odl}}, \bibinfo {author}
  {\bibfnamefont {A.}~\bibnamefont {Sch\"afer}},\ and\ \bibinfo {author}
  {\bibfnamefont {A.}~\bibnamefont {Sternbeck}},\ }\bibfield  {title} {\bibinfo
  {title} {{Nucleon generalized form factors from two-flavor lattice QCD}},\
  }\href {https://doi.org/10.1103/PhysRevD.100.014507} {\bibfield  {journal}
  {\bibinfo  {journal} {Phys. Rev. D}\ }\textbf {\bibinfo {volume} {100}},\
  \bibinfo {pages} {014507} (\bibinfo {year} {2019})},\ \Eprint
  {https://arxiv.org/abs/1812.08256} {arXiv:1812.08256 [hep-lat]} \BibitemShut
  {NoStop}%
\bibitem [{\citenamefont {Won}\ \emph {et~al.}(2023)\citenamefont {Won},
  \citenamefont {Kim},\ and\ \citenamefont {Kim}}]{Won:2023zmf}%
  \BibitemOpen
  \bibfield  {author} {\bibinfo {author} {\bibfnamefont {H.-Y.}\ \bibnamefont
  {Won}}, \bibinfo {author} {\bibfnamefont {H.-C.}\ \bibnamefont {Kim}},\ and\
  \bibinfo {author} {\bibfnamefont {J.-Y.}\ \bibnamefont {Kim}},\ }\bibfield
  {title} {\bibinfo {title} {{Mechanical structure of the nucleon and the
  baryon octet}},\ }\href@noop {} {\  (\bibinfo {year} {2023})},\ \Eprint
  {https://arxiv.org/abs/2310.04670} {arXiv:2310.04670 [hep-ph]} \BibitemShut
  {NoStop}%
\bibitem [{\citenamefont {Guo}\ \emph {et~al.}(2023{\natexlab{a}})\citenamefont
  {Guo}, \citenamefont {Ji},\ and\ \citenamefont {Yuan}}]{Guo:2023qgu}%
  \BibitemOpen
  \bibfield  {author} {\bibinfo {author} {\bibfnamefont {Y.}~\bibnamefont
  {Guo}}, \bibinfo {author} {\bibfnamefont {X.}~\bibnamefont {Ji}},\ and\
  \bibinfo {author} {\bibfnamefont {F.}~\bibnamefont {Yuan}},\ }\bibfield
  {title} {\bibinfo {title} {{Proton's gluon GPDs at large skewness and
  gravitational form factors from near threshold heavy quarkonium
  photo-production}},\ }\href@noop {} {\  (\bibinfo {year}
  {2023}{\natexlab{a}})},\ \Eprint {https://arxiv.org/abs/2308.13006}
  {arXiv:2308.13006 [hep-ph]} \BibitemShut {NoStop}%
\bibitem [{\citenamefont {Burkert}\ \emph {et~al.}(2018)\citenamefont
  {Burkert}, \citenamefont {Elouadrhiri},\ and\ \citenamefont
  {Girod}}]{Burkert:2018bqq}%
  \BibitemOpen
  \bibfield  {author} {\bibinfo {author} {\bibfnamefont {V.}~\bibnamefont
  {Burkert}}, \bibinfo {author} {\bibfnamefont {L.}~\bibnamefont
  {Elouadrhiri}},\ and\ \bibinfo {author} {\bibfnamefont {F.}~\bibnamefont
  {Girod}},\ }\bibfield  {title} {\bibinfo {title} {{The pressure distribution
  inside the proton}},\ }\href {https://doi.org/10.1038/s41586-018-0060-z}
  {\bibfield  {journal} {\bibinfo  {journal} {Nature}\ }\textbf {\bibinfo
  {volume} {557}},\ \bibinfo {pages} {396} (\bibinfo {year}
  {2018})}\BibitemShut {NoStop}%
\bibitem [{\citenamefont {Lorc\'e}\ \emph {et~al.}(2019)\citenamefont
  {Lorc\'e}, \citenamefont {Moutarde},\ and\ \citenamefont
  {Trawi\'nski}}]{Lorce:2018egm}%
  \BibitemOpen
  \bibfield  {author} {\bibinfo {author} {\bibfnamefont {C.}~\bibnamefont
  {Lorc\'e}}, \bibinfo {author} {\bibfnamefont {H.}~\bibnamefont {Moutarde}},\
  and\ \bibinfo {author} {\bibfnamefont {A.~P.}\ \bibnamefont {Trawi\'nski}},\
  }\bibfield  {title} {\bibinfo {title} {{Revisiting the mechanical properties
  of the nucleon}},\ }\href {https://doi.org/10.1140/epjc/s10052-019-6572-3}
  {\bibfield  {journal} {\bibinfo  {journal} {Eur. Phys. J. C}\ }\textbf
  {\bibinfo {volume} {79}},\ \bibinfo {pages} {89} (\bibinfo {year} {2019})},\
  \Eprint {https://arxiv.org/abs/1810.09837} {arXiv:1810.09837 [hep-ph]}
  \BibitemShut {NoStop}%
\bibitem [{\citenamefont {Polyakov}(2003)}]{Polyakov:2002yz}%
  \BibitemOpen
  \bibfield  {author} {\bibinfo {author} {\bibfnamefont {M.}~\bibnamefont
  {Polyakov}},\ }\bibfield  {title} {\bibinfo {title} {{Generalized parton
  distributions and strong forces inside nucleons and nuclei}},\ }\href
  {https://doi.org/10.1016/S0370-2693(03)00036-4} {\bibfield  {journal}
  {\bibinfo  {journal} {Phys. Lett. B}\ }\textbf {\bibinfo {volume} {555}},\
  \bibinfo {pages} {57} (\bibinfo {year} {2003})},\ \Eprint
  {https://arxiv.org/abs/hep-ph/0210165} {arXiv:hep-ph/0210165} \BibitemShut
  {NoStop}%
\bibitem [{\citenamefont {Polyakov}\ and\ \citenamefont
  {Schweitzer}(2018)}]{Polyakov:2018zvc}%
  \BibitemOpen
  \bibfield  {author} {\bibinfo {author} {\bibfnamefont {M.~V.}\ \bibnamefont
  {Polyakov}}\ and\ \bibinfo {author} {\bibfnamefont {P.}~\bibnamefont
  {Schweitzer}},\ }\bibfield  {title} {\bibinfo {title} {{Forces inside
  hadrons: pressure, surface tension, mechanical radius, and all that}},\
  }\href {https://doi.org/10.1142/S0217751X18300259} {\bibfield  {journal}
  {\bibinfo  {journal} {Int. J. Mod. Phys. A}\ }\textbf {\bibinfo {volume}
  {33}},\ \bibinfo {pages} {1830025} (\bibinfo {year} {2018})},\ \Eprint
  {https://arxiv.org/abs/1805.06596} {arXiv:1805.06596 [hep-ph]} \BibitemShut
  {NoStop}%
\bibitem [{\citenamefont {Burkert}\ \emph
  {et~al.}(2023{\natexlab{a}})\citenamefont {Burkert}, \citenamefont
  {Elouadrhiri}, \citenamefont {Girod}, \citenamefont {Lorc\'e}, \citenamefont
  {Schweitzer},\ and\ \citenamefont {Shanahan}}]{Burkert:2023wzr}%
  \BibitemOpen
  \bibfield  {author} {\bibinfo {author} {\bibfnamefont {V.~D.}\ \bibnamefont
  {Burkert}}, \bibinfo {author} {\bibfnamefont {L.}~\bibnamefont
  {Elouadrhiri}}, \bibinfo {author} {\bibfnamefont {F.~X.}\ \bibnamefont
  {Girod}}, \bibinfo {author} {\bibfnamefont {C.}~\bibnamefont {Lorc\'e}},
  \bibinfo {author} {\bibfnamefont {P.}~\bibnamefont {Schweitzer}},\ and\
  \bibinfo {author} {\bibfnamefont {P.~E.}\ \bibnamefont {Shanahan}},\
  }\bibfield  {title} {\bibinfo {title} {{Colloquium: Gravitational Form
  Factors of the Proton}},\ }\href@noop {} {\  (\bibinfo {year}
  {2023}{\natexlab{a}})},\ \Eprint {https://arxiv.org/abs/2303.08347}
  {arXiv:2303.08347 [hep-ph]} \BibitemShut {NoStop}%
\bibitem [{\citenamefont {Nielsen}(1977)}]{Nielsen:1977sy}%
  \BibitemOpen
  \bibfield  {author} {\bibinfo {author} {\bibfnamefont {N.~K.}\ \bibnamefont
  {Nielsen}},\ }\bibfield  {title} {\bibinfo {title} {{The Energy Momentum
  Tensor in a Nonabelian Quark Gluon Theory}},\ }\href
  {https://doi.org/10.1016/0550-3213(77)90040-2} {\bibfield  {journal}
  {\bibinfo  {journal} {Nucl. Phys. B}\ }\textbf {\bibinfo {volume} {120}},\
  \bibinfo {pages} {212} (\bibinfo {year} {1977})}\BibitemShut {NoStop}%
\bibitem [{\citenamefont {Belinfante}(1962)}]{Belinfante}%
  \BibitemOpen
  \bibfield  {author} {\bibinfo {author} {\bibfnamefont {F.~J.}\ \bibnamefont
  {Belinfante}},\ }\bibfield  {title} {\bibinfo {title} {Consequences of the
  postulate of a complete commuting set of observables in quantum
  electrodynamics},\ }\href {https://doi.org/10.1103/PhysRev.128.2832}
  {\bibfield  {journal} {\bibinfo  {journal} {Phys. Rev.}\ }\textbf {\bibinfo
  {volume} {128}},\ \bibinfo {pages} {2832} (\bibinfo {year}
  {1962})}\BibitemShut {NoStop}%
\bibitem [{\citenamefont {M\"uller}\ \emph {et~al.}(1994)\citenamefont
  {M\"uller}, \citenamefont {Robaschik}, \citenamefont {Geyer}, \citenamefont
  {Dittes},\ and\ \citenamefont {Ho\v{r}ej\v{s}i}}]{Muller:1994ses}%
  \BibitemOpen
  \bibfield  {author} {\bibinfo {author} {\bibfnamefont {D.}~\bibnamefont
  {M\"uller}}, \bibinfo {author} {\bibfnamefont {D.}~\bibnamefont {Robaschik}},
  \bibinfo {author} {\bibfnamefont {B.}~\bibnamefont {Geyer}}, \bibinfo
  {author} {\bibfnamefont {F.~M.}\ \bibnamefont {Dittes}},\ and\ \bibinfo
  {author} {\bibfnamefont {J.}~\bibnamefont {Ho\v{r}ej\v{s}i}},\ }\bibfield
  {title} {\bibinfo {title} {{Wave functions, evolution equations and evolution
  kernels from light ray operators of QCD}},\ }\href
  {https://doi.org/10.1002/prop.2190420202} {\bibfield  {journal} {\bibinfo
  {journal} {Fortsch. Phys.}\ }\textbf {\bibinfo {volume} {42}},\ \bibinfo
  {pages} {101} (\bibinfo {year} {1994})},\ \Eprint
  {https://arxiv.org/abs/hep-ph/9812448} {arXiv:hep-ph/9812448} \BibitemShut
  {NoStop}%
\bibitem [{\citenamefont {Ji}(1997)}]{Ji:1996ek}%
  \BibitemOpen
  \bibfield  {author} {\bibinfo {author} {\bibfnamefont {X.-D.}\ \bibnamefont
  {Ji}},\ }\bibfield  {title} {\bibinfo {title} {{Gauge-Invariant Decomposition
  of Nucleon Spin}},\ }\href {https://doi.org/10.1103/PhysRevLett.78.610}
  {\bibfield  {journal} {\bibinfo  {journal} {Phys. Rev. Lett.}\ }\textbf
  {\bibinfo {volume} {78}},\ \bibinfo {pages} {610} (\bibinfo {year} {1997})},\
  \Eprint {https://arxiv.org/abs/hep-ph/9603249} {arXiv:hep-ph/9603249}
  \BibitemShut {NoStop}%
\bibitem [{\citenamefont {Radyushkin}(1996)}]{Radyushkin:1996nd}%
  \BibitemOpen
  \bibfield  {author} {\bibinfo {author} {\bibfnamefont {A.~V.}\ \bibnamefont
  {Radyushkin}},\ }\bibfield  {title} {\bibinfo {title} {{Scaling limit of
  deeply virtual Compton scattering}},\ }\href
  {https://doi.org/10.1016/0370-2693(96)00528-X} {\bibfield  {journal}
  {\bibinfo  {journal} {Phys. Lett. B}\ }\textbf {\bibinfo {volume} {380}},\
  \bibinfo {pages} {417} (\bibinfo {year} {1996})},\ \Eprint
  {https://arxiv.org/abs/hep-ph/9604317} {arXiv:hep-ph/9604317} \BibitemShut
  {NoStop}%
\bibitem [{\citenamefont {Bakker}\ \emph {et~al.}(2004)\citenamefont {Bakker},
  \citenamefont {Leader},\ and\ \citenamefont {Trueman}}]{Bakker:2004ib}%
  \BibitemOpen
  \bibfield  {author} {\bibinfo {author} {\bibfnamefont {B.~L.~G.}\
  \bibnamefont {Bakker}}, \bibinfo {author} {\bibfnamefont {E.}~\bibnamefont
  {Leader}},\ and\ \bibinfo {author} {\bibfnamefont {T.~L.}\ \bibnamefont
  {Trueman}},\ }\bibfield  {title} {\bibinfo {title} {{A Critique of the
  angular momentum sum rules and a new angular momentum sum rule}},\ }\href
  {https://doi.org/10.1103/PhysRevD.70.114001} {\bibfield  {journal} {\bibinfo
  {journal} {Phys. Rev. D}\ }\textbf {\bibinfo {volume} {70}},\ \bibinfo
  {pages} {114001} (\bibinfo {year} {2004})},\ \Eprint
  {https://arxiv.org/abs/hep-ph/0406139} {arXiv:hep-ph/0406139} \BibitemShut
  {NoStop}%
\bibitem [{\citenamefont {Kobzarev}\ and\ \citenamefont
  {Okun}(1962)}]{Kobzarev:1962wt}%
  \BibitemOpen
  \bibfield  {author} {\bibinfo {author} {\bibfnamefont {I.~Y.}\ \bibnamefont
  {Kobzarev}}\ and\ \bibinfo {author} {\bibfnamefont {L.~B.}\ \bibnamefont
  {Okun}},\ }\bibfield  {title} {\bibinfo {title} {{GRAVITATIONAL INTERACTION
  OF FERMIONS}},\ }\href@noop {} {\bibfield  {journal} {\bibinfo  {journal}
  {Zh. Eksp. Teor. Fiz.}\ }\textbf {\bibinfo {volume} {43}},\ \bibinfo {pages}
  {1904} (\bibinfo {year} {1962})}\BibitemShut {NoStop}%
\bibitem [{\citenamefont {Pagels}(1966)}]{Pagels:1966zza}%
  \BibitemOpen
  \bibfield  {author} {\bibinfo {author} {\bibfnamefont {H.}~\bibnamefont
  {Pagels}},\ }\bibfield  {title} {\bibinfo {title} {{Energy-Momentum Structure
  Form Factors of Particles}},\ }\href
  {https://doi.org/10.1103/PhysRev.144.1250} {\bibfield  {journal} {\bibinfo
  {journal} {Phys. Rev.}\ }\textbf {\bibinfo {volume} {144}},\ \bibinfo {pages}
  {1250} (\bibinfo {year} {1966})}\BibitemShut {NoStop}%
\bibitem [{\citenamefont {Polyakov}\ and\ \citenamefont
  {Weiss}(1999)}]{Polyakov:1999gs}%
  \BibitemOpen
  \bibfield  {author} {\bibinfo {author} {\bibfnamefont {M.~V.}\ \bibnamefont
  {Polyakov}}\ and\ \bibinfo {author} {\bibfnamefont {C.}~\bibnamefont
  {Weiss}},\ }\bibfield  {title} {\bibinfo {title} {{Skewed and double
  distributions in pion and nucleon}},\ }\href
  {https://doi.org/10.1103/PhysRevD.60.114017} {\bibfield  {journal} {\bibinfo
  {journal} {Phys. Rev. D}\ }\textbf {\bibinfo {volume} {60}},\ \bibinfo
  {pages} {114017} (\bibinfo {year} {1999})},\ \Eprint
  {https://arxiv.org/abs/hep-ph/9902451} {arXiv:hep-ph/9902451} \BibitemShut
  {NoStop}%
\bibitem [{\citenamefont {Fan}\ \emph {et~al.}(2023)\citenamefont {Fan},
  \citenamefont {Lin},\ and\ \citenamefont {Zeilbeck}}]{Fan:2022qve}%
  \BibitemOpen
  \bibfield  {author} {\bibinfo {author} {\bibfnamefont {Z.}~\bibnamefont
  {Fan}}, \bibinfo {author} {\bibfnamefont {H.-W.}\ \bibnamefont {Lin}},\ and\
  \bibinfo {author} {\bibfnamefont {M.}~\bibnamefont {Zeilbeck}},\ }\bibfield
  {title} {\bibinfo {title} {{Nonperturbatively renormalized nucleon gluon
  momentum fraction in the continuum limit of Nf=2+1+1 lattice QCD}},\ }\href
  {https://doi.org/10.1103/PhysRevD.107.034505} {\bibfield  {journal} {\bibinfo
   {journal} {Phys. Rev. D}\ }\textbf {\bibinfo {volume} {107}},\ \bibinfo
  {pages} {034505} (\bibinfo {year} {2023})},\ \Eprint
  {https://arxiv.org/abs/2208.00980} {arXiv:2208.00980 [hep-lat]} \BibitemShut
  {NoStop}%
\bibitem [{\citenamefont {Ethier}\ and\ \citenamefont
  {Nocera}(2020)}]{Ethier:2020way}%
  \BibitemOpen
  \bibfield  {author} {\bibinfo {author} {\bibfnamefont {J.~J.}\ \bibnamefont
  {Ethier}}\ and\ \bibinfo {author} {\bibfnamefont {E.~R.}\ \bibnamefont
  {Nocera}},\ }\bibfield  {title} {\bibinfo {title} {{Parton Distributions in
  Nucleons and Nuclei}},\ }\href
  {https://doi.org/10.1146/annurev-nucl-011720-042725} {\bibfield  {journal}
  {\bibinfo  {journal} {Ann. Rev. Nucl. Part. Sci.}\ }\textbf {\bibinfo
  {volume} {70}},\ \bibinfo {pages} {43} (\bibinfo {year} {2020})},\ \Eprint
  {https://arxiv.org/abs/2001.07722} {arXiv:2001.07722 [hep-ph]} \BibitemShut
  {NoStop}%
\bibitem [{\citenamefont {Detmold}\ \emph {et~al.}(2019)\citenamefont
  {Detmold}, \citenamefont {Edwards}, \citenamefont {Dudek}, \citenamefont
  {Engelhardt}, \citenamefont {Lin}, \citenamefont {Meinel}, \citenamefont
  {Orginos},\ and\ \citenamefont {Shanahan}}]{Detmold:2019ghl}%
  \BibitemOpen
  \bibfield  {author} {\bibinfo {author} {\bibfnamefont {W.}~\bibnamefont
  {Detmold}}, \bibinfo {author} {\bibfnamefont {R.~G.}\ \bibnamefont
  {Edwards}}, \bibinfo {author} {\bibfnamefont {J.~J.}\ \bibnamefont {Dudek}},
  \bibinfo {author} {\bibfnamefont {M.}~\bibnamefont {Engelhardt}}, \bibinfo
  {author} {\bibfnamefont {H.-W.}\ \bibnamefont {Lin}}, \bibinfo {author}
  {\bibfnamefont {S.}~\bibnamefont {Meinel}}, \bibinfo {author} {\bibfnamefont
  {K.}~\bibnamefont {Orginos}},\ and\ \bibinfo {author} {\bibfnamefont
  {P.}~\bibnamefont {Shanahan}} (\bibinfo {collaboration} {USQCD}),\ }\bibfield
   {title} {\bibinfo {title} {{Hadrons and Nuclei}},\ }\href
  {https://doi.org/10.1140/epja/i2019-12902-4} {\bibfield  {journal} {\bibinfo
  {journal} {Eur. Phys. J. A}\ }\textbf {\bibinfo {volume} {55}},\ \bibinfo
  {pages} {193} (\bibinfo {year} {2019})},\ \Eprint
  {https://arxiv.org/abs/1904.09512} {arXiv:1904.09512 [hep-lat]} \BibitemShut
  {NoStop}%
\bibitem [{\citenamefont {Liu}(2022)}]{Liu:2021lke}%
  \BibitemOpen
  \bibfield  {author} {\bibinfo {author} {\bibfnamefont {K.-F.}\ \bibnamefont
  {Liu}},\ }\bibfield  {title} {\bibinfo {title} {{Status on lattice
  calculations of the proton spin decomposition}},\ }\href
  {https://doi.org/10.1007/s43673-022-00037-4} {\bibfield  {journal} {\bibinfo
  {journal} {AAPPS Bull.}\ }\textbf {\bibinfo {volume} {32}},\ \bibinfo {pages}
  {8} (\bibinfo {year} {2022})},\ \Eprint {https://arxiv.org/abs/2112.08416}
  {arXiv:2112.08416 [hep-lat]} \BibitemShut {NoStop}%
\bibitem [{\citenamefont {Duran}\ \emph {et~al.}(2023)\citenamefont {Duran}
  \emph {et~al.}}]{Duran:2022xag}%
  \BibitemOpen
  \bibfield  {author} {\bibinfo {author} {\bibfnamefont {B.}~\bibnamefont
  {Duran}} \emph {et~al.},\ }\bibfield  {title} {\bibinfo {title} {{Determining
  the gluonic gravitational form factors of the proton}},\ }\href
  {https://doi.org/10.1038/s41586-023-05730-4} {\bibfield  {journal} {\bibinfo
  {journal} {Nature}\ }\textbf {\bibinfo {volume} {615}},\ \bibinfo {pages}
  {813} (\bibinfo {year} {2023})},\ \Eprint {https://arxiv.org/abs/2207.05212}
  {arXiv:2207.05212 [nucl-ex]} \BibitemShut {NoStop}%
\bibitem [{\citenamefont {Hagler}\ \emph {et~al.}(2008)\citenamefont {Hagler}
  \emph {et~al.}}]{LHPC:2007blg}%
  \BibitemOpen
  \bibfield  {author} {\bibinfo {author} {\bibfnamefont {P.}~\bibnamefont
  {Hagler}} \emph {et~al.} (\bibinfo {collaboration} {LHPC}),\ }\bibfield
  {title} {\bibinfo {title} {{Nucleon Generalized Parton Distributions from
  Full Lattice QCD}},\ }\href {https://doi.org/10.1103/PhysRevD.77.094502}
  {\bibfield  {journal} {\bibinfo  {journal} {Phys. Rev. D}\ }\textbf {\bibinfo
  {volume} {77}},\ \bibinfo {pages} {094502} (\bibinfo {year} {2008})},\
  \Eprint {https://arxiv.org/abs/0705.4295} {arXiv:0705.4295 [hep-lat]}
  \BibitemShut {NoStop}%
\bibitem [{\citenamefont {Lin}(2021)}]{Lin:2020rxa}%
  \BibitemOpen
  \bibfield  {author} {\bibinfo {author} {\bibfnamefont {H.-W.}\ \bibnamefont
  {Lin}},\ }\bibfield  {title} {\bibinfo {title} {{Nucleon Tomography and
  Generalized Parton Distribution at Physical Pion Mass from Lattice QCD}},\
  }\href {https://doi.org/10.1103/PhysRevLett.127.182001} {\bibfield  {journal}
  {\bibinfo  {journal} {Phys. Rev. Lett.}\ }\textbf {\bibinfo {volume} {127}},\
  \bibinfo {pages} {182001} (\bibinfo {year} {2021})},\ \Eprint
  {https://arxiv.org/abs/2008.12474} {arXiv:2008.12474 [hep-ph]} \BibitemShut
  {NoStop}%
\bibitem [{\citenamefont {Alexandrou}\ \emph {et~al.}(2023)\citenamefont
  {Alexandrou} \emph {et~al.}}]{Alexandrou:2022dtc}%
  \BibitemOpen
  \bibfield  {author} {\bibinfo {author} {\bibfnamefont {C.}~\bibnamefont
  {Alexandrou}} \emph {et~al.},\ }\bibfield  {title} {\bibinfo {title}
  {{Moments of the nucleon transverse quark spin densities using lattice
  QCD}},\ }\href {https://doi.org/10.1103/PhysRevD.107.054504} {\bibfield
  {journal} {\bibinfo  {journal} {Phys. Rev. D}\ }\textbf {\bibinfo {volume}
  {107}},\ \bibinfo {pages} {054504} (\bibinfo {year} {2023})},\ \Eprint
  {https://arxiv.org/abs/2202.09871} {arXiv:2202.09871 [hep-lat]} \BibitemShut
  {NoStop}%
\bibitem [{\citenamefont {Guo}\ \emph {et~al.}(2023{\natexlab{b}})\citenamefont
  {Guo}, \citenamefont {Ji}, \citenamefont {Liu},\ and\ \citenamefont
  {Yang}}]{Guo:2023pqw}%
  \BibitemOpen
  \bibfield  {author} {\bibinfo {author} {\bibfnamefont {Y.}~\bibnamefont
  {Guo}}, \bibinfo {author} {\bibfnamefont {X.}~\bibnamefont {Ji}}, \bibinfo
  {author} {\bibfnamefont {Y.}~\bibnamefont {Liu}},\ and\ \bibinfo {author}
  {\bibfnamefont {J.}~\bibnamefont {Yang}},\ }\bibfield  {title} {\bibinfo
  {title} {{Updated analysis of near-threshold heavy quarkonium production for
  probe of proton\textquoteright{}s gluonic gravitational form factors}},\
  }\href {https://doi.org/10.1103/PhysRevD.108.034003} {\bibfield  {journal}
  {\bibinfo  {journal} {Phys. Rev. D}\ }\textbf {\bibinfo {volume} {108}},\
  \bibinfo {pages} {034003} (\bibinfo {year} {2023}{\natexlab{b}})},\ \Eprint
  {https://arxiv.org/abs/2305.06992} {arXiv:2305.06992 [hep-ph]} \BibitemShut
  {NoStop}%
\bibitem [{\citenamefont {Edwards}\ \emph {et~al.}(2016)\citenamefont
  {Edwards}, \citenamefont {Gupta}, \citenamefont {Joó}, \citenamefont
  {Orginos}, \citenamefont {Richards}, \citenamefont {Winter},\ and\
  \citenamefont {Yoon}}]{ensembles}%
  \BibitemOpen
  \bibfield  {author} {\bibinfo {author} {\bibfnamefont {R.}~\bibnamefont
  {Edwards}}, \bibinfo {author} {\bibfnamefont {R.}~\bibnamefont {Gupta}},
  \bibinfo {author} {\bibfnamefont {N.}~\bibnamefont {Joó}}, \bibinfo {author}
  {\bibfnamefont {K.}~\bibnamefont {Orginos}}, \bibinfo {author} {\bibfnamefont
  {D.}~\bibnamefont {Richards}}, \bibinfo {author} {\bibfnamefont
  {F.}~\bibnamefont {Winter}},\ and\ \bibinfo {author} {\bibfnamefont
  {B.}~\bibnamefont {Yoon}},\ }\bibfield  {title} {\bibinfo {title} {U.s. 2+1
  flavor clover lattice generation program},\ }\href@noop {} {\bibfield
  {journal} {\bibinfo  {journal} {unpublished}\ } (\bibinfo {year}
  {2016})}\BibitemShut {NoStop}%
\bibitem [{\citenamefont {L{\"u}scher}\ and\ \citenamefont
  {Weisz}(1985)}]{Luscher:1984xn}%
  \BibitemOpen
  \bibfield  {author} {\bibinfo {author} {\bibfnamefont {M.}~\bibnamefont
  {L{\"u}scher}}\ and\ \bibinfo {author} {\bibfnamefont {P.}~\bibnamefont
  {Weisz}},\ }\bibfield  {title} {\bibinfo {title} {{On-Shell Improved Lattice
  Gauge Theories}},\ }\href {https://doi.org/10.1007/BF01206178} {\bibfield
  {journal} {\bibinfo  {journal} {Commun. Math. Phys.}\ }\textbf {\bibinfo
  {volume} {97}},\ \bibinfo {pages} {59} (\bibinfo {year} {1985})},\ \bibinfo
  {note} {[Erratum: Commun.Math.Phys. 98, 433 (1985)]}\BibitemShut {NoStop}%
\bibitem [{\citenamefont {Sheikholeslami}\ and\ \citenamefont
  {Wohlert}(1985)}]{Sheikholeslami:1985ij}%
  \BibitemOpen
  \bibfield  {author} {\bibinfo {author} {\bibfnamefont {B.}~\bibnamefont
  {Sheikholeslami}}\ and\ \bibinfo {author} {\bibfnamefont {R.}~\bibnamefont
  {Wohlert}},\ }\bibfield  {title} {\bibinfo {title} {{Improved Continuum Limit
  Lattice Action for QCD with Wilson Fermions}},\ }\href
  {https://doi.org/10.1016/0550-3213(85)90002-1} {\bibfield  {journal}
  {\bibinfo  {journal} {Nucl. Phys. B}\ }\textbf {\bibinfo {volume} {259}},\
  \bibinfo {pages} {572} (\bibinfo {year} {1985})}\BibitemShut {NoStop}%
\bibitem [{\citenamefont {Morningstar}\ and\ \citenamefont
  {Peardon}(2004)}]{Morningstar:2003gk}%
  \BibitemOpen
  \bibfield  {author} {\bibinfo {author} {\bibfnamefont {C.}~\bibnamefont
  {Morningstar}}\ and\ \bibinfo {author} {\bibfnamefont {M.~J.}\ \bibnamefont
  {Peardon}},\ }\bibfield  {title} {\bibinfo {title} {{Analytic smearing of
  SU(3) link variables in lattice QCD}},\ }\href
  {https://doi.org/10.1103/PhysRevD.69.054501} {\bibfield  {journal} {\bibinfo
  {journal} {Phys. Rev. D}\ }\textbf {\bibinfo {volume} {69}},\ \bibinfo
  {pages} {054501} (\bibinfo {year} {2004})},\ \Eprint
  {https://arxiv.org/abs/hep-lat/0311018} {arXiv:hep-lat/0311018} \BibitemShut
  {NoStop}%
\bibitem [{\citenamefont {Park}\ \emph {et~al.}(2022)\citenamefont {Park},
  \citenamefont {Gupta}, \citenamefont {Yoon}, \citenamefont {Mondal},
  \citenamefont {Bhattacharya}, \citenamefont {Jang}, \citenamefont {Joó},\
  and\ \citenamefont {Winter}}]{Park:2021ypf}%
  \BibitemOpen
  \bibfield  {author} {\bibinfo {author} {\bibfnamefont {S.}~\bibnamefont
  {Park}}, \bibinfo {author} {\bibfnamefont {R.}~\bibnamefont {Gupta}},
  \bibinfo {author} {\bibfnamefont {B.}~\bibnamefont {Yoon}}, \bibinfo {author}
  {\bibfnamefont {S.}~\bibnamefont {Mondal}}, \bibinfo {author} {\bibfnamefont
  {T.}~\bibnamefont {Bhattacharya}}, \bibinfo {author} {\bibfnamefont {Y.-C.}\
  \bibnamefont {Jang}}, \bibinfo {author} {\bibfnamefont {B.}~\bibnamefont
  {Joó}},\ and\ \bibinfo {author} {\bibfnamefont {F.}~\bibnamefont {Winter}}
  (\bibinfo {collaboration} {Nucleon Matrix Elements (NME)}),\ }\bibfield
  {title} {\bibinfo {title} {{Precision nucleon charges and form factors using
  (2+1)-flavor lattice QCD}},\ }\href
  {https://doi.org/10.1103/PhysRevD.105.054505} {\bibfield  {journal} {\bibinfo
   {journal} {Phys. Rev. D}\ }\textbf {\bibinfo {volume} {105}},\ \bibinfo
  {pages} {054505} (\bibinfo {year} {2022})},\ \Eprint
  {https://arxiv.org/abs/2103.05599} {arXiv:2103.05599 [hep-lat]} \BibitemShut
  {NoStop}%
\bibitem [{\citenamefont {Borsanyi}\ \emph {et~al.}(2012)\citenamefont
  {Borsanyi} \emph {et~al.}}]{BMW:2012hcm}%
  \BibitemOpen
  \bibfield  {author} {\bibinfo {author} {\bibfnamefont {S.}~\bibnamefont
  {Borsanyi}} \emph {et~al.} (\bibinfo {collaboration} {BMW}),\ }\bibfield
  {title} {\bibinfo {title} {{High-precision scale setting in lattice QCD}},\
  }\href {https://doi.org/10.1007/JHEP09(2012)010} {\bibfield  {journal}
  {\bibinfo  {journal} {JHEP}\ }\textbf {\bibinfo {volume} {09}},\ \bibinfo
  {pages} {010}},\ \Eprint {https://arxiv.org/abs/1203.4469} {arXiv:1203.4469
  [hep-lat]} \BibitemShut {NoStop}%
\bibitem [{\citenamefont {Hackett}\ \emph {et~al.}(2023)\citenamefont
  {Hackett}, \citenamefont {Oare}, \citenamefont {Pefkou},\ and\ \citenamefont
  {Shanahan}}]{Hackett:2023nkr}%
  \BibitemOpen
  \bibfield  {author} {\bibinfo {author} {\bibfnamefont {D.~C.}\ \bibnamefont
  {Hackett}}, \bibinfo {author} {\bibfnamefont {P.~R.}\ \bibnamefont {Oare}},
  \bibinfo {author} {\bibfnamefont {D.~A.}\ \bibnamefont {Pefkou}},\ and\
  \bibinfo {author} {\bibfnamefont {P.~E.}\ \bibnamefont {Shanahan}},\
  }\bibfield  {title} {\bibinfo {title} {{Gravitational form factors of the
  pion from lattice QCD}},\ }\href
  {https://doi.org/10.1103/PhysRevD.108.114504} {\bibfield  {journal} {\bibinfo
   {journal} {Phys. Rev. D}\ }\textbf {\bibinfo {volume} {108}},\ \bibinfo
  {pages} {114504} (\bibinfo {year} {2023})},\ \Eprint
  {https://arxiv.org/abs/2307.11707} {arXiv:2307.11707 [hep-lat]} \BibitemShut
  {NoStop}%
\bibitem [{\citenamefont {L\"uscher}(2010)}]{Luscher:2010iy}%
  \BibitemOpen
  \bibfield  {author} {\bibinfo {author} {\bibfnamefont {M.}~\bibnamefont
  {L\"uscher}},\ }\bibfield  {title} {\bibinfo {title} {{Properties and uses of
  the Wilson flow in lattice QCD}},\ }\href
  {https://doi.org/10.1007/JHEP08(2010)071} {\bibfield  {journal} {\bibinfo
  {journal} {JHEP}\ }\textbf {\bibinfo {volume} {08}},\ \bibinfo {pages}
  {071}},\ \bibinfo {note} {[Erratum: JHEP 03, 092 (2014)]},\ \Eprint
  {https://arxiv.org/abs/1006.4518} {arXiv:1006.4518 [hep-lat]} \BibitemShut
  {NoStop}%
\bibitem [{\citenamefont {Narayanan}\ and\ \citenamefont
  {Neuberger}(2006)}]{Narayanan:2006rf}%
  \BibitemOpen
  \bibfield  {author} {\bibinfo {author} {\bibfnamefont {R.}~\bibnamefont
  {Narayanan}}\ and\ \bibinfo {author} {\bibfnamefont {H.}~\bibnamefont
  {Neuberger}},\ }\bibfield  {title} {\bibinfo {title} {{Infinite N phase
  transitions in continuum Wilson loop operators}},\ }\href
  {https://doi.org/10.1088/1126-6708/2006/03/064} {\bibfield  {journal}
  {\bibinfo  {journal} {JHEP}\ }\textbf {\bibinfo {volume} {03}},\ \bibinfo
  {pages} {064}},\ \Eprint {https://arxiv.org/abs/hep-th/0601210}
  {arXiv:hep-th/0601210} \BibitemShut {NoStop}%
\bibitem [{\citenamefont {Lohmayer}\ and\ \citenamefont
  {Neuberger}(2012)}]{Lohmayer:2012hs}%
  \BibitemOpen
  \bibfield  {author} {\bibinfo {author} {\bibfnamefont {R.}~\bibnamefont
  {Lohmayer}}\ and\ \bibinfo {author} {\bibfnamefont {H.}~\bibnamefont
  {Neuberger}},\ }\bibfield  {title} {\bibinfo {title} {{Continuous smearing of
  Wilson Loops.}},\ }\href {https://doi.org/10.22323/1.139.0249} {\bibfield
  {journal} {\bibinfo  {journal} {Proc. Sci.}\ }\textbf {\bibinfo {volume}
  {Lattice 2011}},\ \bibinfo {pages} {249} (\bibinfo {year} {2012})},\ \Eprint
  {https://arxiv.org/abs/1110.3522} {arXiv:1110.3522 [hep-lat]} \BibitemShut
  {NoStop}%
\bibitem [{\citenamefont {Hutchinson}(1990)}]{doi:10.1080/03610919008812866}%
  \BibitemOpen
  \bibfield  {author} {\bibinfo {author} {\bibfnamefont {M.}~\bibnamefont
  {Hutchinson}},\ }\bibfield  {title} {\bibinfo {title} {A stochastic estimator
  of the trace of the influence matrix for laplacian smoothing splines},\
  }\href {https://doi.org/10.1080/03610919008812866} {\bibfield  {journal}
  {\bibinfo  {journal} {Communications in Statistics - Simulation and
  Computation}\ }\textbf {\bibinfo {volume} {19}},\ \bibinfo {pages} {433}
  (\bibinfo {year} {1990})},\ \Eprint
  {https://arxiv.org/abs/https://doi.org/10.1080/03610919008812866}
  {https://doi.org/10.1080/03610919008812866} \BibitemShut {NoStop}%
\bibitem [{\citenamefont {Stathopoulos}\ \emph {et~al.}(2013)\citenamefont
  {Stathopoulos}, \citenamefont {Laeuchli},\ and\ \citenamefont
  {Orginos}}]{Stathopoulos:2013aci}%
  \BibitemOpen
  \bibfield  {author} {\bibinfo {author} {\bibfnamefont {A.}~\bibnamefont
  {Stathopoulos}}, \bibinfo {author} {\bibfnamefont {J.}~\bibnamefont
  {Laeuchli}},\ and\ \bibinfo {author} {\bibfnamefont {K.}~\bibnamefont
  {Orginos}},\ }\bibfield  {title} {\bibinfo {title} {{Hierarchical probing for
  estimating the trace of the matrix inverse on toroidal lattices}},\
  }\href@noop {} {\bibfield  {journal} {\bibinfo  {journal} {SIAM J. Sci.
  Comput.}\ }\textbf {\bibinfo {volume} {35}},\ \bibinfo {pages} {S299–S322}
  (\bibinfo {year} {2013})},\ \Eprint {https://arxiv.org/abs/1302.4018}
  {arXiv:1302.4018 [hep-lat]} \BibitemShut {NoStop}%
\bibitem [{\citenamefont {Gambhir}(2017)}]{Gambhir:2017the}%
  \BibitemOpen
  \bibfield  {author} {\bibinfo {author} {\bibfnamefont {A.~S.}\ \bibnamefont
  {Gambhir}},\ }\emph {\bibinfo {title} {{Disconnected Diagrams in Lattice
  QCD}}},\ \href {https://doi.org/10.21220/S2108D} {Ph.D. thesis},\ \bibinfo
  {school} {William-Mary Coll.} (\bibinfo {year} {2017})\BibitemShut {NoStop}%
\bibitem [{\citenamefont {Capitani}\ \emph {et~al.}(2012)\citenamefont
  {Capitani}, \citenamefont {Della~Morte}, \citenamefont {von Hippel},
  \citenamefont {J{\"a}ger}, \citenamefont {Juttner}, \citenamefont
  {Knippschild}, \citenamefont {Meyer},\ and\ \citenamefont
  {Wittig}}]{Capitani:2012gj}%
  \BibitemOpen
  \bibfield  {author} {\bibinfo {author} {\bibfnamefont {S.}~\bibnamefont
  {Capitani}}, \bibinfo {author} {\bibfnamefont {M.}~\bibnamefont
  {Della~Morte}}, \bibinfo {author} {\bibfnamefont {G.}~\bibnamefont {von
  Hippel}}, \bibinfo {author} {\bibfnamefont {B.}~\bibnamefont {J{\"a}ger}},
  \bibinfo {author} {\bibfnamefont {A.}~\bibnamefont {Juttner}}, \bibinfo
  {author} {\bibfnamefont {B.}~\bibnamefont {Knippschild}}, \bibinfo {author}
  {\bibfnamefont {H.~B.}\ \bibnamefont {Meyer}},\ and\ \bibinfo {author}
  {\bibfnamefont {H.}~\bibnamefont {Wittig}},\ }\bibfield  {title} {\bibinfo
  {title} {{The nucleon axial charge from lattice QCD with controlled
  errors}},\ }\href {https://doi.org/10.1103/PhysRevD.86.074502} {\bibfield
  {journal} {\bibinfo  {journal} {Phys. Rev. D}\ }\textbf {\bibinfo {volume}
  {86}},\ \bibinfo {pages} {074502} (\bibinfo {year} {2012})},\ \Eprint
  {https://arxiv.org/abs/1205.0180} {arXiv:1205.0180 [hep-lat]} \BibitemShut
  {NoStop}%
\bibitem [{\citenamefont {Maiani}\ \emph {et~al.}(1987)\citenamefont {Maiani},
  \citenamefont {Martinelli}, \citenamefont {Paciello},\ and\ \citenamefont
  {Taglienti}}]{Maiani:1987by}%
  \BibitemOpen
  \bibfield  {author} {\bibinfo {author} {\bibfnamefont {L.}~\bibnamefont
  {Maiani}}, \bibinfo {author} {\bibfnamefont {G.}~\bibnamefont {Martinelli}},
  \bibinfo {author} {\bibfnamefont {M.~L.}\ \bibnamefont {Paciello}},\ and\
  \bibinfo {author} {\bibfnamefont {B.}~\bibnamefont {Taglienti}},\ }\bibfield
  {title} {\bibinfo {title} {{Scalar Densities and Baryon Mass Differences in
  Lattice {QCD} With Wilson Fermions}},\ }\href
  {https://doi.org/10.1016/0550-3213(87)90078-2} {\bibfield  {journal}
  {\bibinfo  {journal} {Nucl. Phys. B}\ }\textbf {\bibinfo {volume} {293}},\
  \bibinfo {pages} {420} (\bibinfo {year} {1987})}\BibitemShut {NoStop}%
\bibitem [{\citenamefont {Dong}\ \emph {et~al.}(1998)\citenamefont {Dong},
  \citenamefont {Liu},\ and\ \citenamefont {Williams}}]{Dong:1997xr}%
  \BibitemOpen
  \bibfield  {author} {\bibinfo {author} {\bibfnamefont {S.~J.}\ \bibnamefont
  {Dong}}, \bibinfo {author} {\bibfnamefont {K.~F.}\ \bibnamefont {Liu}},\ and\
  \bibinfo {author} {\bibfnamefont {A.~G.}\ \bibnamefont {Williams}},\
  }\bibfield  {title} {\bibinfo {title} {{Lattice calculation of the
  strangeness magnetic moment of the nucleon}},\ }\href
  {https://doi.org/10.1103/PhysRevD.58.074504} {\bibfield  {journal} {\bibinfo
  {journal} {Phys. Rev. D}\ }\textbf {\bibinfo {volume} {58}},\ \bibinfo
  {pages} {074504} (\bibinfo {year} {1998})},\ \Eprint
  {https://arxiv.org/abs/hep-ph/9712483} {arXiv:hep-ph/9712483} \BibitemShut
  {NoStop}%
\bibitem [{\citenamefont {Djukanovic}\ \emph {et~al.}(2021)\citenamefont
  {Djukanovic}, \citenamefont {Harris}, \citenamefont {von Hippel},
  \citenamefont {Junnarkar}, \citenamefont {Meyer}, \citenamefont {Mohler},
  \citenamefont {Ottnad}, \citenamefont {Schulz}, \citenamefont {Wilhelm},\
  and\ \citenamefont {Wittig}}]{Djukanovic:2021cgp}%
  \BibitemOpen
  \bibfield  {author} {\bibinfo {author} {\bibfnamefont {D.}~\bibnamefont
  {Djukanovic}}, \bibinfo {author} {\bibfnamefont {T.}~\bibnamefont {Harris}},
  \bibinfo {author} {\bibfnamefont {G.}~\bibnamefont {von Hippel}}, \bibinfo
  {author} {\bibfnamefont {P.~M.}\ \bibnamefont {Junnarkar}}, \bibinfo {author}
  {\bibfnamefont {H.~B.}\ \bibnamefont {Meyer}}, \bibinfo {author}
  {\bibfnamefont {D.}~\bibnamefont {Mohler}}, \bibinfo {author} {\bibfnamefont
  {K.}~\bibnamefont {Ottnad}}, \bibinfo {author} {\bibfnamefont
  {T.}~\bibnamefont {Schulz}}, \bibinfo {author} {\bibfnamefont
  {J.}~\bibnamefont {Wilhelm}},\ and\ \bibinfo {author} {\bibfnamefont
  {H.}~\bibnamefont {Wittig}},\ }\bibfield  {title} {\bibinfo {title}
  {{Isovector electromagnetic form factors of the nucleon from lattice QCD and
  the proton radius puzzle}},\ }\href
  {https://doi.org/10.1103/PhysRevD.103.094522} {\bibfield  {journal} {\bibinfo
   {journal} {Phys. Rev. D}\ }\textbf {\bibinfo {volume} {103}},\ \bibinfo
  {pages} {094522} (\bibinfo {year} {2021})},\ \Eprint
  {https://arxiv.org/abs/2102.07460} {arXiv:2102.07460 [hep-lat]} \BibitemShut
  {NoStop}%
\bibitem [{\citenamefont {Akaike}(1998)}]{Akaike:1998zah}%
  \BibitemOpen
  \bibfield  {author} {\bibinfo {author} {\bibfnamefont {H.}~\bibnamefont
  {Akaike}},\ }\bibinfo {title} {{Information Theory and an Extension of the
  Maximum Likelihood Principle}}\ (\bibinfo  {publisher} {Springer
  Science+Business Media},\ \bibinfo {address} {New York},\ \bibinfo {year}
  {1998})\BibitemShut {NoStop}%
\bibitem [{\citenamefont {Jay}\ and\ \citenamefont {Neil}(2021)}]{Jay:2020jkz}%
  \BibitemOpen
  \bibfield  {author} {\bibinfo {author} {\bibfnamefont {W.~I.}\ \bibnamefont
  {Jay}}\ and\ \bibinfo {author} {\bibfnamefont {E.~T.}\ \bibnamefont {Neil}},\
  }\bibfield  {title} {\bibinfo {title} {{Bayesian model averaging for analysis
  of lattice field theory results}},\ }\href
  {https://doi.org/10.1103/PhysRevD.103.114502} {\bibfield  {journal} {\bibinfo
   {journal} {Phys. Rev. D}\ }\textbf {\bibinfo {volume} {103}},\ \bibinfo
  {pages} {114502} (\bibinfo {year} {2021})},\ \Eprint
  {https://arxiv.org/abs/2008.01069} {arXiv:2008.01069 [stat.ME]} \BibitemShut
  {NoStop}%
\bibitem [{\citenamefont {Rinaldi}\ \emph {et~al.}(2019)\citenamefont
  {Rinaldi}, \citenamefont {Syritsyn}, \citenamefont {Wagman}, \citenamefont
  {Buchoff}, \citenamefont {Schroeder},\ and\ \citenamefont
  {Wasem}}]{Rinaldi:2018osy}%
  \BibitemOpen
  \bibfield  {author} {\bibinfo {author} {\bibfnamefont {E.}~\bibnamefont
  {Rinaldi}}, \bibinfo {author} {\bibfnamefont {S.}~\bibnamefont {Syritsyn}},
  \bibinfo {author} {\bibfnamefont {M.~L.}\ \bibnamefont {Wagman}}, \bibinfo
  {author} {\bibfnamefont {M.~I.}\ \bibnamefont {Buchoff}}, \bibinfo {author}
  {\bibfnamefont {C.}~\bibnamefont {Schroeder}},\ and\ \bibinfo {author}
  {\bibfnamefont {J.}~\bibnamefont {Wasem}},\ }\bibfield  {title} {\bibinfo
  {title} {{Neutron-antineutron oscillations from lattice QCD}},\ }\href
  {https://doi.org/10.1103/PhysRevLett.122.162001} {\bibfield  {journal}
  {\bibinfo  {journal} {Phys. Rev. Lett.}\ }\textbf {\bibinfo {volume} {122}},\
  \bibinfo {pages} {162001} (\bibinfo {year} {2019})},\ \Eprint
  {https://arxiv.org/abs/1809.00246} {arXiv:1809.00246 [hep-lat]} \BibitemShut
  {NoStop}%
\bibitem [{\citenamefont {Beane}\ \emph {et~al.}(2015)\citenamefont {Beane},
  \citenamefont {Chang}, \citenamefont {Detmold}, \citenamefont {Orginos},
  \citenamefont {Parre\~no}, \citenamefont {Savage},\ and\ \citenamefont
  {Tiburzi}}]{Beane:2015yha}%
  \BibitemOpen
  \bibfield  {author} {\bibinfo {author} {\bibfnamefont {S.~R.}\ \bibnamefont
  {Beane}}, \bibinfo {author} {\bibfnamefont {E.}~\bibnamefont {Chang}},
  \bibinfo {author} {\bibfnamefont {W.}~\bibnamefont {Detmold}}, \bibinfo
  {author} {\bibfnamefont {K.}~\bibnamefont {Orginos}}, \bibinfo {author}
  {\bibfnamefont {A.}~\bibnamefont {Parre\~no}}, \bibinfo {author}
  {\bibfnamefont {M.~J.}\ \bibnamefont {Savage}},\ and\ \bibinfo {author}
  {\bibfnamefont {B.~C.}\ \bibnamefont {Tiburzi}} (\bibinfo {collaboration}
  {NPLQCD}),\ }\bibfield  {title} {\bibinfo {title} {{Ab initio Calculation of
  the np\textrightarrow{}d\ensuremath{\gamma} Radiative Capture Process}},\
  }\href {https://doi.org/10.1103/PhysRevLett.115.132001} {\bibfield  {journal}
  {\bibinfo  {journal} {Phys. Rev. Lett.}\ }\textbf {\bibinfo {volume} {115}},\
  \bibinfo {pages} {132001} (\bibinfo {year} {2015})},\ \Eprint
  {https://arxiv.org/abs/1505.02422} {arXiv:1505.02422 [hep-lat]} \BibitemShut
  {NoStop}%
\bibitem [{\citenamefont {Steinberg}(2019)}]{kmeans1d}%
  \BibitemOpen
  \bibfield  {author} {\bibinfo {author} {\bibfnamefont {D.}~\bibnamefont
  {Steinberg}},\ }\href@noop {} {\bibinfo {title} {kmeans1d}},\ \bibinfo
  {howpublished} {\url{https://github.com/dstein64/kmeans1d}} (\bibinfo {year}
  {2019})\BibitemShut {NoStop}%
\bibitem [{\citenamefont {Hou}\ \emph {et~al.}(2021)\citenamefont {Hou} \emph
  {et~al.}}]{Hou:2019efy}%
  \BibitemOpen
  \bibfield  {author} {\bibinfo {author} {\bibfnamefont {T.-J.}\ \bibnamefont
  {Hou}} \emph {et~al.},\ }\bibfield  {title} {\bibinfo {title} {{New CTEQ
  global analysis of quantum chromodynamics with high-precision data from the
  LHC}},\ }\href {https://doi.org/10.1103/PhysRevD.103.014013} {\bibfield
  {journal} {\bibinfo  {journal} {Phys. Rev. D}\ }\textbf {\bibinfo {volume}
  {103}},\ \bibinfo {pages} {014013} (\bibinfo {year} {2021})},\ \Eprint
  {https://arxiv.org/abs/1912.10053} {arXiv:1912.10053 [hep-ph]} \BibitemShut
  {NoStop}%
\bibitem [{\citenamefont {Hill}\ and\ \citenamefont {Paz}(2010)}]{Hill:2010yb}%
  \BibitemOpen
  \bibfield  {author} {\bibinfo {author} {\bibfnamefont {R.~J.}\ \bibnamefont
  {Hill}}\ and\ \bibinfo {author} {\bibfnamefont {G.}~\bibnamefont {Paz}},\
  }\bibfield  {title} {\bibinfo {title} {{Model independent extraction of the
  proton charge radius from electron scattering}},\ }\href
  {https://doi.org/10.1103/PhysRevD.82.113005} {\bibfield  {journal} {\bibinfo
  {journal} {Phys. Rev. D}\ }\textbf {\bibinfo {volume} {82}},\ \bibinfo
  {pages} {113005} (\bibinfo {year} {2010})},\ \Eprint
  {https://arxiv.org/abs/1008.4619} {arXiv:1008.4619 [hep-ph]} \BibitemShut
  {NoStop}%
\bibitem [{\citenamefont {Wang}\ \emph {et~al.}(2022)\citenamefont {Wang},
  \citenamefont {Yang}, \citenamefont {Liang}, \citenamefont {Draper},\ and\
  \citenamefont {Liu}}]{Wang:2021vqy}%
  \BibitemOpen
  \bibfield  {author} {\bibinfo {author} {\bibfnamefont {G.}~\bibnamefont
  {Wang}}, \bibinfo {author} {\bibfnamefont {Y.-B.}\ \bibnamefont {Yang}},
  \bibinfo {author} {\bibfnamefont {J.}~\bibnamefont {Liang}}, \bibinfo
  {author} {\bibfnamefont {T.}~\bibnamefont {Draper}},\ and\ \bibinfo {author}
  {\bibfnamefont {K.-F.}\ \bibnamefont {Liu}} (\bibinfo {collaboration}
  {\ensuremath{\chi}QCD}),\ }\bibfield  {title} {\bibinfo {title} {{Proton
  momentum and angular momentum decompositions with overlap fermions}},\ }\href
  {https://doi.org/10.1103/PhysRevD.106.014512} {\bibfield  {journal} {\bibinfo
   {journal} {Phys. Rev. D}\ }\textbf {\bibinfo {volume} {106}},\ \bibinfo
  {pages} {014512} (\bibinfo {year} {2022})},\ \Eprint
  {https://arxiv.org/abs/2111.09329} {arXiv:2111.09329 [hep-lat]} \BibitemShut
  {NoStop}%
\bibitem [{\citenamefont {Alexandrou}\ \emph
  {et~al.}(2020{\natexlab{b}})\citenamefont {Alexandrou}, \citenamefont
  {Bacchio}, \citenamefont {Constantinou}, \citenamefont {Finkenrath},
  \citenamefont {Hadjiyiannakou}, \citenamefont {Jansen}, \citenamefont
  {Koutsou}, \citenamefont {Panagopoulos},\ and\ \citenamefont
  {Spanoudes}}]{Alexandrou:2020sml}%
  \BibitemOpen
  \bibfield  {author} {\bibinfo {author} {\bibfnamefont {C.}~\bibnamefont
  {Alexandrou}}, \bibinfo {author} {\bibfnamefont {S.}~\bibnamefont {Bacchio}},
  \bibinfo {author} {\bibfnamefont {M.}~\bibnamefont {Constantinou}}, \bibinfo
  {author} {\bibfnamefont {J.}~\bibnamefont {Finkenrath}}, \bibinfo {author}
  {\bibfnamefont {K.}~\bibnamefont {Hadjiyiannakou}}, \bibinfo {author}
  {\bibfnamefont {K.}~\bibnamefont {Jansen}}, \bibinfo {author} {\bibfnamefont
  {G.}~\bibnamefont {Koutsou}}, \bibinfo {author} {\bibfnamefont
  {H.}~\bibnamefont {Panagopoulos}},\ and\ \bibinfo {author} {\bibfnamefont
  {G.}~\bibnamefont {Spanoudes}},\ }\bibfield  {title} {\bibinfo {title}
  {{Complete flavor decomposition of the spin and momentum fraction of the
  proton using lattice QCD simulations at physical pion mass}},\ }\href
  {https://doi.org/10.1103/PhysRevD.101.094513} {\bibfield  {journal} {\bibinfo
   {journal} {Phys. Rev. D}\ }\textbf {\bibinfo {volume} {101}},\ \bibinfo
  {pages} {094513} (\bibinfo {year} {2020}{\natexlab{b}})},\ \Eprint
  {https://arxiv.org/abs/2003.08486} {arXiv:2003.08486 [hep-lat]} \BibitemShut
  {NoStop}%
\bibitem [{\citenamefont {Gegelia}\ and\ \citenamefont
  {Polyakov}(2021)}]{Gegelia:2021wnj}%
  \BibitemOpen
  \bibfield  {author} {\bibinfo {author} {\bibfnamefont {J.}~\bibnamefont
  {Gegelia}}\ and\ \bibinfo {author} {\bibfnamefont {M.~V.}\ \bibnamefont
  {Polyakov}},\ }\bibfield  {title} {\bibinfo {title} {{A bound on the nucleon
  Druck-term from chiral EFT in curved space-time and mechanical stability
  conditions}},\ }\href {https://doi.org/10.1016/j.physletb.2021.136572}
  {\bibfield  {journal} {\bibinfo  {journal} {Phys. Lett. B}\ }\textbf
  {\bibinfo {volume} {820}},\ \bibinfo {pages} {136572} (\bibinfo {year}
  {2021})},\ \Eprint {https://arxiv.org/abs/2104.13954} {arXiv:2104.13954
  [hep-ph]} \BibitemShut {NoStop}%
\bibitem [{\citenamefont {Goeke}\ \emph
  {et~al.}(2007{\natexlab{a}})\citenamefont {Goeke}, \citenamefont {Grabis},
  \citenamefont {Ossmann}, \citenamefont {Polyakov}, \citenamefont
  {Schweitzer}, \citenamefont {Silva},\ and\ \citenamefont
  {Urbano}}]{Goeke:2007fp}%
  \BibitemOpen
  \bibfield  {author} {\bibinfo {author} {\bibfnamefont {K.}~\bibnamefont
  {Goeke}}, \bibinfo {author} {\bibfnamefont {J.}~\bibnamefont {Grabis}},
  \bibinfo {author} {\bibfnamefont {J.}~\bibnamefont {Ossmann}}, \bibinfo
  {author} {\bibfnamefont {M.~V.}\ \bibnamefont {Polyakov}}, \bibinfo {author}
  {\bibfnamefont {P.}~\bibnamefont {Schweitzer}}, \bibinfo {author}
  {\bibfnamefont {A.}~\bibnamefont {Silva}},\ and\ \bibinfo {author}
  {\bibfnamefont {D.}~\bibnamefont {Urbano}},\ }\bibfield  {title} {\bibinfo
  {title} {{Nucleon form-factors of the energy momentum tensor in the chiral
  quark-soliton model}},\ }\href {https://doi.org/10.1103/PhysRevD.75.094021}
  {\bibfield  {journal} {\bibinfo  {journal} {Phys. Rev. D}\ }\textbf {\bibinfo
  {volume} {75}},\ \bibinfo {pages} {094021} (\bibinfo {year}
  {2007}{\natexlab{a}})},\ \Eprint {https://arxiv.org/abs/hep-ph/0702030}
  {arXiv:hep-ph/0702030} \BibitemShut {NoStop}%
\bibitem [{\citenamefont {Schweitzer}\ \emph {et~al.}(2002)\citenamefont
  {Schweitzer}, \citenamefont {Boffi},\ and\ \citenamefont
  {Radici}}]{Schweitzer:2002nm}%
  \BibitemOpen
  \bibfield  {author} {\bibinfo {author} {\bibfnamefont {P.}~\bibnamefont
  {Schweitzer}}, \bibinfo {author} {\bibfnamefont {S.}~\bibnamefont {Boffi}},\
  and\ \bibinfo {author} {\bibfnamefont {M.}~\bibnamefont {Radici}},\
  }\bibfield  {title} {\bibinfo {title} {{Polynomiality of unpolarized off
  forward distribution functions and the D term in the chiral quark soliton
  model}},\ }\href {https://doi.org/10.1103/PhysRevD.66.114004} {\bibfield
  {journal} {\bibinfo  {journal} {Phys. Rev. D}\ }\textbf {\bibinfo {volume}
  {66}},\ \bibinfo {pages} {114004} (\bibinfo {year} {2002})},\ \Eprint
  {https://arxiv.org/abs/hep-ph/0207230} {arXiv:hep-ph/0207230} \BibitemShut
  {NoStop}%
\bibitem [{\citenamefont {Ossmann}\ \emph {et~al.}(2005)\citenamefont
  {Ossmann}, \citenamefont {Polyakov}, \citenamefont {Schweitzer},
  \citenamefont {Urbano},\ and\ \citenamefont {Goeke}}]{Ossmann:2004bp}%
  \BibitemOpen
  \bibfield  {author} {\bibinfo {author} {\bibfnamefont {J.}~\bibnamefont
  {Ossmann}}, \bibinfo {author} {\bibfnamefont {M.~V.}\ \bibnamefont
  {Polyakov}}, \bibinfo {author} {\bibfnamefont {P.}~\bibnamefont
  {Schweitzer}}, \bibinfo {author} {\bibfnamefont {D.}~\bibnamefont {Urbano}},\
  and\ \bibinfo {author} {\bibfnamefont {K.}~\bibnamefont {Goeke}},\ }\bibfield
   {title} {\bibinfo {title} {{The Generalized parton distribution function
  (E**u + E**d)(x,xi,t) of the nucleon in the chiral quark soliton model}},\
  }\href {https://doi.org/10.1103/PhysRevD.71.034011} {\bibfield  {journal}
  {\bibinfo  {journal} {Phys. Rev. D}\ }\textbf {\bibinfo {volume} {71}},\
  \bibinfo {pages} {034011} (\bibinfo {year} {2005})},\ \Eprint
  {https://arxiv.org/abs/hep-ph/0411172} {arXiv:hep-ph/0411172} \BibitemShut
  {NoStop}%
\bibitem [{\citenamefont {Goeke}\ \emph
  {et~al.}(2007{\natexlab{b}})\citenamefont {Goeke}, \citenamefont {Grabis},
  \citenamefont {Ossmann}, \citenamefont {Schweitzer}, \citenamefont {Silva},\
  and\ \citenamefont {Urbano}}]{Goeke:2007fq}%
  \BibitemOpen
  \bibfield  {author} {\bibinfo {author} {\bibfnamefont {K.}~\bibnamefont
  {Goeke}}, \bibinfo {author} {\bibfnamefont {J.}~\bibnamefont {Grabis}},
  \bibinfo {author} {\bibfnamefont {J.}~\bibnamefont {Ossmann}}, \bibinfo
  {author} {\bibfnamefont {P.}~\bibnamefont {Schweitzer}}, \bibinfo {author}
  {\bibfnamefont {A.}~\bibnamefont {Silva}},\ and\ \bibinfo {author}
  {\bibfnamefont {D.}~\bibnamefont {Urbano}},\ }\bibfield  {title} {\bibinfo
  {title} {{The pion mass dependence of the nucleon form-factors of the energy
  momentum tensor in the chiral quark-soliton model}},\ }\href
  {https://doi.org/10.1103/PhysRevC.75.055207} {\bibfield  {journal} {\bibinfo
  {journal} {Phys. Rev. C}\ }\textbf {\bibinfo {volume} {75}},\ \bibinfo
  {pages} {055207} (\bibinfo {year} {2007}{\natexlab{b}})},\ \Eprint
  {https://arxiv.org/abs/hep-ph/0702031} {arXiv:hep-ph/0702031} \BibitemShut
  {NoStop}%
\bibitem [{\citenamefont {Wakamatsu}(2007)}]{Wakamatsu:2007uc}%
  \BibitemOpen
  \bibfield  {author} {\bibinfo {author} {\bibfnamefont {M.}~\bibnamefont
  {Wakamatsu}},\ }\bibfield  {title} {\bibinfo {title} {{On the D-term of the
  nucleon generalized parton distributions}},\ }\href
  {https://doi.org/10.1016/j.physletb.2007.03.013} {\bibfield  {journal}
  {\bibinfo  {journal} {Phys. Lett. B}\ }\textbf {\bibinfo {volume} {648}},\
  \bibinfo {pages} {181} (\bibinfo {year} {2007})},\ \Eprint
  {https://arxiv.org/abs/hep-ph/0701057} {arXiv:hep-ph/0701057} \BibitemShut
  {NoStop}%
\bibitem [{\citenamefont {Cebulla}\ \emph {et~al.}(2007)\citenamefont
  {Cebulla}, \citenamefont {Goeke}, \citenamefont {Ossmann},\ and\
  \citenamefont {Schweitzer}}]{Cebulla:2007ei}%
  \BibitemOpen
  \bibfield  {author} {\bibinfo {author} {\bibfnamefont {C.}~\bibnamefont
  {Cebulla}}, \bibinfo {author} {\bibfnamefont {K.}~\bibnamefont {Goeke}},
  \bibinfo {author} {\bibfnamefont {J.}~\bibnamefont {Ossmann}},\ and\ \bibinfo
  {author} {\bibfnamefont {P.}~\bibnamefont {Schweitzer}},\ }\bibfield  {title}
  {\bibinfo {title} {{The Nucleon form-factors of the energy momentum tensor in
  the Skyrme model}},\ }\href {https://doi.org/10.1016/j.nuclphysa.2007.08.004}
  {\bibfield  {journal} {\bibinfo  {journal} {Nucl. Phys. A}\ }\textbf
  {\bibinfo {volume} {794}},\ \bibinfo {pages} {87} (\bibinfo {year} {2007})},\
  \Eprint {https://arxiv.org/abs/hep-ph/0703025} {arXiv:hep-ph/0703025}
  \BibitemShut {NoStop}%
\bibitem [{\citenamefont {Jung}\ \emph {et~al.}(2014)\citenamefont {Jung},
  \citenamefont {Yakhshiev},\ and\ \citenamefont {Kim}}]{Jung:2013bya}%
  \BibitemOpen
  \bibfield  {author} {\bibinfo {author} {\bibfnamefont {J.-H.}\ \bibnamefont
  {Jung}}, \bibinfo {author} {\bibfnamefont {U.}~\bibnamefont {Yakhshiev}},\
  and\ \bibinfo {author} {\bibfnamefont {H.-C.}\ \bibnamefont {Kim}},\
  }\bibfield  {title} {\bibinfo {title} {{Energy\textendash{}momentum tensor
  form factors of the nucleon within a
  \ensuremath{\pi}\textendash{}\ensuremath{\rho}\textendash{}\ensuremath{\omega}
  soliton model}},\ }\href {https://doi.org/10.1088/0954-3899/41/5/055107}
  {\bibfield  {journal} {\bibinfo  {journal} {J. Phys. G}\ }\textbf {\bibinfo
  {volume} {41}},\ \bibinfo {pages} {055107} (\bibinfo {year} {2014})},\
  \Eprint {https://arxiv.org/abs/1310.8064} {arXiv:1310.8064 [hep-ph]}
  \BibitemShut {NoStop}%
\bibitem [{\citenamefont {Chatagnon}\ \emph {et~al.}(2021)\citenamefont
  {Chatagnon} \emph {et~al.}}]{CLAS:2021lky}%
  \BibitemOpen
  \bibfield  {author} {\bibinfo {author} {\bibfnamefont {P.}~\bibnamefont
  {Chatagnon}} \emph {et~al.} (\bibinfo {collaboration} {CLAS}),\ }\bibfield
  {title} {\bibinfo {title} {{First Measurement of Timelike Compton
  Scattering}},\ }\href {https://doi.org/10.1103/PhysRevLett.127.262501}
  {\bibfield  {journal} {\bibinfo  {journal} {Phys. Rev. Lett.}\ }\textbf
  {\bibinfo {volume} {127}},\ \bibinfo {pages} {262501} (\bibinfo {year}
  {2021})},\ \Eprint {https://arxiv.org/abs/2108.11746} {arXiv:2108.11746
  [hep-ex]} \BibitemShut {NoStop}%
\bibitem [{\citenamefont {Guo}\ \emph {et~al.}(2021)\citenamefont {Guo},
  \citenamefont {Ji},\ and\ \citenamefont {Liu}}]{Guo:2021ibg}%
  \BibitemOpen
  \bibfield  {author} {\bibinfo {author} {\bibfnamefont {Y.}~\bibnamefont
  {Guo}}, \bibinfo {author} {\bibfnamefont {X.}~\bibnamefont {Ji}},\ and\
  \bibinfo {author} {\bibfnamefont {Y.}~\bibnamefont {Liu}},\ }\bibfield
  {title} {\bibinfo {title} {{QCD Analysis of Near-Threshold Photon-Proton
  Production of Heavy Quarkonium}},\ }\href
  {https://doi.org/10.1103/PhysRevD.103.096010} {\bibfield  {journal} {\bibinfo
   {journal} {Phys. Rev. D}\ }\textbf {\bibinfo {volume} {103}},\ \bibinfo
  {pages} {096010} (\bibinfo {year} {2021})},\ \Eprint
  {https://arxiv.org/abs/2103.11506} {arXiv:2103.11506 [hep-ph]} \BibitemShut
  {NoStop}%
\bibitem [{\citenamefont {Adhikari}\ \emph {et~al.}(2023)\citenamefont
  {Adhikari} \emph {et~al.}}]{GlueX:2023pev}%
  \BibitemOpen
  \bibfield  {author} {\bibinfo {author} {\bibfnamefont {S.}~\bibnamefont
  {Adhikari}} \emph {et~al.} (\bibinfo {collaboration} {GlueX}),\ }\bibfield
  {title} {\bibinfo {title} {{Measurement of the J/$\psi $ photoproduction
  cross section over the full near-threshold kinematic region}},\ }\href
  {https://doi.org/10.1103/PhysRevC.108.025201} {\bibfield  {journal} {\bibinfo
   {journal} {Phys. Rev. C}\ }\textbf {\bibinfo {volume} {108}},\ \bibinfo
  {pages} {025201} (\bibinfo {year} {2023})},\ \Eprint
  {https://arxiv.org/abs/2304.03845} {arXiv:2304.03845 [nucl-ex]} \BibitemShut
  {NoStop}%
\bibitem [{\citenamefont {Ji}\ and\ \citenamefont {Liu}(2022)}]{Ji:2021mfb}%
  \BibitemOpen
  \bibfield  {author} {\bibinfo {author} {\bibfnamefont {X.}~\bibnamefont
  {Ji}}\ and\ \bibinfo {author} {\bibfnamefont {Y.}~\bibnamefont {Liu}},\
  }\bibfield  {title} {\bibinfo {title} {{Momentum-Current Gravitational
  Multipoles of Hadrons}},\ }\href
  {https://doi.org/10.1103/PhysRevD.106.034028} {\bibfield  {journal} {\bibinfo
   {journal} {Phys. Rev. D}\ }\textbf {\bibinfo {volume} {106}},\ \bibinfo
  {pages} {034028} (\bibinfo {year} {2022})},\ \Eprint
  {https://arxiv.org/abs/2110.14781} {arXiv:2110.14781 [hep-ph]} \BibitemShut
  {NoStop}%
\bibitem [{\citenamefont {Jaffe}(2021)}]{Jaffe:2020ebz}%
  \BibitemOpen
  \bibfield  {author} {\bibinfo {author} {\bibfnamefont {R.~L.}\ \bibnamefont
  {Jaffe}},\ }\bibfield  {title} {\bibinfo {title} {{Ambiguities in the
  definition of local spatial densities in light hadrons}},\ }\href
  {https://doi.org/10.1103/PhysRevD.103.016017} {\bibfield  {journal} {\bibinfo
   {journal} {Phys. Rev. D}\ }\textbf {\bibinfo {volume} {103}},\ \bibinfo
  {pages} {016017} (\bibinfo {year} {2021})},\ \Eprint
  {https://arxiv.org/abs/2010.15887} {arXiv:2010.15887 [hep-ph]} \BibitemShut
  {NoStop}%
\bibitem [{\citenamefont {Panteleeva}\ \emph {et~al.}(2023)\citenamefont
  {Panteleeva}, \citenamefont {Epelbaum}, \citenamefont {Gegelia},\ and\
  \citenamefont {Mei\ss{}ner}}]{Panteleeva:2022uii}%
  \BibitemOpen
  \bibfield  {author} {\bibinfo {author} {\bibfnamefont {J.~Y.}\ \bibnamefont
  {Panteleeva}}, \bibinfo {author} {\bibfnamefont {E.}~\bibnamefont
  {Epelbaum}}, \bibinfo {author} {\bibfnamefont {J.}~\bibnamefont {Gegelia}},\
  and\ \bibinfo {author} {\bibfnamefont {U.~G.}\ \bibnamefont {Mei\ss{}ner}},\
  }\bibfield  {title} {\bibinfo {title} {{Definition of gravitational local
  spatial densities for spin-0 and spin-1/2 systems}},\ }\href
  {https://doi.org/10.1140/epjc/s10052-023-11746-x} {\bibfield  {journal}
  {\bibinfo  {journal} {Eur. Phys. J. C}\ }\textbf {\bibinfo {volume} {83}},\
  \bibinfo {pages} {617} (\bibinfo {year} {2023})},\ \Eprint
  {https://arxiv.org/abs/2211.09596} {arXiv:2211.09596 [hep-ph]} \BibitemShut
  {NoStop}%
\bibitem [{\citenamefont {Freese}\ and\ \citenamefont
  {Miller}(2022)}]{Freese:2021mzg}%
  \BibitemOpen
  \bibfield  {author} {\bibinfo {author} {\bibfnamefont {A.}~\bibnamefont
  {Freese}}\ and\ \bibinfo {author} {\bibfnamefont {G.~A.}\ \bibnamefont
  {Miller}},\ }\bibfield  {title} {\bibinfo {title} {{Unified formalism for
  electromagnetic and gravitational probes: Densities}},\ }\href
  {https://doi.org/10.1103/PhysRevD.105.014003} {\bibfield  {journal} {\bibinfo
   {journal} {Phys. Rev. D}\ }\textbf {\bibinfo {volume} {105}},\ \bibinfo
  {pages} {014003} (\bibinfo {year} {2022})},\ \Eprint
  {https://arxiv.org/abs/2108.03301} {arXiv:2108.03301 [hep-ph]} \BibitemShut
  {NoStop}%
\bibitem [{\citenamefont {Burkert}\ \emph
  {et~al.}(2023{\natexlab{b}})\citenamefont {Burkert}, \citenamefont
  {Elouadrhiri},\ and\ \citenamefont {Girod}}]{BEG:2023mech}%
  \BibitemOpen
  \bibfield  {author} {\bibinfo {author} {\bibfnamefont {V.~D.}\ \bibnamefont
  {Burkert}}, \bibinfo {author} {\bibfnamefont {L.}~\bibnamefont
  {Elouadrhiri}},\ and\ \bibinfo {author} {\bibfnamefont {F.~X.}\ \bibnamefont
  {Girod}},\ }\bibfield  {title} {\bibinfo {title} {{The mechanical radius of
  the proton}},\ }\href@noop {} {\  (\bibinfo {year} {2023}{\natexlab{b}})},\
  \Eprint {https://arxiv.org/abs/2310.11568} {arXiv:2310.11568 [hep-ph]}
  \BibitemShut {NoStop}%
\bibitem [{\citenamefont {Workman}\ \emph {et~al.}(2022)\citenamefont {Workman}
  \emph {et~al.}}]{ParticleDataGroup:2022pth}%
  \BibitemOpen
  \bibfield  {author} {\bibinfo {author} {\bibfnamefont {R.~L.}\ \bibnamefont
  {Workman}} \emph {et~al.} (\bibinfo {collaboration} {Particle Data Group}),\
  }\bibfield  {title} {\bibinfo {title} {{Review of Particle Physics}},\ }\href
  {https://doi.org/10.1093/ptep/ptac097} {\bibfield  {journal} {\bibinfo
  {journal} {PTEP}\ }\textbf {\bibinfo {volume} {2022}},\ \bibinfo {pages}
  {083C01} (\bibinfo {year} {2022})}\BibitemShut {NoStop}%
\bibitem [{\citenamefont {Costa}\ \emph {et~al.}(2021)\citenamefont {Costa},
  \citenamefont {Karpasitis}, \citenamefont {Pafitis}, \citenamefont
  {Panagopoulos}, \citenamefont {Panagopoulos}, \citenamefont {Skouroupathis},\
  and\ \citenamefont {Spanoudes}}]{Costa:2021iyv}%
  \BibitemOpen
  \bibfield  {author} {\bibinfo {author} {\bibfnamefont {M.}~\bibnamefont
  {Costa}}, \bibinfo {author} {\bibfnamefont {I.}~\bibnamefont {Karpasitis}},
  \bibinfo {author} {\bibfnamefont {T.}~\bibnamefont {Pafitis}}, \bibinfo
  {author} {\bibfnamefont {G.}~\bibnamefont {Panagopoulos}}, \bibinfo {author}
  {\bibfnamefont {H.}~\bibnamefont {Panagopoulos}}, \bibinfo {author}
  {\bibfnamefont {A.}~\bibnamefont {Skouroupathis}},\ and\ \bibinfo {author}
  {\bibfnamefont {G.}~\bibnamefont {Spanoudes}},\ }\bibfield  {title} {\bibinfo
  {title} {{Gauge-invariant renormalization scheme in QCD: Application to
  fermion bilinears and the energy-momentum tensor}},\ }\href
  {https://doi.org/10.1103/PhysRevD.103.094509} {\bibfield  {journal} {\bibinfo
   {journal} {Phys. Rev. D}\ }\textbf {\bibinfo {volume} {103}},\ \bibinfo
  {pages} {094509} (\bibinfo {year} {2021})},\ \Eprint
  {https://arxiv.org/abs/2102.00858} {arXiv:2102.00858 [hep-lat]} \BibitemShut
  {NoStop}%
\bibitem [{\citenamefont {Arrington}\ \emph {et~al.}(2023)\citenamefont
  {Arrington} \emph {et~al.}}]{JeffersonLabSoLID:2022iod}%
  \BibitemOpen
  \bibfield  {author} {\bibinfo {author} {\bibfnamefont {J.}~\bibnamefont
  {Arrington}} \emph {et~al.} (\bibinfo {collaboration} {Jefferson Lab
  SoLID}),\ }\bibfield  {title} {\bibinfo {title} {{The solenoidal large
  intensity device (SoLID) for JLab 12 GeV}},\ }\href
  {https://doi.org/10.1088/1361-6471/acda21} {\bibfield  {journal} {\bibinfo
  {journal} {J. Phys. G}\ }\textbf {\bibinfo {volume} {50}},\ \bibinfo {pages}
  {110501} (\bibinfo {year} {2023})},\ \Eprint
  {https://arxiv.org/abs/2209.13357} {arXiv:2209.13357 [nucl-ex]} \BibitemShut
  {NoStop}%
\bibitem [{\citenamefont
  {Burkert}(2018)}]{doi:10.1146/annurev-nucl-101917-021129}%
  \BibitemOpen
  \bibfield  {author} {\bibinfo {author} {\bibfnamefont {V.~D.}\ \bibnamefont
  {Burkert}},\ }\bibfield  {title} {\bibinfo {title} {Jefferson lab at 12 gev:
  The science program},\ }\href
  {https://doi.org/10.1146/annurev-nucl-101917-021129} {\bibfield  {journal}
  {\bibinfo  {journal} {Annual Review of Nuclear and Particle Science}\
  }\textbf {\bibinfo {volume} {68}},\ \bibinfo {pages} {405} (\bibinfo {year}
  {2018})},\ \Eprint
  {https://arxiv.org/abs/https://doi.org/10.1146/annurev-nucl-101917-021129}
  {https://doi.org/10.1146/annurev-nucl-101917-021129} \BibitemShut {NoStop}%
\bibitem [{\citenamefont {Arrington}\ \emph {et~al.}(2022)\citenamefont
  {Arrington} \emph {et~al.}}]{Arrington:2021alx}%
  \BibitemOpen
  \bibfield  {author} {\bibinfo {author} {\bibfnamefont {J.}~\bibnamefont
  {Arrington}} \emph {et~al.},\ }\bibfield  {title} {\bibinfo {title} {{Physics
  with CEBAF at 12 GeV and future opportunities}},\ }\href
  {https://doi.org/10.1016/j.ppnp.2022.103985} {\bibfield  {journal} {\bibinfo
  {journal} {Prog. Part. Nucl. Phys.}\ }\textbf {\bibinfo {volume} {127}},\
  \bibinfo {pages} {103985} (\bibinfo {year} {2022})},\ \Eprint
  {https://arxiv.org/abs/2112.00060} {arXiv:2112.00060 [nucl-ex]} \BibitemShut
  {NoStop}%
\bibitem [{\citenamefont {Abdul~Khalek}\ \emph {et~al.}(2022)\citenamefont
  {Abdul~Khalek} \emph {et~al.}}]{AbdulKhalek:2021gbh}%
  \BibitemOpen
  \bibfield  {author} {\bibinfo {author} {\bibfnamefont {R.}~\bibnamefont
  {Abdul~Khalek}} \emph {et~al.},\ }\bibfield  {title} {\bibinfo {title}
  {{Science Requirements and Detector Concepts for the Electron-Ion Collider}:
  {EIC Yellow Report}},\ }\href
  {https://doi.org/10.1016/j.nuclphysa.2022.122447} {\bibfield  {journal}
  {\bibinfo  {journal} {Nucl. Phys. A}\ }\textbf {\bibinfo {volume} {1026}},\
  \bibinfo {pages} {122447} (\bibinfo {year} {2022})},\ \Eprint
  {https://arxiv.org/abs/2103.05419} {arXiv:2103.05419 [physics.ins-det]}
  \BibitemShut {NoStop}%
\bibitem [{\citenamefont {Edwards}\ and\ \citenamefont
  {Joó}(2005)}]{Edwards:2004sx}%
  \BibitemOpen
  \bibfield  {author} {\bibinfo {author} {\bibfnamefont {R.~G.}\ \bibnamefont
  {Edwards}}\ and\ \bibinfo {author} {\bibfnamefont {B.}~\bibnamefont {Joó}}
  (\bibinfo {collaboration} {SciDAC Collaboration, LHPC Collaboration, UKQCD
  Collaboration}),\ }\bibfield  {title} {\bibinfo {title} {{The Chroma software
  system for lattice QCD}},\ }\href
  {https://doi.org/10.1016/j.nuclphysbps.2004.11.254} {\bibfield  {journal}
  {\bibinfo  {journal} {Nucl.Phys.Proc.Suppl.}\ }\textbf {\bibinfo {volume}
  {140}},\ \bibinfo {pages} {832} (\bibinfo {year} {2005})},\ \Eprint
  {https://arxiv.org/abs/hep-lat/0409003} {arXiv:hep-lat/0409003 [hep-lat]}
  \BibitemShut {NoStop}%
\bibitem [{\citenamefont {Pochinsky}()}]{qlua}%
  \BibitemOpen
  \bibfield  {author} {\bibinfo {author} {\bibfnamefont {A.}~\bibnamefont
  {Pochinsky}},\ }\href@noop {} {\bibinfo {title} {Qlua,
  https://usqcd.lns.mit.edu/qlua.}}\BibitemShut {Stop}%
\bibitem [{\citenamefont {Clark}\ \emph {et~al.}(2010)\citenamefont {Clark},
  \citenamefont {Babich}, \citenamefont {Barros}, \citenamefont {Brower},\ and\
  \citenamefont {Rebbi}}]{Clark:2009wm}%
  \BibitemOpen
  \bibfield  {author} {\bibinfo {author} {\bibfnamefont {M.}~\bibnamefont
  {Clark}}, \bibinfo {author} {\bibfnamefont {R.}~\bibnamefont {Babich}},
  \bibinfo {author} {\bibfnamefont {K.}~\bibnamefont {Barros}}, \bibinfo
  {author} {\bibfnamefont {R.}~\bibnamefont {Brower}},\ and\ \bibinfo {author}
  {\bibfnamefont {C.}~\bibnamefont {Rebbi}},\ }\bibfield  {title} {\bibinfo
  {title} {{Solving Lattice QCD systems of equations using mixed precision
  solvers on GPUs}},\ }\href {https://doi.org/10.1016/j.cpc.2010.05.002}
  {\bibfield  {journal} {\bibinfo  {journal} {Comput. Phys. Commun.}\ }\textbf
  {\bibinfo {volume} {181}},\ \bibinfo {pages} {1517} (\bibinfo {year}
  {2010})},\ \Eprint {https://arxiv.org/abs/0911.3191} {arXiv:0911.3191
  [hep-lat]} \BibitemShut {NoStop}%
\bibitem [{\citenamefont {Babich}\ \emph {et~al.}(2011)\citenamefont {Babich},
  \citenamefont {Clark}, \citenamefont {Joó}, \citenamefont {Shi},
  \citenamefont {Brower},\ and\ \citenamefont {Gottlieb}}]{Babich:2011np}%
  \BibitemOpen
  \bibfield  {author} {\bibinfo {author} {\bibfnamefont {R.}~\bibnamefont
  {Babich}}, \bibinfo {author} {\bibfnamefont {M.}~\bibnamefont {Clark}},
  \bibinfo {author} {\bibfnamefont {B.}~\bibnamefont {Joó}}, \bibinfo {author}
  {\bibfnamefont {G.}~\bibnamefont {Shi}}, \bibinfo {author} {\bibfnamefont
  {R.}~\bibnamefont {Brower}},\ and\ \bibinfo {author} {\bibfnamefont
  {S.}~\bibnamefont {Gottlieb}},\ }\bibfield  {title} {\bibinfo {title}
  {{Scaling Lattice QCD beyond 100 GPUs}},\ }in\ \href
  {https://doi.org/10.1145/2063384.2063478} {\emph {\bibinfo {booktitle} {{SC11
  International Conference for High Performance Computing, Networking, Storage
  and Analysis}}}}\ (\bibinfo {year} {2011})\ \Eprint
  {https://arxiv.org/abs/1109.2935} {arXiv:1109.2935 [hep-lat]} \BibitemShut
  {NoStop}%
\bibitem [{\citenamefont {Clark}\ \emph {et~al.}(2016)\citenamefont {Clark},
  \citenamefont {Joó}, \citenamefont {Strelchenko}, \citenamefont {Cheng},
  \citenamefont {Gambhir},\ and\ \citenamefont {Brower}}]{Clark:2016rdz}%
  \BibitemOpen
  \bibfield  {author} {\bibinfo {author} {\bibfnamefont {M.~A.}\ \bibnamefont
  {Clark}}, \bibinfo {author} {\bibfnamefont {B.}~\bibnamefont {Joó}},
  \bibinfo {author} {\bibfnamefont {A.}~\bibnamefont {Strelchenko}}, \bibinfo
  {author} {\bibfnamefont {M.}~\bibnamefont {Cheng}}, \bibinfo {author}
  {\bibfnamefont {A.}~\bibnamefont {Gambhir}},\ and\ \bibinfo {author}
  {\bibfnamefont {R.}~\bibnamefont {Brower}},\ }\bibfield  {title} {\bibinfo
  {title} {{Accelerating Lattice QCD Multigrid on GPUs Using Fine-Grained
  Parallelization}},\ }\href@noop {} {\  (\bibinfo {year} {2016})},\ \Eprint
  {https://arxiv.org/abs/1612.07873} {arXiv:1612.07873 [hep-lat]} \BibitemShut
  {NoStop}%
\bibitem [{\citenamefont {{Winter}}\ \emph {et~al.}(2014)\citenamefont
  {{Winter}}, \citenamefont {{Clark}}, \citenamefont {{Edwards}},\ and\
  \citenamefont {Joó}}]{6877336}%
  \BibitemOpen
  \bibfield  {author} {\bibinfo {author} {\bibfnamefont {F.~T.}\ \bibnamefont
  {{Winter}}}, \bibinfo {author} {\bibfnamefont {M.~A.}\ \bibnamefont
  {{Clark}}}, \bibinfo {author} {\bibfnamefont {R.~G.}\ \bibnamefont
  {{Edwards}}},\ and\ \bibinfo {author} {\bibfnamefont {B.}~\bibnamefont
  {Joó}},\ }\bibfield  {title} {\bibinfo {title} {A framework for lattice qcd
  calculations on gpus},\ }in\ \href {https://doi.org/10.1109/IPDPS.2014.112}
  {\emph {\bibinfo {booktitle} {2014 IEEE 28th International Parallel and
  Distributed Processing Symposium}}}\ (\bibinfo {year} {2014})\ pp.\ \bibinfo
  {pages} {1073--1082}\BibitemShut {NoStop}%
\bibitem [{\citenamefont {Joó}\ \emph {et~al.}(2016)\citenamefont {Joó},
  \citenamefont {Kalamkar}, \citenamefont {Kurth}, \citenamefont
  {Vaidyanathan},\ and\ \citenamefont {Walden}}]{10.1007/978-3-319-46079-6_30}%
  \BibitemOpen
  \bibfield  {author} {\bibinfo {author} {\bibfnamefont {B.}~\bibnamefont
  {Joó}}, \bibinfo {author} {\bibfnamefont {D.~D.}\ \bibnamefont {Kalamkar}},
  \bibinfo {author} {\bibfnamefont {T.}~\bibnamefont {Kurth}}, \bibinfo
  {author} {\bibfnamefont {K.}~\bibnamefont {Vaidyanathan}},\ and\ \bibinfo
  {author} {\bibfnamefont {A.}~\bibnamefont {Walden}},\ }\bibfield  {title}
  {\bibinfo {title} {Optimizing wilson-dirac operator and linear solvers for
  intel® knl},\ }in\ \href@noop {} {\emph {\bibinfo {booktitle} {High
  Performance Computing}}},\ \bibinfo {editor} {edited by\ \bibinfo {editor}
  {\bibfnamefont {M.}~\bibnamefont {Taufer}}, \bibinfo {editor} {\bibfnamefont
  {B.}~\bibnamefont {Mohr}},\ and\ \bibinfo {editor} {\bibfnamefont {J.~M.}\
  \bibnamefont {Kunkel}}}\ (\bibinfo  {publisher} {Springer International
  Publishing},\ \bibinfo {address} {Cham},\ \bibinfo {year} {2016})\ pp.\
  \bibinfo {pages} {415--427}\BibitemShut {NoStop}%
\bibitem [{\citenamefont {Romero}\ and\ \citenamefont
  {Kallidonis}()}]{chromaform}%
  \BibitemOpen
  \bibfield  {author} {\bibinfo {author} {\bibfnamefont {E.}~\bibnamefont
  {Romero}}\ and\ \bibinfo {author} {\bibfnamefont {C.}~\bibnamefont
  {Kallidonis}},\ }\bibfield  {title} {\bibinfo {title} {chromaform},\
  }\href@noop {} {\ }\Eprint
  {https://arxiv.org/abs/https://github.com/eromero-vlc/chromaform}
  {https://github.com/eromero-vlc/chromaform} \BibitemShut {NoStop}%
\bibitem [{\citenamefont {Gambhir}\ \emph {et~al.}(2018)\citenamefont
  {Gambhir}, \citenamefont {Brantley}, \citenamefont {Chang}, \citenamefont
  {Hörz}, \citenamefont {Monge-Camacho}, \citenamefont {Vranas},\ and\
  \citenamefont {Walker-Loud}}]{lalibe}%
  \BibitemOpen
  \bibfield  {author} {\bibinfo {author} {\bibfnamefont {A.}~\bibnamefont
  {Gambhir}}, \bibinfo {author} {\bibfnamefont {D.}~\bibnamefont {Brantley}},
  \bibinfo {author} {\bibfnamefont {J.}~\bibnamefont {Chang}}, \bibinfo
  {author} {\bibfnamefont {B.}~\bibnamefont {Hörz}}, \bibinfo {author}
  {\bibfnamefont {H.}~\bibnamefont {Monge-Camacho}}, \bibinfo {author}
  {\bibfnamefont {P.}~\bibnamefont {Vranas}},\ and\ \bibinfo {author}
  {\bibfnamefont {A.}~\bibnamefont {Walker-Loud}},\ }\href@noop {} {\bibinfo
  {title} {lalibe}},\ \bibinfo {howpublished}
  {\url{https://github.com/callat-qcd/lalibe}} (\bibinfo {year}
  {2018})\BibitemShut {NoStop}%
\bibitem [{\citenamefont {Harris}\ \emph {et~al.}(2020)\citenamefont {Harris},
  \citenamefont {Millman}, \citenamefont {van~der Walt}, \citenamefont
  {Gommers}, \citenamefont {Virtanen}, \citenamefont {Cournapeau},
  \citenamefont {Wieser}, \citenamefont {Taylor}, \citenamefont {Berg},
  \citenamefont {Smith}, \citenamefont {Kern}, \citenamefont {Picus},
  \citenamefont {Hoyer}, \citenamefont {van Kerkwijk}, \citenamefont {Brett},
  \citenamefont {Haldane}, \citenamefont {del R{\'{i}}o}, \citenamefont
  {Wiebe}, \citenamefont {Peterson}, \citenamefont {G{\'{e}}rard-Marchant},
  \citenamefont {Sheppard}, \citenamefont {Reddy}, \citenamefont {Weckesser},
  \citenamefont {Abbasi}, \citenamefont {Gohlke},\ and\ \citenamefont
  {Oliphant}}]{harris2020array}%
  \BibitemOpen
  \bibfield  {author} {\bibinfo {author} {\bibfnamefont {C.~R.}\ \bibnamefont
  {Harris}}, \bibinfo {author} {\bibfnamefont {K.~J.}\ \bibnamefont {Millman}},
  \bibinfo {author} {\bibfnamefont {S.~J.}\ \bibnamefont {van~der Walt}},
  \bibinfo {author} {\bibfnamefont {R.}~\bibnamefont {Gommers}}, \bibinfo
  {author} {\bibfnamefont {P.}~\bibnamefont {Virtanen}}, \bibinfo {author}
  {\bibfnamefont {D.}~\bibnamefont {Cournapeau}}, \bibinfo {author}
  {\bibfnamefont {E.}~\bibnamefont {Wieser}}, \bibinfo {author} {\bibfnamefont
  {J.}~\bibnamefont {Taylor}}, \bibinfo {author} {\bibfnamefont
  {S.}~\bibnamefont {Berg}}, \bibinfo {author} {\bibfnamefont {N.~J.}\
  \bibnamefont {Smith}}, \bibinfo {author} {\bibfnamefont {R.}~\bibnamefont
  {Kern}}, \bibinfo {author} {\bibfnamefont {M.}~\bibnamefont {Picus}},
  \bibinfo {author} {\bibfnamefont {S.}~\bibnamefont {Hoyer}}, \bibinfo
  {author} {\bibfnamefont {M.~H.}\ \bibnamefont {van Kerkwijk}}, \bibinfo
  {author} {\bibfnamefont {M.}~\bibnamefont {Brett}}, \bibinfo {author}
  {\bibfnamefont {A.}~\bibnamefont {Haldane}}, \bibinfo {author} {\bibfnamefont
  {J.~F.}\ \bibnamefont {del R{\'{i}}o}}, \bibinfo {author} {\bibfnamefont
  {M.}~\bibnamefont {Wiebe}}, \bibinfo {author} {\bibfnamefont
  {P.}~\bibnamefont {Peterson}}, \bibinfo {author} {\bibfnamefont
  {P.}~\bibnamefont {G{\'{e}}rard-Marchant}}, \bibinfo {author} {\bibfnamefont
  {K.}~\bibnamefont {Sheppard}}, \bibinfo {author} {\bibfnamefont
  {T.}~\bibnamefont {Reddy}}, \bibinfo {author} {\bibfnamefont
  {W.}~\bibnamefont {Weckesser}}, \bibinfo {author} {\bibfnamefont
  {H.}~\bibnamefont {Abbasi}}, \bibinfo {author} {\bibfnamefont
  {C.}~\bibnamefont {Gohlke}},\ and\ \bibinfo {author} {\bibfnamefont {T.~E.}\
  \bibnamefont {Oliphant}},\ }\bibfield  {title} {\bibinfo {title} {Array
  programming with {NumPy}},\ }\href
  {https://doi.org/10.1038/s41586-020-2649-2} {\bibfield  {journal} {\bibinfo
  {journal} {Nature}\ }\textbf {\bibinfo {volume} {585}},\ \bibinfo {pages}
  {357} (\bibinfo {year} {2020})}\BibitemShut {NoStop}%
\bibitem [{\citenamefont {Virtanen}\ \emph {et~al.}(2020)\citenamefont
  {Virtanen}, \citenamefont {Gommers}, \citenamefont {Oliphant}, \citenamefont
  {Haberland}, \citenamefont {Reddy}, \citenamefont {Cournapeau}, \citenamefont
  {Burovski}, \citenamefont {Peterson}, \citenamefont {Weckesser},
  \citenamefont {Bright}, \citenamefont {{van der Walt}}, \citenamefont
  {Brett}, \citenamefont {Wilson}, \citenamefont {Millman}, \citenamefont
  {Mayorov}, \citenamefont {Nelson}, \citenamefont {Jones}, \citenamefont
  {Kern}, \citenamefont {Larson}, \citenamefont {Carey}, \citenamefont {Polat},
  \citenamefont {Feng}, \citenamefont {Moore}, \citenamefont {{VanderPlas}},
  \citenamefont {Laxalde}, \citenamefont {Perktold}, \citenamefont {Cimrman},
  \citenamefont {Henriksen}, \citenamefont {Quintero}, \citenamefont {Harris},
  \citenamefont {Archibald}, \citenamefont {Ribeiro}, \citenamefont
  {Pedregosa}, \citenamefont {{van Mulbregt}},\ and\ \citenamefont {{SciPy 1.0
  Contributors}}}]{2020SciPy-NMeth}%
  \BibitemOpen
  \bibfield  {author} {\bibinfo {author} {\bibfnamefont {P.}~\bibnamefont
  {Virtanen}}, \bibinfo {author} {\bibfnamefont {R.}~\bibnamefont {Gommers}},
  \bibinfo {author} {\bibfnamefont {T.~E.}\ \bibnamefont {Oliphant}}, \bibinfo
  {author} {\bibfnamefont {M.}~\bibnamefont {Haberland}}, \bibinfo {author}
  {\bibfnamefont {T.}~\bibnamefont {Reddy}}, \bibinfo {author} {\bibfnamefont
  {D.}~\bibnamefont {Cournapeau}}, \bibinfo {author} {\bibfnamefont
  {E.}~\bibnamefont {Burovski}}, \bibinfo {author} {\bibfnamefont
  {P.}~\bibnamefont {Peterson}}, \bibinfo {author} {\bibfnamefont
  {W.}~\bibnamefont {Weckesser}}, \bibinfo {author} {\bibfnamefont
  {J.}~\bibnamefont {Bright}}, \bibinfo {author} {\bibfnamefont {S.~J.}\
  \bibnamefont {{van der Walt}}}, \bibinfo {author} {\bibfnamefont
  {M.}~\bibnamefont {Brett}}, \bibinfo {author} {\bibfnamefont
  {J.}~\bibnamefont {Wilson}}, \bibinfo {author} {\bibfnamefont {K.~J.}\
  \bibnamefont {Millman}}, \bibinfo {author} {\bibfnamefont {N.}~\bibnamefont
  {Mayorov}}, \bibinfo {author} {\bibfnamefont {A.~R.~J.}\ \bibnamefont
  {Nelson}}, \bibinfo {author} {\bibfnamefont {E.}~\bibnamefont {Jones}},
  \bibinfo {author} {\bibfnamefont {R.}~\bibnamefont {Kern}}, \bibinfo {author}
  {\bibfnamefont {E.}~\bibnamefont {Larson}}, \bibinfo {author} {\bibfnamefont
  {C.~J.}\ \bibnamefont {Carey}}, \bibinfo {author} {\bibfnamefont
  {{\.I}.}~\bibnamefont {Polat}}, \bibinfo {author} {\bibfnamefont
  {Y.}~\bibnamefont {Feng}}, \bibinfo {author} {\bibfnamefont {E.~W.}\
  \bibnamefont {Moore}}, \bibinfo {author} {\bibfnamefont {J.}~\bibnamefont
  {{VanderPlas}}}, \bibinfo {author} {\bibfnamefont {D.}~\bibnamefont
  {Laxalde}}, \bibinfo {author} {\bibfnamefont {J.}~\bibnamefont {Perktold}},
  \bibinfo {author} {\bibfnamefont {R.}~\bibnamefont {Cimrman}}, \bibinfo
  {author} {\bibfnamefont {I.}~\bibnamefont {Henriksen}}, \bibinfo {author}
  {\bibfnamefont {E.~A.}\ \bibnamefont {Quintero}}, \bibinfo {author}
  {\bibfnamefont {C.~R.}\ \bibnamefont {Harris}}, \bibinfo {author}
  {\bibfnamefont {A.~M.}\ \bibnamefont {Archibald}}, \bibinfo {author}
  {\bibfnamefont {A.~H.}\ \bibnamefont {Ribeiro}}, \bibinfo {author}
  {\bibfnamefont {F.}~\bibnamefont {Pedregosa}}, \bibinfo {author}
  {\bibfnamefont {P.}~\bibnamefont {{van Mulbregt}}},\ and\ \bibinfo {author}
  {\bibnamefont {{SciPy 1.0 Contributors}}},\ }\bibfield  {title} {\bibinfo
  {title} {{{SciPy} 1.0: Fundamental Algorithms for Scientific Computing in
  Python}},\ }\href {https://doi.org/10.1038/s41592-019-0686-2} {\bibfield
  {journal} {\bibinfo  {journal} {Nature Methods}\ }\textbf {\bibinfo {volume}
  {17}},\ \bibinfo {pages} {261} (\bibinfo {year} {2020})}\BibitemShut
  {NoStop}%
\bibitem [{\citenamefont {Reback}\ \emph {et~al.}(2020)\citenamefont {Reback},
  \citenamefont {McKinney}, \citenamefont {jbrockmendel}, \citenamefont {den
  Bossche}, \citenamefont {Augspurger}, \citenamefont {Cloud}, \citenamefont
  {gfyoung}, \citenamefont {Sinhrks}, \citenamefont {Klein}, \citenamefont
  {Roeschke}, \citenamefont {Hawkins}, \citenamefont {Tratner}, \citenamefont
  {She}, \citenamefont {Ayd}, \citenamefont {Petersen}, \citenamefont {Garcia},
  \citenamefont {Schendel}, \citenamefont {Hayden}, \citenamefont
  {MomIsBestFriend}, \citenamefont {Jancauskas}, \citenamefont {Battiston},
  \citenamefont {Seabold}, \citenamefont {chris b1}, \citenamefont
  {h~vetinari}, \citenamefont {Hoyer}, \citenamefont {Overmeire}, \citenamefont
  {alimcmaster1}, \citenamefont {Dong}, \citenamefont {Whelan},\ and\
  \citenamefont {Mehyar}}]{jeff_reback_2020_3715232}%
  \BibitemOpen
  \bibfield  {author} {\bibinfo {author} {\bibfnamefont {J.}~\bibnamefont
  {Reback}}, \bibinfo {author} {\bibfnamefont {W.}~\bibnamefont {McKinney}},
  \bibinfo {author} {\bibnamefont {jbrockmendel}}, \bibinfo {author}
  {\bibfnamefont {J.~V.}\ \bibnamefont {den Bossche}}, \bibinfo {author}
  {\bibfnamefont {T.}~\bibnamefont {Augspurger}}, \bibinfo {author}
  {\bibfnamefont {P.}~\bibnamefont {Cloud}}, \bibinfo {author} {\bibnamefont
  {gfyoung}}, \bibinfo {author} {\bibnamefont {Sinhrks}}, \bibinfo {author}
  {\bibfnamefont {A.}~\bibnamefont {Klein}}, \bibinfo {author} {\bibfnamefont
  {M.}~\bibnamefont {Roeschke}}, \bibinfo {author} {\bibfnamefont
  {S.}~\bibnamefont {Hawkins}}, \bibinfo {author} {\bibfnamefont
  {J.}~\bibnamefont {Tratner}}, \bibinfo {author} {\bibfnamefont
  {C.}~\bibnamefont {She}}, \bibinfo {author} {\bibfnamefont {W.}~\bibnamefont
  {Ayd}}, \bibinfo {author} {\bibfnamefont {T.}~\bibnamefont {Petersen}},
  \bibinfo {author} {\bibfnamefont {M.}~\bibnamefont {Garcia}}, \bibinfo
  {author} {\bibfnamefont {J.}~\bibnamefont {Schendel}}, \bibinfo {author}
  {\bibfnamefont {A.}~\bibnamefont {Hayden}}, \bibinfo {author} {\bibnamefont
  {MomIsBestFriend}}, \bibinfo {author} {\bibfnamefont {V.}~\bibnamefont
  {Jancauskas}}, \bibinfo {author} {\bibfnamefont {P.}~\bibnamefont
  {Battiston}}, \bibinfo {author} {\bibfnamefont {S.}~\bibnamefont {Seabold}},
  \bibinfo {author} {\bibnamefont {chris b1}}, \bibinfo {author} {\bibnamefont
  {h~vetinari}}, \bibinfo {author} {\bibfnamefont {S.}~\bibnamefont {Hoyer}},
  \bibinfo {author} {\bibfnamefont {W.}~\bibnamefont {Overmeire}}, \bibinfo
  {author} {\bibnamefont {alimcmaster1}}, \bibinfo {author} {\bibfnamefont
  {K.}~\bibnamefont {Dong}}, \bibinfo {author} {\bibfnamefont {C.}~\bibnamefont
  {Whelan}},\ and\ \bibinfo {author} {\bibfnamefont {M.}~\bibnamefont
  {Mehyar}},\ }\href {https://doi.org/10.5281/zenodo.3715232} {\bibinfo {title}
  {pandas-dev/pandas: Pandas 1.0.3}} (\bibinfo {year} {2020})\BibitemShut
  {NoStop}%
\bibitem [{\citenamefont {{W}es
  {M}c{K}inney}(2010)}]{mckinney-proc-scipy-2010}%
  \BibitemOpen
  \bibfield  {author} {\bibinfo {author} {\bibnamefont {{W}es {M}c{K}inney}},\
  }\bibfield  {title} {\bibinfo {title} {{D}ata {S}tructures for {S}tatistical
  {C}omputing in {P}ython},\ }in\ \href
  {https://doi.org/10.25080/Majora-92bf1922-00a} {\emph {\bibinfo {booktitle}
  {{P}roceedings of the 9th {P}ython in {S}cience {C}onference}}},\ \bibinfo
  {editor} {edited by\ \bibinfo {editor} {\bibnamefont {{S}t\'efan van~der
  {W}alt}}\ and\ \bibinfo {editor} {\bibnamefont {{J}arrod {M}illman}}}\
  (\bibinfo {year} {2010})\ pp.\ \bibinfo {pages} {56 -- 61}\BibitemShut
  {NoStop}%
\bibitem [{\citenamefont
  {Lepage}(2020{\natexlab{a}})}]{peter_lepage_2020_4037174}%
  \BibitemOpen
  \bibfield  {author} {\bibinfo {author} {\bibfnamefont {G.~P.}\ \bibnamefont
  {Lepage}},\ }\bibfield  {title} {\bibinfo {title} {{lsqfit v. 11.7}}\ }\href
  {https://doi.org/doi:10.5281/zenodo.4037174} {doi:10.5281/zenodo.4037174}
  (\bibinfo {year} {2020}{\natexlab{a}}),\ \Eprint
  {https://arxiv.org/abs/https://github.com/gplepage/lsqfit}
  {https://github.com/gplepage/lsqfit} \BibitemShut {NoStop}%
\bibitem [{\citenamefont
  {Lepage}(2020{\natexlab{b}})}]{peter_lepage_2020_4290884}%
  \BibitemOpen
  \bibfield  {author} {\bibinfo {author} {\bibfnamefont {G.~P.}\ \bibnamefont
  {Lepage}},\ }\bibfield  {title} {\bibinfo {title} {{gvar v. 11.9.1}}\ }\href
  {https://doi.org/doi:10.5281/zenodo.4290884} {doi:10.5281/zenodo.4290884}
  (\bibinfo {year} {2020}{\natexlab{b}}),\ \Eprint
  {https://arxiv.org/abs/https://github.com/gplepage/gvar}
  {https://github.com/gplepage/gvar} \BibitemShut {NoStop}%
\bibitem [{\citenamefont {Hunter}(2007)}]{Hunter:2007}%
  \BibitemOpen
  \bibfield  {author} {\bibinfo {author} {\bibfnamefont {J.~D.}\ \bibnamefont
  {Hunter}},\ }\bibfield  {title} {\bibinfo {title} {Matplotlib: A 2d graphics
  environment},\ }\href {https://doi.org/10.1109/MCSE.2007.55} {\bibfield
  {journal} {\bibinfo  {journal} {Computing in Science \& Engineering}\
  }\textbf {\bibinfo {volume} {9}},\ \bibinfo {pages} {90} (\bibinfo {year}
  {2007})}\BibitemShut {NoStop}%
\bibitem [{\citenamefont {Mandula}\ \emph {et~al.}(1983)\citenamefont
  {Mandula}, \citenamefont {Zweig},\ and\ \citenamefont
  {Govaerts}}]{Mandula:1983ut}%
  \BibitemOpen
  \bibfield  {author} {\bibinfo {author} {\bibfnamefont {J.~E.}\ \bibnamefont
  {Mandula}}, \bibinfo {author} {\bibfnamefont {G.}~\bibnamefont {Zweig}},\
  and\ \bibinfo {author} {\bibfnamefont {J.}~\bibnamefont {Govaerts}},\
  }\bibfield  {title} {\bibinfo {title} {{Representations of the Rotation
  Reflection Symmetry Group of the Four-dimensional Cubic Lattice}},\ }\href
  {https://doi.org/10.1016/0550-3213(83)90399-1} {\bibfield  {journal}
  {\bibinfo  {journal} {Nucl. Phys. B}\ }\textbf {\bibinfo {volume} {228}},\
  \bibinfo {pages} {91} (\bibinfo {year} {1983})}\BibitemShut {NoStop}%
\bibitem [{\citenamefont {G{\"o}ckeler}\ \emph {et~al.}(1996)\citenamefont
  {G{\"o}ckeler}, \citenamefont {Horsley}, \citenamefont {Ilgenfritz},
  \citenamefont {Perlt}, \citenamefont {Rakow}, \citenamefont {Schierholz},\
  and\ \citenamefont {Schiller}}]{Gockeler:1996mu}%
  \BibitemOpen
  \bibfield  {author} {\bibinfo {author} {\bibfnamefont {M.}~\bibnamefont
  {G{\"o}ckeler}}, \bibinfo {author} {\bibfnamefont {R.}~\bibnamefont
  {Horsley}}, \bibinfo {author} {\bibfnamefont {E.-M.}\ \bibnamefont
  {Ilgenfritz}}, \bibinfo {author} {\bibfnamefont {H.}~\bibnamefont {Perlt}},
  \bibinfo {author} {\bibfnamefont {P.~E.}\ \bibnamefont {Rakow}}, \bibinfo
  {author} {\bibfnamefont {G.}~\bibnamefont {Schierholz}},\ and\ \bibinfo
  {author} {\bibfnamefont {A.}~\bibnamefont {Schiller}},\ }\bibfield  {title}
  {\bibinfo {title} {{Lattice operators for moments of the structure functions
  and their transformation under the hypercubic group}},\ }\href
  {https://doi.org/10.1103/PhysRevD.54.5705} {\bibfield  {journal} {\bibinfo
  {journal} {Phys. Rev. D}\ }\textbf {\bibinfo {volume} {54}},\ \bibinfo
  {pages} {5705} (\bibinfo {year} {1996})},\ \Eprint
  {https://arxiv.org/abs/hep-lat/9602029} {arXiv:hep-lat/9602029} \BibitemShut
  {NoStop}%
\bibitem [{\citenamefont {Mondal}\ \emph {et~al.}(2020)\citenamefont {Mondal},
  \citenamefont {Gupta}, \citenamefont {Park}, \citenamefont {Yoon},
  \citenamefont {Bhattacharya}, \citenamefont {Jo\'o},\ and\ \citenamefont
  {Winter}}]{Mondal:2020ela}%
  \BibitemOpen
  \bibfield  {author} {\bibinfo {author} {\bibfnamefont {S.}~\bibnamefont
  {Mondal}}, \bibinfo {author} {\bibfnamefont {R.}~\bibnamefont {Gupta}},
  \bibinfo {author} {\bibfnamefont {S.}~\bibnamefont {Park}}, \bibinfo {author}
  {\bibfnamefont {B.}~\bibnamefont {Yoon}}, \bibinfo {author} {\bibfnamefont
  {T.}~\bibnamefont {Bhattacharya}}, \bibinfo {author} {\bibfnamefont
  {B.}~\bibnamefont {Jo\'o}},\ and\ \bibinfo {author} {\bibfnamefont
  {F.}~\bibnamefont {Winter}} (\bibinfo {collaboration} {Nucleon Matrix
  Elements (NME)}),\ }\bibfield  {title} {\bibinfo {title} {{Nucleon momentum
  fraction, helicity and transversity from 2+1-flavor lattice QCD}},\ }\href
  {https://doi.org/10.1007/JHEP04(2021)044} {\bibfield  {journal} {\bibinfo
  {journal} {JHEP}\ }\textbf {\bibinfo {volume} {21}},\ \bibinfo {pages}
  {004}},\ \Eprint {https://arxiv.org/abs/2011.12787} {arXiv:2011.12787
  [hep-lat]} \BibitemShut {NoStop}%
\bibitem [{\citenamefont {Dorati}\ \emph {et~al.}(2008)\citenamefont {Dorati},
  \citenamefont {Gail},\ and\ \citenamefont {Hemmert}}]{Dorati:2007bk}%
  \BibitemOpen
  \bibfield  {author} {\bibinfo {author} {\bibfnamefont {M.}~\bibnamefont
  {Dorati}}, \bibinfo {author} {\bibfnamefont {T.~A.}\ \bibnamefont {Gail}},\
  and\ \bibinfo {author} {\bibfnamefont {T.~R.}\ \bibnamefont {Hemmert}},\
  }\bibfield  {title} {\bibinfo {title} {{Chiral perturbation theory and the
  first moments of the generalized parton distributions in a nucleon}},\ }\href
  {https://doi.org/10.1016/j.nuclphysa.2007.10.012} {\bibfield  {journal}
  {\bibinfo  {journal} {Nucl. Phys. A}\ }\textbf {\bibinfo {volume} {798}},\
  \bibinfo {pages} {96} (\bibinfo {year} {2008})},\ \Eprint
  {https://arxiv.org/abs/nucl-th/0703073} {arXiv:nucl-th/0703073} \BibitemShut
  {NoStop}%
\bibitem [{\citenamefont {Wein}\ \emph {et~al.}(2014)\citenamefont {Wein},
  \citenamefont {Bruns},\ and\ \citenamefont {Sch\"afer}}]{Wein:2014wma}%
  \BibitemOpen
  \bibfield  {author} {\bibinfo {author} {\bibfnamefont {P.}~\bibnamefont
  {Wein}}, \bibinfo {author} {\bibfnamefont {P.~C.}\ \bibnamefont {Bruns}},\
  and\ \bibinfo {author} {\bibfnamefont {A.}~\bibnamefont {Sch\"afer}},\
  }\bibfield  {title} {\bibinfo {title} {{First moments of nucleon generalized
  parton distributions in chiral perturbation theory at full one-loop order}},\
  }\href {https://doi.org/10.1103/PhysRevD.89.116002} {\bibfield  {journal}
  {\bibinfo  {journal} {Phys. Rev. D}\ }\textbf {\bibinfo {volume} {89}},\
  \bibinfo {pages} {116002} (\bibinfo {year} {2014})},\ \Eprint
  {https://arxiv.org/abs/1402.4979} {arXiv:1402.4979 [hep-ph]} \BibitemShut
  {NoStop}%
\bibitem [{\citenamefont {Alharazin}\ \emph {et~al.}(2020)\citenamefont
  {Alharazin}, \citenamefont {Djukanovic}, \citenamefont {Gegelia},\ and\
  \citenamefont {Polyakov}}]{Alharazin:2020yjv}%
  \BibitemOpen
  \bibfield  {author} {\bibinfo {author} {\bibfnamefont {H.}~\bibnamefont
  {Alharazin}}, \bibinfo {author} {\bibfnamefont {D.}~\bibnamefont
  {Djukanovic}}, \bibinfo {author} {\bibfnamefont {J.}~\bibnamefont
  {Gegelia}},\ and\ \bibinfo {author} {\bibfnamefont {M.~V.}\ \bibnamefont
  {Polyakov}},\ }\bibfield  {title} {\bibinfo {title} {{Chiral theory of
  nucleons and pions in the presence of an external gravitational field}},\
  }\href {https://doi.org/10.1103/PhysRevD.102.076023} {\bibfield  {journal}
  {\bibinfo  {journal} {Phys. Rev. D}\ }\textbf {\bibinfo {volume} {102}},\
  \bibinfo {pages} {076023} (\bibinfo {year} {2020})},\ \Eprint
  {https://arxiv.org/abs/2006.05890} {arXiv:2006.05890 [hep-ph]} \BibitemShut
  {NoStop}%
\bibitem [{\citenamefont {L\"offler}\ \emph {et~al.}(2022)\citenamefont
  {L\"offler}, \citenamefont {Wein}, \citenamefont {Wurm}, \citenamefont
  {Weish\"aupl}, \citenamefont {Jenkins}, \citenamefont {R\"odl}, \citenamefont
  {Sch\"afer},\ and\ \citenamefont {Walter}}]{Loffler:2021afv}%
  \BibitemOpen
  \bibfield  {author} {\bibinfo {author} {\bibfnamefont {M.}~\bibnamefont
  {L\"offler}}, \bibinfo {author} {\bibfnamefont {P.}~\bibnamefont {Wein}},
  \bibinfo {author} {\bibfnamefont {T.}~\bibnamefont {Wurm}}, \bibinfo {author}
  {\bibfnamefont {S.}~\bibnamefont {Weish\"aupl}}, \bibinfo {author}
  {\bibfnamefont {D.}~\bibnamefont {Jenkins}}, \bibinfo {author} {\bibfnamefont
  {R.}~\bibnamefont {R\"odl}}, \bibinfo {author} {\bibfnamefont
  {A.}~\bibnamefont {Sch\"afer}},\ and\ \bibinfo {author} {\bibfnamefont
  {L.}~\bibnamefont {Walter}} (\bibinfo {collaboration} {RQCD}),\ }\bibfield
  {title} {\bibinfo {title} {{Mellin moments of spin dependent and independent
  PDFs of the pion and rho meson}},\ }\href
  {https://doi.org/10.1103/PhysRevD.105.014505} {\bibfield  {journal} {\bibinfo
   {journal} {Phys. Rev. D}\ }\textbf {\bibinfo {volume} {105}},\ \bibinfo
  {pages} {014505} (\bibinfo {year} {2022})},\ \Eprint
  {https://arxiv.org/abs/2108.07544} {arXiv:2108.07544 [hep-lat]} \BibitemShut
  {NoStop}%
\bibitem [{\citenamefont {Jang}\ \emph {et~al.}(2020)\citenamefont {Jang},
  \citenamefont {Gupta}, \citenamefont {Yoon},\ and\ \citenamefont
  {Bhattacharya}}]{Jang:2019vkm}%
  \BibitemOpen
  \bibfield  {author} {\bibinfo {author} {\bibfnamefont {Y.-C.}\ \bibnamefont
  {Jang}}, \bibinfo {author} {\bibfnamefont {R.}~\bibnamefont {Gupta}},
  \bibinfo {author} {\bibfnamefont {B.}~\bibnamefont {Yoon}},\ and\ \bibinfo
  {author} {\bibfnamefont {T.}~\bibnamefont {Bhattacharya}},\ }\bibfield
  {title} {\bibinfo {title} {{Axial Vector Form Factors from Lattice QCD that
  Satisfy the PCAC Relation}},\ }\href
  {https://doi.org/10.1103/PhysRevLett.124.072002} {\bibfield  {journal}
  {\bibinfo  {journal} {Phys. Rev. Lett.}\ }\textbf {\bibinfo {volume} {124}},\
  \bibinfo {pages} {072002} (\bibinfo {year} {2020})},\ \Eprint
  {https://arxiv.org/abs/1905.06470} {arXiv:1905.06470 [hep-lat]} \BibitemShut
  {NoStop}%
\end{thebibliography}%

\onecolumngrid
\setcounter{page}{1} 

\section{Further details of lattice QCD calculation}

\begin{figure}[b]
    \centering
    \includegraphics[width=0.48\textwidth]{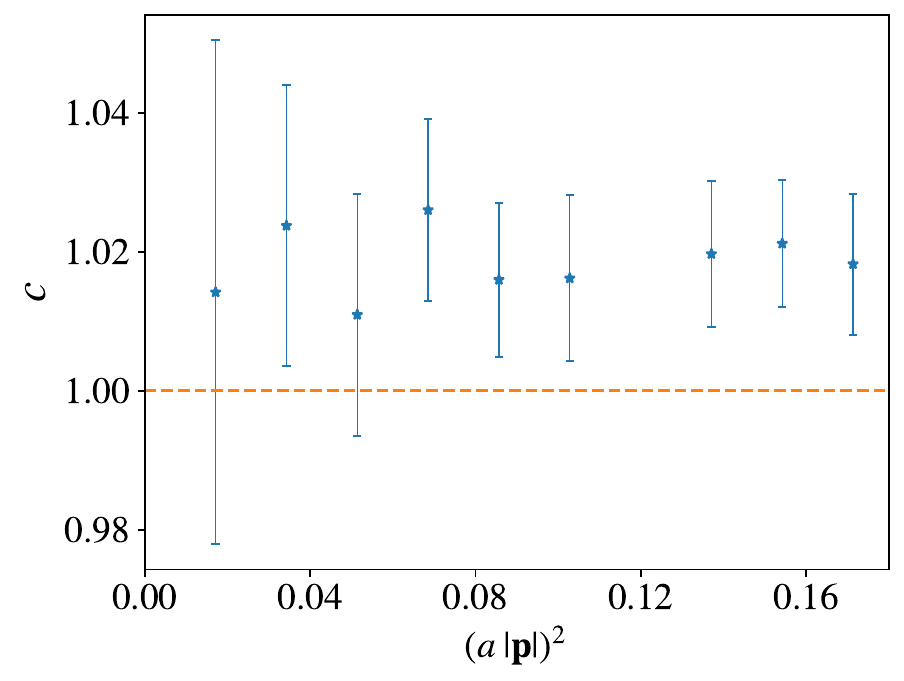}
    \caption{
        The values of $c$ obtained from the dispersion relation $E_{\vec{p}} = \sqrt{m^2+|c\vec{p}|^2}$ using the ground state energies extracted from fits to nucleon two-point correlation functions. }
    \label{fig:dispersion}
\end{figure}

The momentum-projected two-point correlation function of the proton for spin channel $s \rightarrow s'$ is defined as
\begin{equation} \label{eq:twopoint_nuc}
C^{\text{2pt}}_{ss'}(\vec{p},t_s; \vec{x}_0,t_0) =
\sum_{\vec{x}} e^{-i\vec{p} \cdot (\vec{x} - \vec{x}_0) }
\mathrm{tr}\big[\Gamma_{s's}\vev{\chi(\vec{x},t_s+t_0)  \bar{\chi}(\vec{x}_0,t_0) }\big],
\end{equation}
where the interpolating operator $\chi(x)$ is
\begin{equation}\label{eq:interp}
\chi(x) = \epsilon^{abc}\left[ \psi_{u}^{T,b}(x) C \gamma_5 \psi_{d}^c(x) \right] \psi_{u}^a(x) \;,
\end{equation}
with $a,b,c$ being color indices and $C$ the charge conjugation matrix, and the spin-projection matrices are
\begin{equation}
\Gamma_{s' s} =
\begin{pmatrix}
P_+(1+\gamma_x\gamma_y) & P_+\gamma_z(\gamma_x+i\gamma_y) \\
P_+\gamma_z(\gamma_x-i\gamma_y) & P_+(1-\gamma_x\gamma_y)
\end{pmatrix}
_{s's} \;,
\end{equation}
with $P_+ \equiv \frac{1}{2}(1+\gamma_t)$ being the positive-energy projector. 
$C^{\text{2pt}}_{ss'}$ is computed on quark sources smeared by gauge-invariant Gaussian smearing up to radius $4.5a$, defined using spatially stout-smeared~\cite{Morningstar:2003gk} link fields, for $1024$ source positions on each of $2511$ configurations. The 1024 sources are arranged in two $4^3 \times 8$ grids offset by $(6,6,6,6)$ lattice units, with an overall random offset for each configuration.  The two-point functions are projected to all momenta satisfying $|\vec{p}|^2\leq 10(2\pi/L)^2$.
To extract the proton energies, the two-point functions are averaged over all momenta of equivalent magnitude, all sources, and the two diagonal spin channels; the results are fit to the functional form
\begin{equation}
C^{\text{2pt}}(\vec{p},t_s) 
\sim
\sum_{n=0} |Z_{\vec{p}}^n|^2
e^{-E^n_\vec{p} t_s},
\end{equation}
where $n=0$ is the ground state.
The momentum-averaged correlator for each distinct $|\vec{p}|$ is analyzed independently.
The parameters $E^{0}_{\vec{p}}$ are determined by model averaging~\cite{Jay:2020jkz} over the results of fits to models with 2 and 3 states, and various ranges $t_{s,\text{min}} \leq t_s < t_{s,\text{max}}(|\vec{p}|)$.
All $t_{s,\text{min}} \in [1,15]$ are used, and $t_{s,\text{max}}(|\vec{p}|)$ is taken to be the last point before the noise-to-signal ratio exceeds 5\%, ranging between 27 for $\vec{p}^2 = 0$ to 21 for $\vec{p}^2 = 10$.
The energies are parameterized as a tower of positive gaps $\Delta^{n}$, fitted with a log-space prior $\log \Delta^{n} \sim \log[0.5(1)]$ for each.
Wide, uninformative priors are used for the $Z^n_\vec{p}$.
Uncertainties are propagated by bootstrapping as described in the main text, which is reconciled with model averaging as described in Ref.~\cite{Pefkou:2021fni}.

Figure~\ref{fig:dispersion} shows the effective speed of light $c$ computed from the energies obtained as described above, and the dispersion relation $E_{\vec{p}}=\sqrt{m^2+|c\vec{p}|^2}$. The 2\% deviation of the results from $c=1$ is small compared to the uncertainties of the GFFs, and therefore energies obtained from the dispersion relation using the fit to the nucleon mass $am=0.4169$ are used for the remainder of the analysis.

The quark and gluon symmetric EMT contributions in Euclidean space can be defined as
\begin{align}
\hat{T}_{f,\mu\nu}(x) &= \bar{\psi}_f(x)\overleftrightarrow{D}_{\{\mu}\gamma_{\nu\}}\psi_f(x), \\
\hat{T}_{g,\mu\nu}(x) &= 2\,\text{Tr}\left[F_{\mu\rho}(x)F_{\nu\rho}(x)-\frac{1}{4}\delta_{\mu\nu}F_{\alpha\beta}(x)F_{\alpha\beta}(x)\right] ,
\end{align}
where, up to discretization effects, the Euclidean gluon field strength tensor can be expressed as
\begin{equation}
F_{\mu\nu}(x) = \frac{i}{8g_0}(Q_{\mu\nu}(x)-Q^{\dagger}_{\mu\nu}(x)),
\end{equation}
with $Q_{\mu\nu}$ defined in terms of the link fields as
\begin{equation}
\begin{split}
Q_{\mu\nu}(x)= &U_{\mu}(x)U_{\nu}(x+\hat{\mu})U_{\mu}^{\dagger}(x+\hat{\nu})U^{\dagger}_{\nu}(x)+U_{\nu}(x)U_{\mu}^{\dagger}(x-\hat{\mu}-\hat{\nu})U_{\nu}^{\dagger}(x-\hat{\mu})U_{\mu}(x-\hat{\mu})\\
+&U_{\mu}^{\dagger}(x-\hat{\mu})U_{\nu}^{\dagger}(x-\hat{\mu}-\hat{\nu})U_{\mu}(x-\hat{\mu}-\hat{\nu})U_{\nu}(x-\hat{\nu})+U_{\nu}^{\dagger}(x-\hat{\nu})U_{\mu}(x-\hat{\nu})U_{\nu}(x-\hat{\mu}-\hat{\nu})U_{\mu}^{\dagger},
\end{split}
\end{equation}
and where $g_0$ is the bare coupling for a tadpole-improved Luscher-Weisz gauge action with tadpole parameter $u_0$,
\begin{equation}
g_0 = \sqrt{\frac{2N_c}{\beta(1-\frac{2}{5u_0^2})}}.
\end{equation}
The symmetric covariant derivative can be expressed as
\begin{equation}
\overleftrightarrow{D}_{\mu}=\frac{1}{2}\left(\overrightarrow{D}_{\mu}-\overleftarrow{D}_{\mu}\right) ,
\end{equation}
where
\begin{align}
\overrightarrow{D}_{\mu}^E\psi(x)&=\frac{1}{2}\left(U_{\mu}(x)\psi(x+\hat{\mu})-U_{\mu}^{\dagger}(x-\hat{\mu})\psi(x-\hat{\mu})\right) \\ 
\bar{\psi}(x)\overleftarrow{D}_{\mu}^E &= \frac{1}{2}\left(\bar{\psi}(x+\hat{\mu})U^{\dagger}_{\mu}(x)-\bar{\psi}(x-\hat{\mu})U_{\mu}(x-\hat{\mu})\right).
\end{align}
On an isotropic hypercubic lattice, the traceless diagonal and off-diagonal components of the EMT transform under two different irreducible representations (irreps) of the hypercubic group, $\tau_1^{(3)}$ and $\tau_3^{(6)}$~\cite{Mandula:1983ut,Gockeler:1996mu}. The choice of basis for each irrep that is used in this work is
\begin{equation}\label{eq:rep3}
\hat{T}_{\tau_{1,1}^{(3)}} =  \frac{1}{2}(\hat{T}_{11}+\hat{T}_{22}-\hat{T}_{33}-\hat{T}_{44}),\; 
\hat{T}_{\tau_{1,2}^{(3)}} = \frac{1}{\sqrt{2}}(\hat{T}_{11}-\hat{T}_{22}),\; \hat{T}_{\tau_{1,3}^{(3)}} = \frac{1}{\sqrt{2}}(\hat{T}_{33}-\hat{T}_{44}),
\end{equation}
\begin{equation}
\label{eq:rep6}
\begin{split}
\hat{T}_{\tau_{3,1}^{(6)}} = \frac{1}{\sqrt{2}}(\hat{T}_{12}+\hat{T}_{21}),\;\hat{T}_{\tau_{3,2}^{(6)}} =\frac{1}{\sqrt{2}}(\hat{T}_{13}+\hat{T}_{31}) \;,
\hat{T}_{\tau_{3,3}^{(6)}} =\frac{1}{\sqrt{2}}(\hat{T}_{14}+\hat{T}_{41}), \\
\hat{T}_{\tau_{3,4}^{(6)}} =\frac{1}{\sqrt{2}}(\hat{T}_{23}+\hat{T}_{32}), \;
\hat{T}_{\tau_{3,5}^{(6)}} =\frac{1}{\sqrt{2}}(\hat{T}_{24}+\hat{T}_{42}),\;\hat{T}_{\tau_{3,6}^{(6)}} =\frac{1}{\sqrt{2}}(\hat{T}_{34}+\hat{T}_{43}),
\end{split}
\end{equation}
which can be transformed to Minkowski space via
\begin{equation}
\mathrm{Euclidean}\rightarrow\mathrm{Minkowski} : 
\hat{T}_{44}\rightarrow \hat{T}_{00},\;\hat{T}_{4j}\rightarrow-i\hat{T}_{0j},\;\hat{T}_{jk}\rightarrow-\hat{T}_{jk}.
\end{equation}

\begin{table}
    \begin{ruledtabular}
    \begin{tabular}{c|rrrrrrrrrrr}
    $\; t_s \;$ & 6 & 7 & 8 & 9 & 10 & 11 & 12 & 13 & 14 & 16 & 18 \\ 
    $\; N_s  \;$ & 9 & 16 & 16 & 16 & 16 & 16 & 16 & 16 & 32 & 32 & 32
    \end{tabular}
    \end{ruledtabular}
    \caption{Number of sources $N_s$ for which the connected quark contributions to the three-point function, Eq.~\eqref{eq:sm_3pt_def}, are computed for each sink time $t_s$.}
    \label{tab:conn-counts}
\end{table}

The three-point functions needed to extract the matrix elements of operator $\hat{T}_{i\mathcal{R}\ell}$, where $\mathcal{R}\in\{\tau_1^{(3)},\tau_3^{(6)}\}$, $\ell$ denotes the vector in the basis of the irreps defined in Eqs.~\eqref{eq:rep3} and~\eqref{eq:rep6}, and $i\in\{q,g,v_1,v_2\}$,  are defined as
\begin{equation}\label{eq:sm_3pt_def}
C_{i\mathcal{R}\ell ss'}^{3\text{pt}}(\vec{p}',t_s; \vec{\Delta},\tau;\vec{x_0},t_0) = \sum_{\vec{x},\vec{y}}e^{-i\vec{p}'\cdot(\vec{x}-\vec{x}_0)}e^{i\vec{\Delta}\cdot(\vec{y}-\vec{x}_0)}\text{tr}\left[\vev{\Gamma_{s's}\chi(\vec{x},t_s+t_0)\hat{T}_{i\mathcal{R}\ell}\bar{\chi}(\vec{x}_0,t_0)}\right].
\end{equation}
In the limit where $(t_s-t_0)\rightarrow \infty$ and $(t_s-\tau)\rightarrow \infty$, the three-point functions approach the desired matrix elements as:
\begin{equation}
C_{i\mathcal{R}\ell ss'}^{3\text{pt}}(\vec{p}',t_s; \vec{\Delta},\tau;\vec{x_0},t_0)\xrightarrow[(t_s-t_0)\rightarrow \infty]{(t_s-\tau)\rightarrow \infty} Z^*_{\vec{p}}Z_{\vec{p}'}\frac{e^{-E_{\vec{p'}}(t_s-t_0)}e^{-(E_{\vec{p}}-E_{\vec{p'}})(\tau-t_0)}}{4E_{\vec{p'}}E_{\vec{p}}}\bra{N(\vec{p'},s')}\hat{T}_{i\mathcal{R}\ell}\ket{N(\vec{p},s)},
\end{equation}
where $\vec{p}=\vec{p}'-\vec{\Delta}$.
Three-point functions for the disconnected quark and gluon contributions are constructed by correlating the operator measurements with the grid of 1024 two-point functions described above.
The connected quark contributions are measured for a subset of source-sink separations, with different numbers of sources for each as tabulated in Table~\ref{tab:conn-counts}.
These measurements are averaged over all source positions and 1000 bootstrap ensembles are formed. At the bootstrap level, the following ratios of three- and two-point functions are constructed:
\begin{equation}
\begin{split}
R_{i\mathcal{R}\ell ss'}(\vec{p}', t_s; \vec{\Delta}, \tau)
&= \frac{C^{\mathrm{3pt}}_{i\mathcal{R}\ell ss'}(\vec{p}',t_s; \vec{\Delta},\tau)}{C_{s's'}^{\mathrm{2pt}}(\vec{p}',t_s)}
\sqrt{\frac{C^{\mathrm{2pt}}_{ss}(\vec{p},t_s-\tau) C^{\mathrm{2pt}}_{s's'}(\vec{p}',t_s)C^{\mathrm{2pt}}_{s's'}(\vec{p}',\tau)}{C^{\mathrm{2pt}}_{s's'}(\vec{p}',t_s-\tau)C^{\mathrm{2pt}}_{ss}(\vec{p},t_s)C^{\mathrm{2pt}}_{ss}(\vec{p},\tau)}}\\
&\xrightarrow[t_s\rightarrow \infty]{(t_s-\tau)\rightarrow \infty}\frac{\text{tr}\left[\Gamma_{s's}(\cancel{p}'+m)\bra{N(\vec{p'},s')}\hat{T}_{i\mathcal{R}\ell}\ket{N(\vec{p},s)}(\cancel{p}+m)\right]}{4\sqrt{E_{\vec{p}}E_{\vec{p'}}(E_{\vec{p}}+m)(E_{\vec{p'}}+m)}}.\label{eq:3ptspec}
\end{split}
\end{equation}
The above expression is a linear combination of the GFFs, with known kinematical coefficients. Within each irrep, all ratios for choices $(\ell,\vec{p}',\vec{\Delta},s,s')$ that yield identical coefficients up to an overall minus sign are averaged, yielding averaged ratios $\bar{R}_{i\mathcal{R}c}(t; t_s,\tau)$, where $c$ denotes the unique set of coefficients and $t$ the squared momentum transfer associated with it. The averaged ratios yield the matrix elements of interest as
\begin{equation}
\bar{R}_{i\mathcal{R}c}(t; t_s,\tau) \xrightarrow[t_s\rightarrow \infty]{(t_s-\tau)\rightarrow \infty} \text{ME}_{i\mathcal{R}c}(t),
\end{equation}
where $\text{ME}_{i\mathcal{R}c}(t)$ is the rescaled matrix element to be extracted by fitting the Euclidean time dependence of the ratios. This is done using the summation method, wherein one sums over the operator insertion time $\tau$ to form summed ratios~\cite{Capitani:2012gj,Maiani:1987by,Dong:1997xr}
\begin{equation}
\label{eq:sigmabar}
\bar{\Sigma}_{i\mathcal{R}c}(t; t_s, \tau_{\text{cut}}) = \sum_{\tau=\tau_{\text{cut}}}^{t_s-\tau_{\text{cut}}}\bar{R}_{i\mathcal{R}c}(t; t_s,\tau)  \xrightarrow[t_s\rightarrow \infty]{(t_s-\tau)\rightarrow \infty} (t_s-\tau_{\text{cut}}+1) \text{ME}_{i\mathcal{R}c}(t)+\Lambda_{i\mathcal{R}c}(t; \tau_{\text{cut}}),
\end{equation}
where $\Lambda_{i\mathcal{R}c}(t; \tau_{\text{cut}})$ is a $t_s$-independent constant. $\text{ME}_{i\mathcal{R}c}(t)$ is then extracted by fitting the slope of $\bar{\Sigma}_{i\mathcal{R}c}(t; t_s, \tau_{\text{cut}})$ with respect to $t_s$.

\begin{figure}[p]
\includegraphics[width=0.998\textwidth]{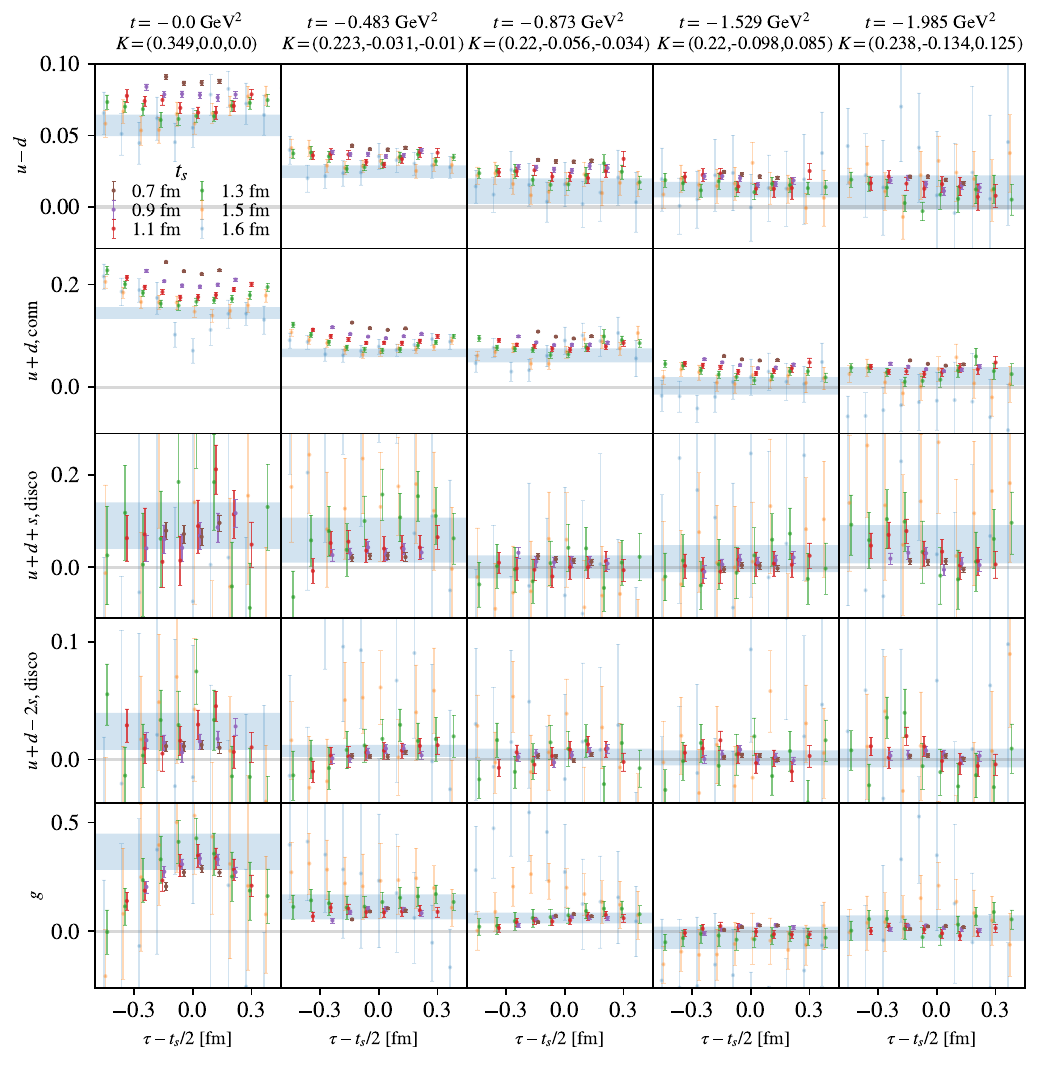}
    \caption{Examples of averaged ratios, defined following Eq.~\eqref{eq:3ptspec}, for $\tau_1^{(3)}$. Each column represents a single ratio, with the corresponding $t$ value and $(K^A,K^J,K^D)$ coefficients shown as column titles. The rows represent the bare $u-d$, the connected part of $u+d$, the disconnected part of $u+d+s$, the disconnected part of $u+d-2s$, and the gluon contributions to the ratio. The overlaid bands show fits to the corresponding ratios obtained via the summation method as described in the text.
         }
    \label{fig:bumps_irrep1}
\end{figure}

\begin{figure}[p]
\includegraphics[width=0.998\textwidth]{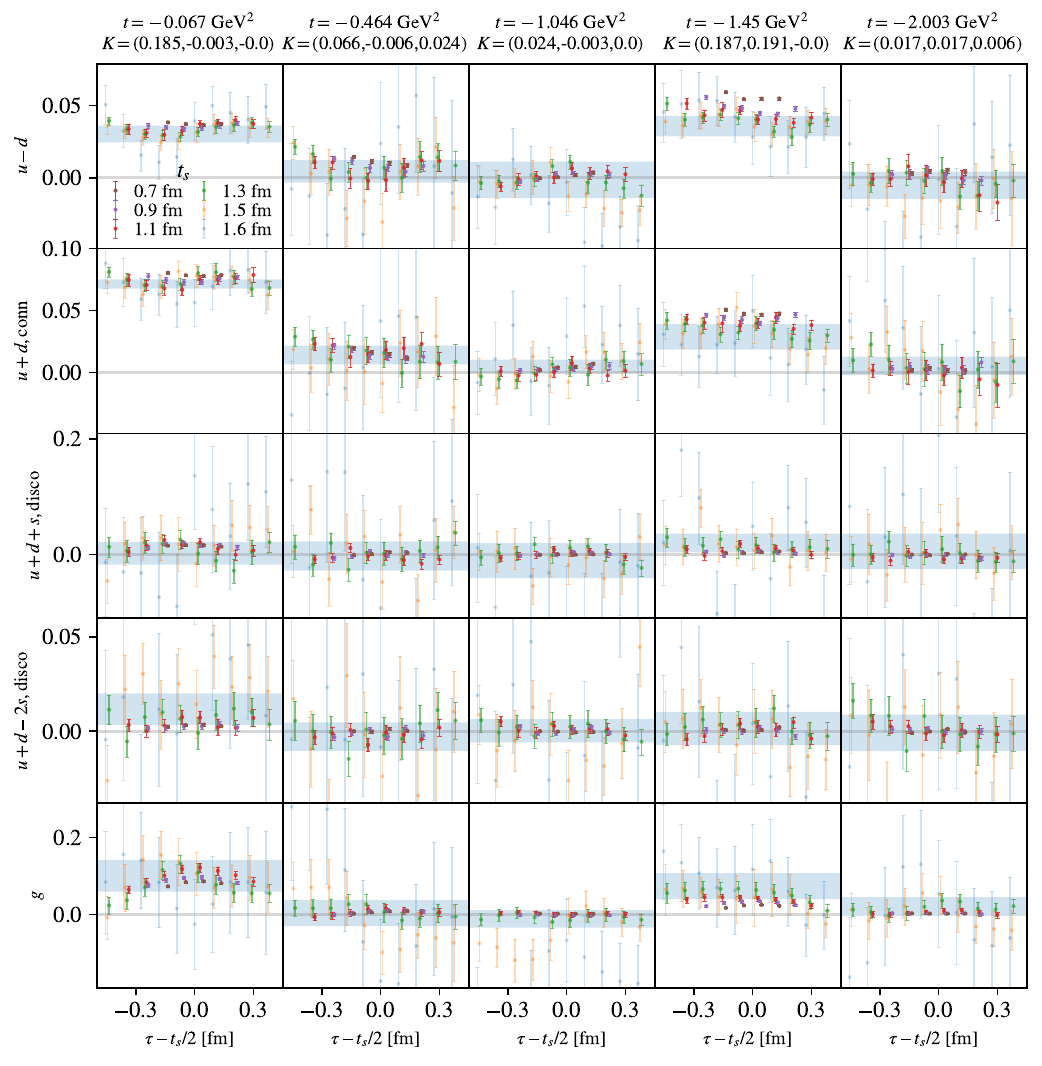}
    \caption{As in Fig.~\ref{fig:bumps_irrep1}, but for irrep $\tau_3^{(6)}$.
         }
    \label{fig:bumps_irrep2}
\end{figure}

For the connected data, summed ratios are constructed and fit for all $\tau_\text{cut} \in [2,6]$. Linear summation fits are performed for all ranges $t_{s,\text{min}} < t_s \leq 18$, where $t_{s,\text{min}} \in [6,14]$ and $t_s=18$ is the largest available in the dataset. The resulting pool of fits are model-averaged to obtain the final estimate of the matrix element.
For the disconnected data, summation fits are performed for $\tau_{\text{cut}}\geq 2$ to all $t_s$-ranges extending over $5$ or more timeslices in the window $[t_{s,\text{min}},t_{s,\text{max}}]$, where $t_{s,\text{min}}=6$ for $\tau_1^{(3)}$ and $t_{s,\text{min}}=10$ for $\tau_3^{(6)}$, which yield the highest $p$-values for most of the $t$-bins in the bare GFF fits of both singlet and non-singlet disconnected quark contributions. The maximum sink time is set to $t_{s,\text{max}}=20$, after which the $c$-bin ratios become consistent with non-Gaussian noise. For the gluon contribution, in order to avoid the effect of contact terms due to the gradient flow, cuts are made such that $\tau_{\text{cut}} \geq 4$ and $t_{\text{min}} \geq 9$ for the summation fits, while the rest of the fitting details are the same as for the disconnected quark data. Examples of averaged ratios and corresponding summation method fits for all different contributions are shown in Figs.~\ref{fig:bumps_irrep1} and~\ref{fig:bumps_irrep2}.

\begin{figure}[p]
    \includegraphics[width=0.998\textwidth]{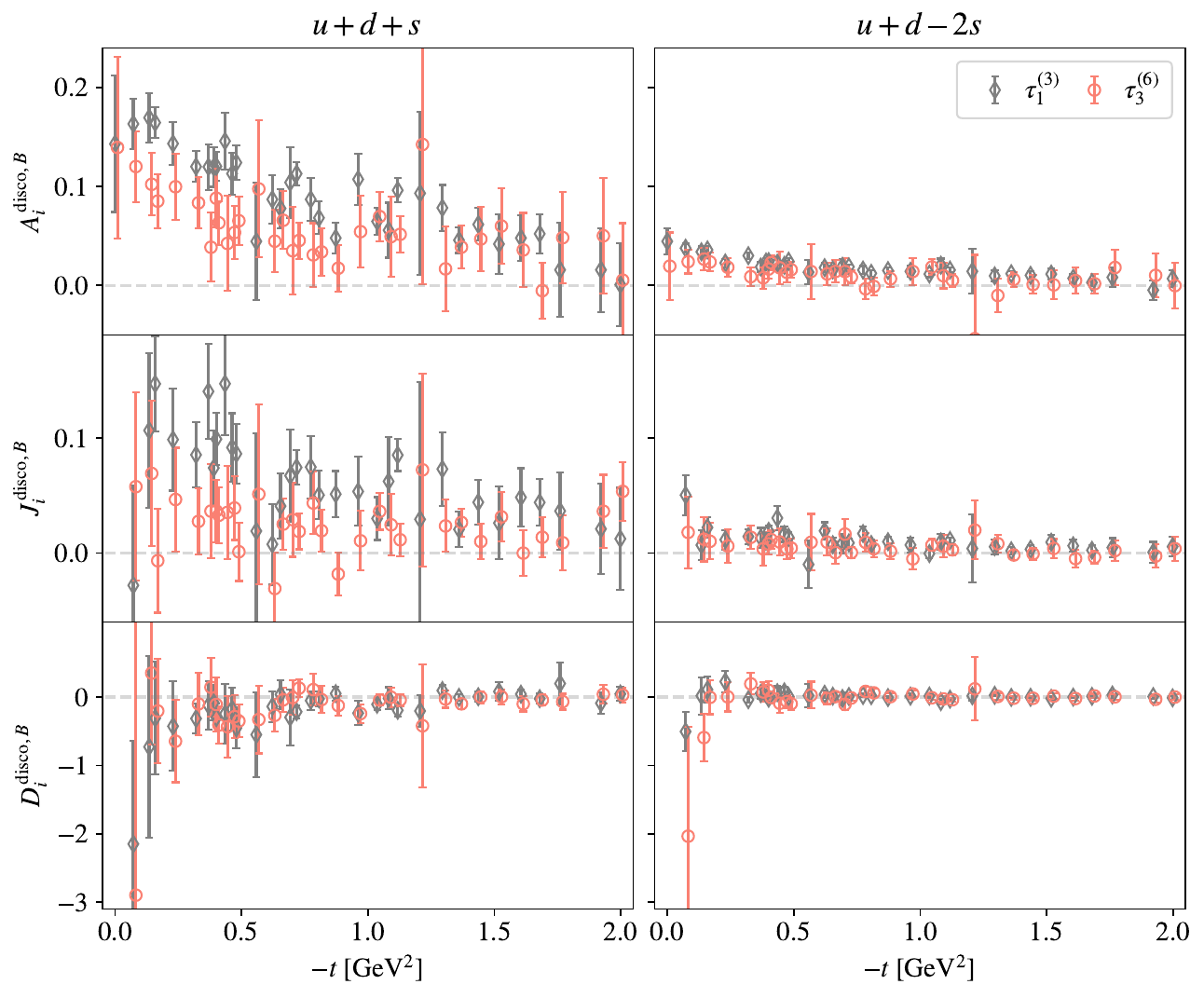} 
    \caption{Bare disconnected quark GFFs for each of the two irreps studied.}
    \label{fig:barediscoGFF}
\end{figure}

The matrix elements of the singlet $u+d+s$ and non-singlet $u+d-2s$ currents contain both a connected and a disconnected contribution. However, data for the connected contribution are only available for a subset of the kinematics compared to the disconnected contribution, as three-point functions were computed for only three sink momenta and one spin orientation via the sequential source method, inverting through the sink. Discarding a considerable fraction of the disconnected contribution measurements is undesirable, since constraining it is already challenging due to poor signal-to-noise. Instead, the bare disconnected singlet and non-singlet contributions to the GFFs are first fit, by separating the complete set of rescaled matrix elements obtained by the summation fits into 34 momentum $t$-bins (the same $t$-bins used for the renormalized GFFs presented in the main text), and inverting the system of equations
\begin{equation}\label{eq:MEdecomp}
\vec{K}^A_{\mathcal{R}t} A_{i\mathcal{R}t}^{\text{bare}}+\vec{K}^J_{\mathcal{R}t} J_{i\mathcal{R}t}^{\text{bare}} + \vec{K}^D_{\mathcal{R}t} D_{i\mathcal{R}t}^{\text{bare}} = \text{\bf{ME}}_{i\mathcal{R}t}
\end{equation}
for each $t$-bin. Here, bold symbols are vectors in the space of $c$-bins, $\vec{K}$ are the kinematic coefficients multiplying the GFFs in the expansion of the rescaled matrix element, as defined by Eqs.~\eqref{eq:3ptspec} and~\eqref{eq:protonME}, and $i\in\{q^{\text{disco}},v_2^{\text{disco}}\}$. The resulting bare GFFs are shown in Fig.~\ref{fig:barediscoGFF}. These are then used to obtain predictions for the disconnected contributions to the smaller subset of matrix elements for which the connected measurements are available.

\begin{figure}[p]
    \centering
    \subfloat[\centering  ]
    {{\includegraphics[width=0.9\textwidth]{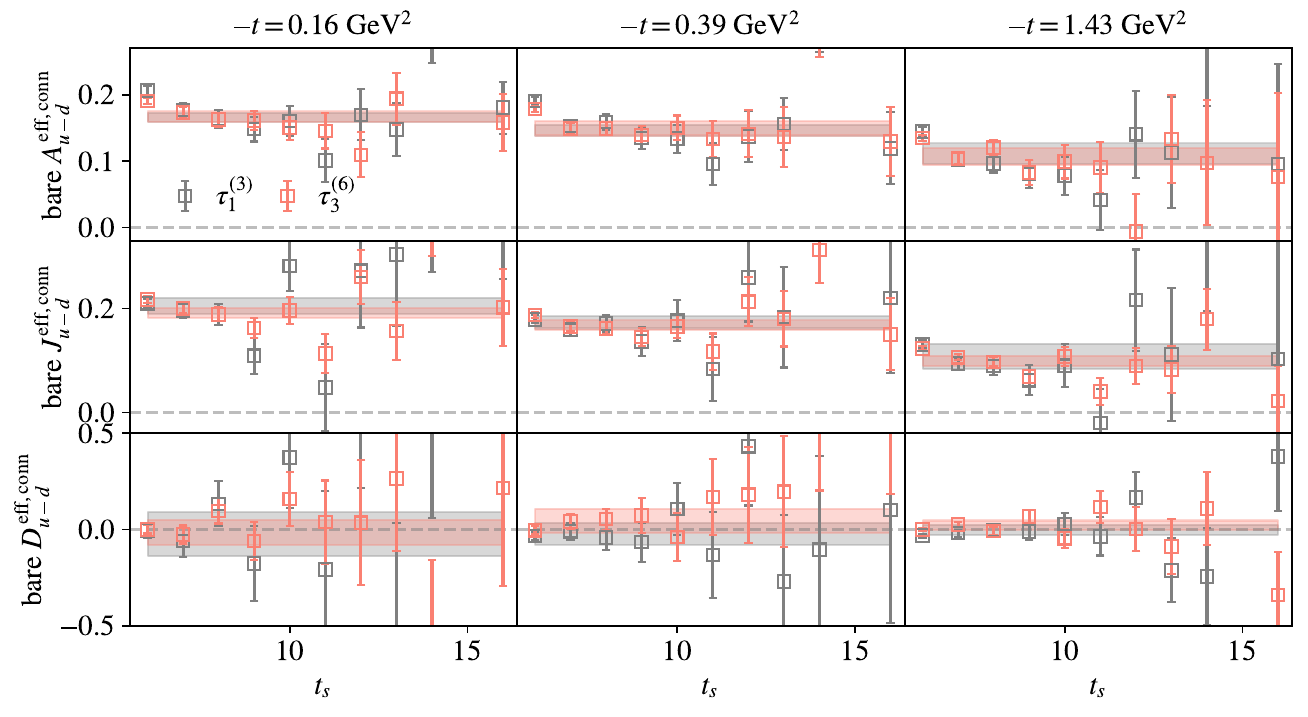}\label{fig:conn_eff_umd}}} \\
    \subfloat[\centering  ]
    {{\includegraphics[width=0.9\textwidth]{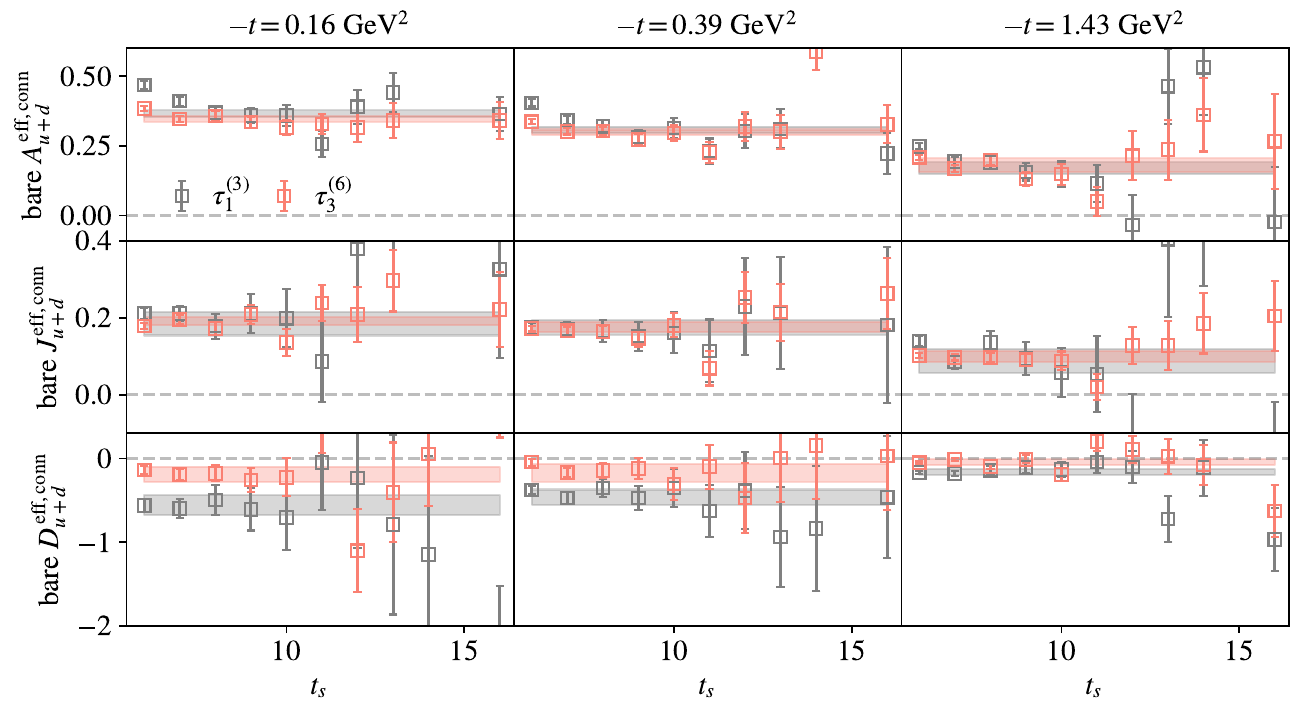}\label{fig:conn_eff_upd} }}
    \caption{Examples of effective GFFs for three different $t$-bins of (a) the purely connected $u-d$ and (b) the connected part of the $u+d$ contribution, computed from Eq.~\eqref{eq:effectiveGFF} using summed ratios with $\tau_{\text{cut}} = 2$. The bands are not fits to the data shown but correspond to bare GFFs obtained by fitting the bare matrix elements used to compute the results from the main text.}
    \label{fig:conn_eff}
\end{figure}

\begin{figure}
    \centering
    \subfloat[\centering  ]
    {{\includegraphics[width=0.9\textwidth]{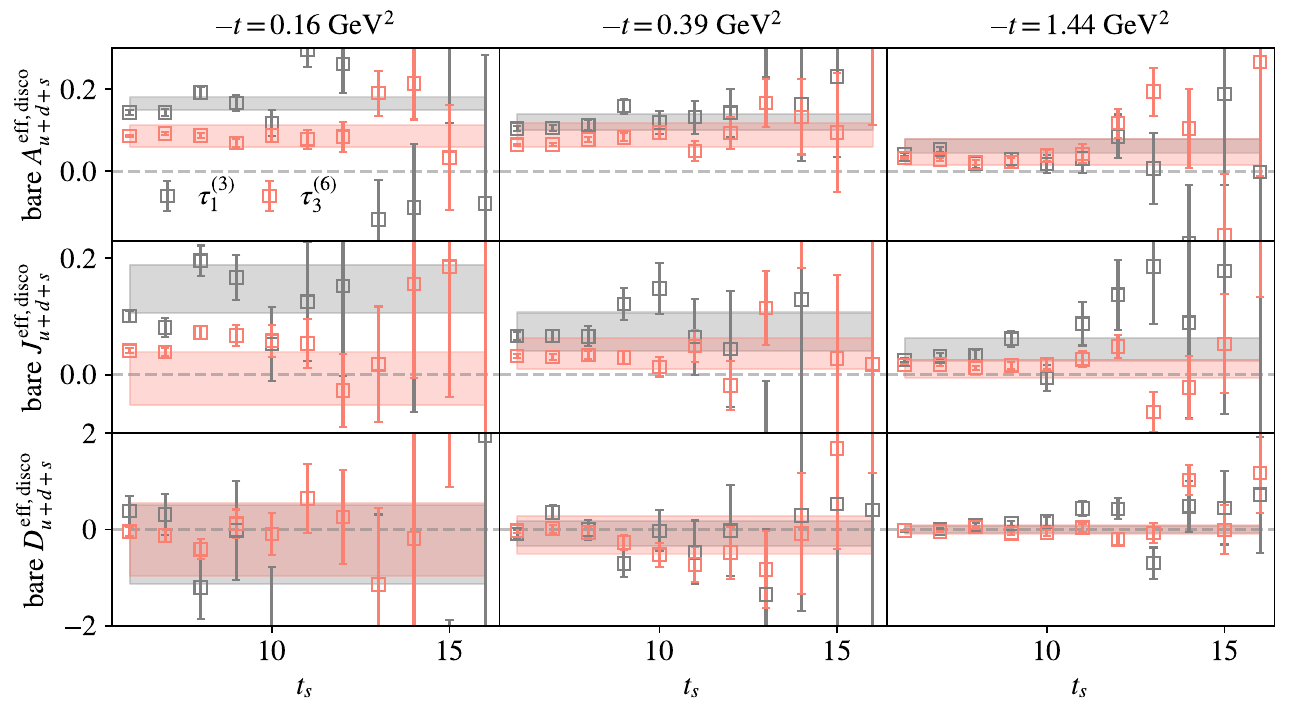}\label{fig:disco_eff_isos}}} \\
    \subfloat[\centering  ]
    {{\includegraphics[width=0.9\textwidth]{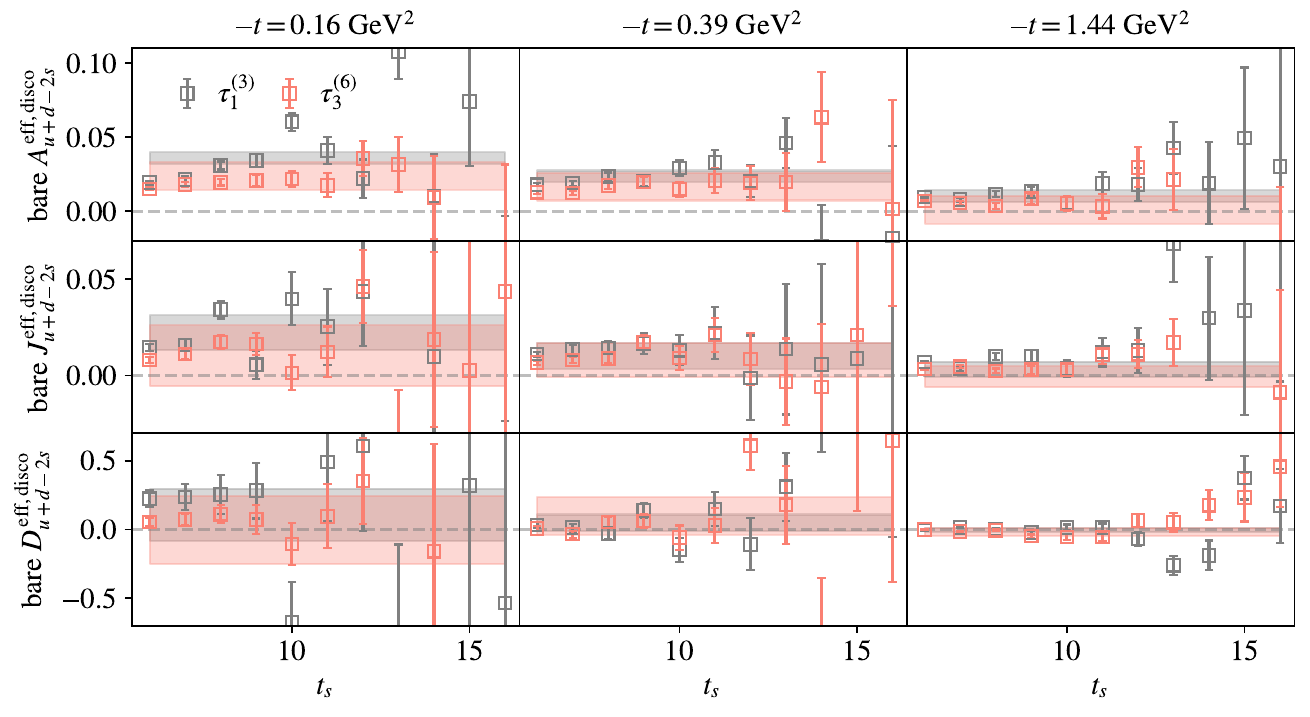}\label{fig:disco_eff_isov} }}
    \caption{Examples of effective GFFs for three different $t$-bins of the disconnected part of the (a) $u+d+s$ and (b) $u+d-2s$ contribution, computed from Eq.~\eqref{eq:effectiveGFF} using summed ratios with $\tau_{\text{cut}} = 2$. The bands are not fits to the data shown but correspond to bare GFFs obtained by fitting the bare matrix elements used to compute the results from the main text.}
    \label{fig:disco_eff}
\end{figure}

\begin{figure}
    \includegraphics[width=0.9\textwidth]{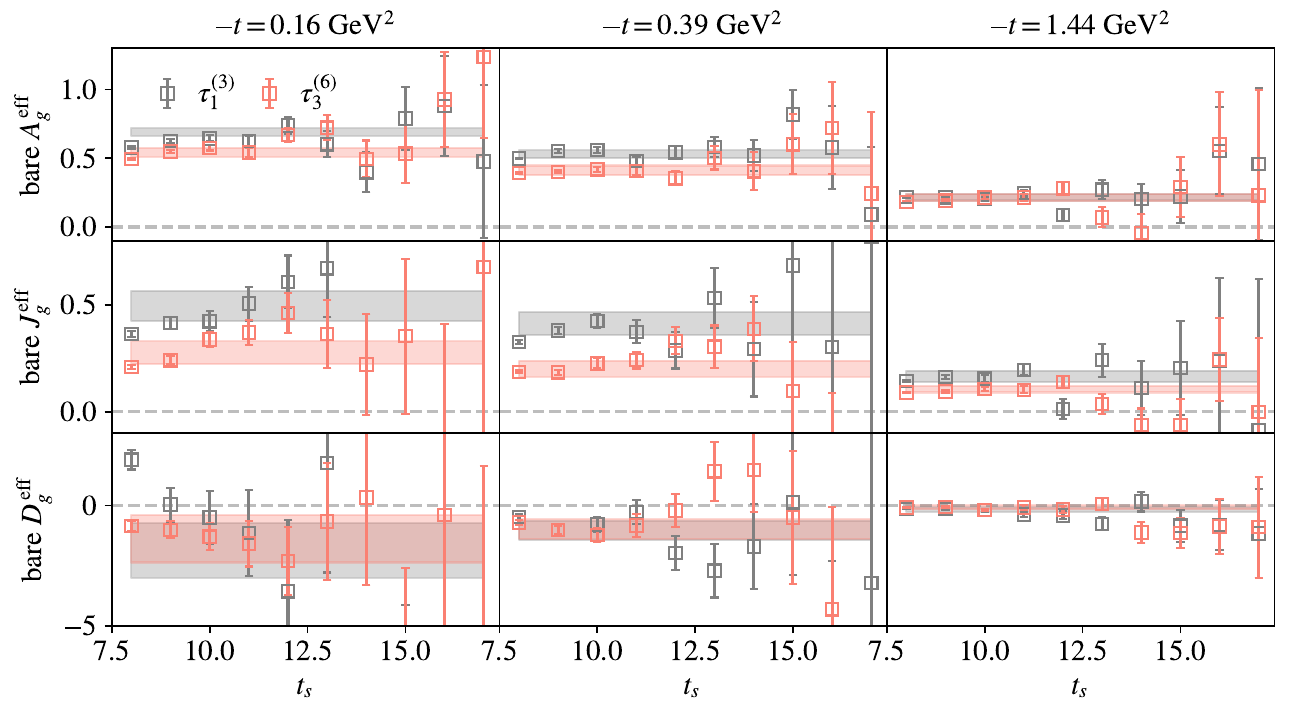}
    \caption{Examples of effective GFFs for three different $t$-bins of the gluon contribution, computed from Eq.~\eqref{eq:effectiveGFF} using summed ratios with $\tau_{\text{cut}} = 4$. The bands are not fits to the data shown but correspond to bare GFFs obtained by fitting the bare matrix elements used to compute the results from the main text.}
    \label{fig:gluon_eff}
\end{figure}

In order to investigate how well the summation fits describe the data cumulatively, one may compare with ``effective GFFs'', i.e., the results of a simpler $t_s$-dependent extraction of the bare GFFs.
First, ``effective matrix elements'' may be defined as functions of $t_s$ by
\begin{equation} \label{eq:effectiveME}
\text{ME}_{i\mathcal{R}c}^{\text{eff}}(t; t_s) = \partial_{t_s}\bar{\Sigma}_{i\mathcal{R}c}(t; t_s)  \approx \frac{1}{\delta {t_s}} \left[ \bar{\Sigma}_{i\mathcal{R}c}(t; t_s+\delta {t_s}) - \bar{\Sigma}_{i\mathcal{R}c}(t; t_s) \right] \;,
\end{equation}
with $\delta {t_s}=1$ for all the gluon and quark disconnected data, and for the majority of the quark connected data\footnote{$\delta {t_s}$ is set to $2$ for the larger sink times of the connected data, since $t_s=15$ and $t_s=17$ were not computed.}. The $\tau_{\text{cut}}$ dependence of the summed ratios is fixed to the minimum value used for the specific contribution, as described above, and is not explicitly shown. The effective matrix elements, which are formed directly from the data, are grouped into the same $t$-bins used for extraction of the GFFs. One can then obtain effective bare GFFs $A^{\text{eff}}_{i\mathcal{R}t}(t_s)$, $J^{\text{eff}}_{i\mathcal{R}t}(t_s)$, and $D^{\text{eff}}_{i\mathcal{R}t}(t_s)$ for each flavor $i$, irrep $\mathcal{R}$, momentum bin $t$, and sink time $t_s$ by fitting the overconstrained system of linear equations,
\begin{equation} \label{eq:effectiveGFF}
\vec{K}^A_{\mathcal{R}t} A_{i\mathcal{R}t}^{\text{eff}}(t_s)+\vec{K}^J_{\mathcal{R}t} J_{i\mathcal{R}t}^{\text{eff}}(t_s) + \vec{K}^D_{\mathcal{R}t} D_{i\mathcal{R}t}^{\text{eff}}(t_s) = \text{\bf{ME}}^{\text{eff}}_{i\mathcal{R}t}(t_s) \;,
\end{equation}
where $\vec{K}^A_{\mathcal{R}t}$, $\vec{K}^J_{\mathcal{R}t}$ and $\vec{K}^D_{\mathcal{R}t}$ are as in Eq.~\eqref{eq:MEdecomp}. Figs.~\ref{fig:conn_eff},~\ref{fig:disco_eff}, and~\ref{fig:gluon_eff} show examples of the bare effective GFFs for all the different contributions and several different $t$-bins, and compare them to bare GFFs computed by fitting the matrix elements obtained from summation fits.
The two methods provide broadly consistent results, up to deviations in the effective GFFs at small $t_s$ which may be attributed to excited-state effects, and at large $t_s$ due to degradation of signal-to-noise.

\begin{figure}
\centering
\subfloat[\centering  ]
{{\includegraphics[width=0.8\textwidth]{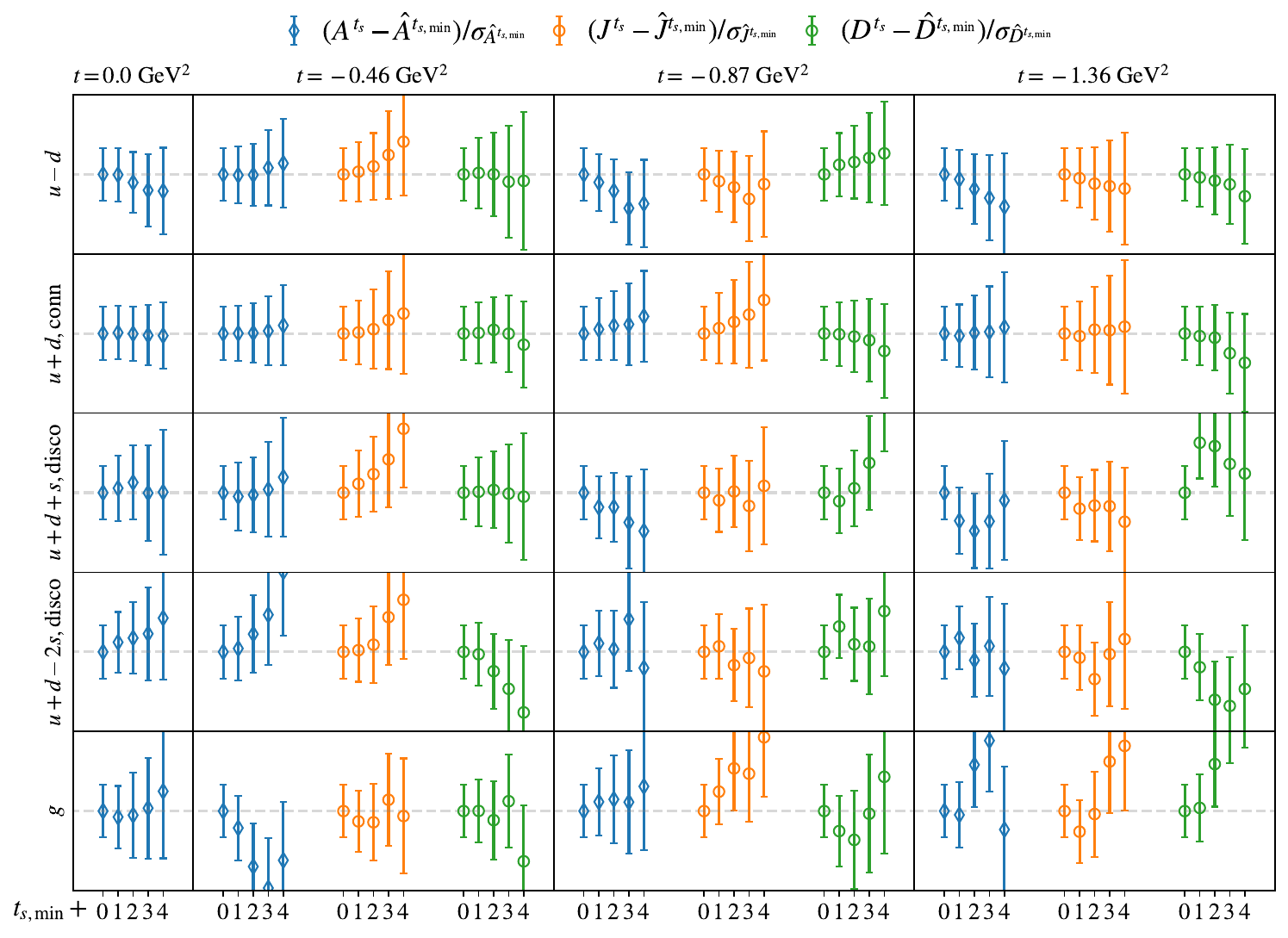}}} \\
\centering
\subfloat[\centering  ]
{{\includegraphics[width=0.8\textwidth]{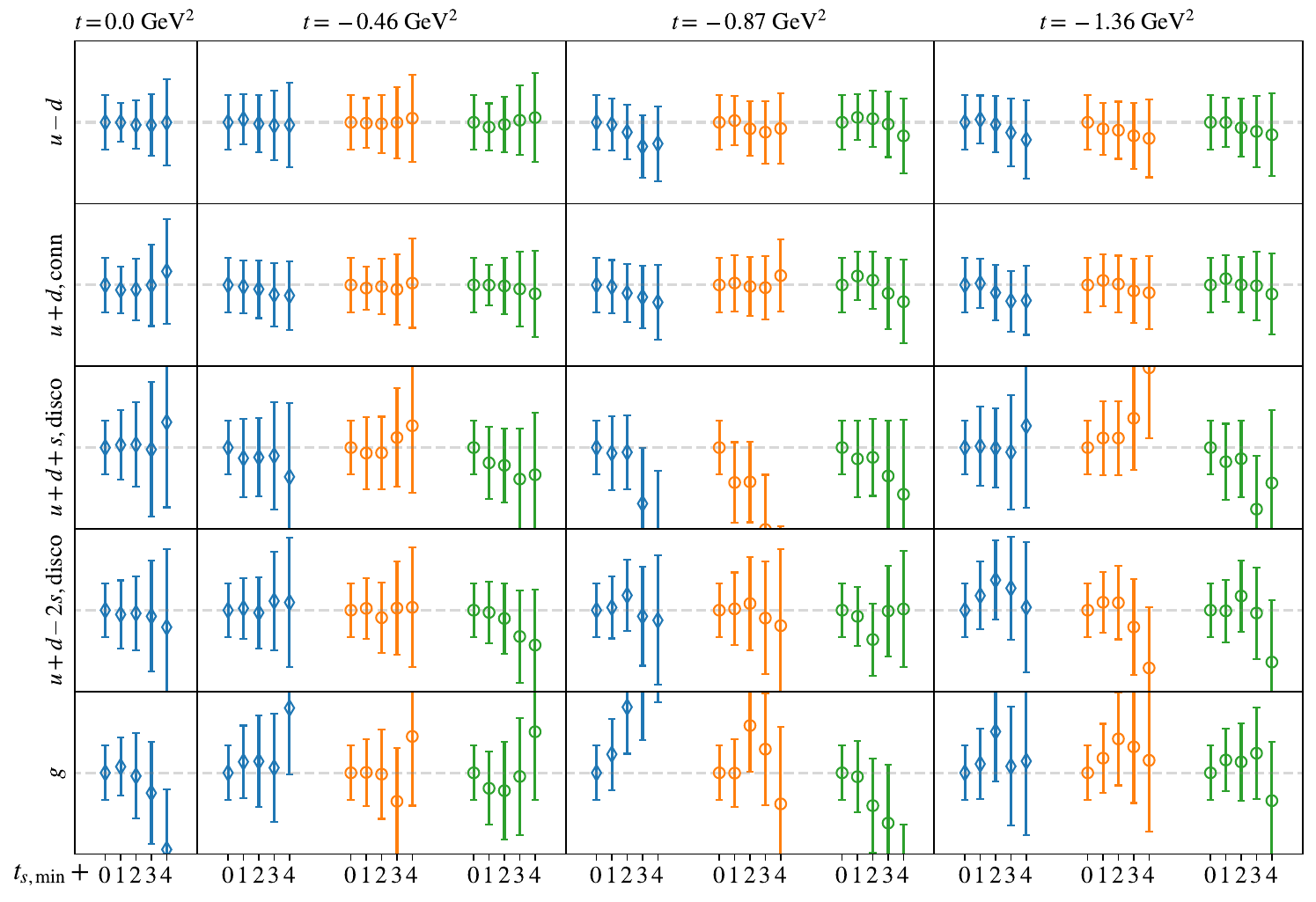}}}
 \caption{To illustrate the effect of the choice of $t_{s,\text{min}}$ on the summation fits, the $\tau_1^{(3)}$ (a) and $\tau_1^{(6)}$ (b) bare GFFs are shown for increasing $t_{s,\text{min}}$ at four different values of $t$. The starting $t_{s,\text{min}}$ for each bare flavor contribution is as defined below Eq.~\eqref{eq:sigmabar}.\label{fig:stab_irrep}}   
\end{figure}

Fig.~\ref{fig:stab_irrep} shows the bare GFFs for each flavor and four different choices of momentum transfer $t$, obtained by inverting Eq.~\eqref{eq:MEdecomp} using matrix elements fit with increasing values of the hyperparameter $t_{s,\text{min}}$. The stability of the results provides a test of remnant excited state contamination; in most cases the values are stable within statistical noise and no apparent pattern of drift is observed. This suggests that excited-state contamination is well-treated by the analysis.


\section{GFF model fit parameters}

The renormalized GFFs presented in Fig.~\ref{fig:GFFflavors} include fits using the $n$-pole model
\begin{equation}
\text{F}_n(t) =\frac{\alpha}{(1-t/\Lambda^2)^n},
\end{equation}
with $\alpha$ and $\Lambda$ being free parameters. All choices $1\leq n \leq 4$ yield consistent results. The dipole model, $n=2$, was chosen for the results presented in the main text based on the $\chi^2$/d.o.f.. Results are also presented using the $z$-expansion~\cite{Hill:2010yb}
\begin{equation} 
\text{F}_{z}(t) = \sum_{k=0}^{k_{\text{max}}}
\alpha_k[z(t)]^k,
\end{equation}
where $\alpha_k$ are free parameters, and
\begin{equation}
z(t) = \frac{\sqrt{t_{\text{cut}}-t}-\sqrt{t_{\text{cut}}-t_0}}{\sqrt{t_{\text{cut}}-t}+\sqrt{t_{\text{cut}}-t_0}},
\end{equation}
where
\begin{equation}
t_0 = t_{\text{cut}}\left(1-\sqrt{1+(2~\text{GeV})^2/t_{\text{cut}}}\right),
\end{equation}
and $t_{\text{cut}}=4m_{\pi}^2$. Fits were performed with $k_{\text{max}}$ varied between $2$ and $4$, and the smallest $\chi^2/$d.o.f.\ is obtained in fits with $k_{\text{max}}=2$, which is used for the results in the main text. Table~\ref{tab:fitparams} presents the resulting parameters of the dipole and $z$-expansion fits to $A(t)$, $J(t)$, and $D(t)$, which are used to produce the bands in Fig.~\ref{fig:GFFflavors}, along with the corresponding $\chi^2$ per degree of freedom of each fit.

\begin{table}[h] 
\begin{center}
\begin{ruledtabular}
\begin{tabular}{SSSSSSSSS}
&& \multicolumn{3}{c}{dipole} & \multicolumn{4}{c}{$z$-expansion}  \\ \midrule
&& {$\alpha$} & {$\Lambda$} & {$\chi^2$/d.o.f} & {$\alpha_0$} & {$\alpha_1$} & {$\alpha_2$} & {$\chi^2$/d.o.f}\\ \midrule
{$A_i$} & {$g$} & 0.501(27) & 1.262(18) & 1.5 & 0.271(14) & -0.648(35) & -0.089(78) & 1.5  \\[2pt]
&{$q$} & 0.510(25) & 1.477(44) & 0.8 & 0.314(13) & -0.591(39) & -0.06(13) & 0.8  \\[2pt]
&{$v_1$} & 0.1665(56) & 1.997(80) & 0.7 & 0.1249(41) & -0.141(11) & -0.029(38) & 0.4 \\[2pt]
&{$v_2$} & 0.433(13) & 1.524(35) & 0.7 & 0.2768(78) & -0.493(22) & -0.156(77) & 0.6 \\ \midrule
{$J_i$} & {$g$} & 0.255(13) & 1.399(49) & 1.1 & 0.1539(54) & -0.301(24) & -0.26(11) & 1.0 \\[2pt]
&{$q$} & 0.251(21) & 1.62(13) & 0.6 & 0.1658(85) & -0.290(42) & -0.01(23) & 0.6 \\[2pt]
&{$v_1$} & 0.2016(86) & 1.698(70) & 0.3 & 0.1399(49) & -0.219(15) & -0.150(68) & 0.3 \\[2pt]
&{$v_2$} & 0.222(13) & 1.65(10) & 0.5 & 0.1492(61) & -0.250(26) & -0.07(13) & 0.5 \\ \midrule
{$D_i$} & {$g$} & -2.57(84) & 0.538(65) & 1.2 & -0.303(28) & 2.20(30) & -5.2(1.1) & 1.1 \\[2pt]
&{$q$} & -1.30(49) & 0.81(14) & 0.5 & -0.378(43) & 1.49(45) & -1.1(1.7) & 0.6 \\[2pt]
&{$v_1$} & 0.009(23) & 1.7(5.0) & 0.4 & 0.0068(83) & -0.003(54) & -0.08(24) & 0.4 \\[2pt]
&{$v_2$} & -0.77(15) & 0.932(98) & 0.9 & -0.272(22) & 1.06(18) & -1.37(70) & 0.9 \\
\end{tabular}
\end{ruledtabular}
\end{center}
\caption{\label{tab:fitparams}Fit parameters of the dipole and $z$-expansion parametrizations of the $t$-dependence of the proton GFFs, $A_i(t)$, $J_i(t)$, and $D_i(t)$, renormalized in the $\overline{\text{MS}}$ scheme at scale $\mu=2~\text{GeV}$.}
\end{table}

\section{Single-irrep renormalized GFFs}
\label{app:renorm}

The results presented in the main text are the result of a simultaneous fit to both irreps using the procedure and renormalization coefficients of Ref.~\cite{Hackett:2023nkr}. The renormalized results for the two irreps must agree in the continuum limit, but can have different discretization artifacts. Since a continuum extrapolation is not performed, Fig.~\ref{fig:single-irrep} presents the renormalized GFFs obtained by fitting the two irreps separately. Although some tension is observed between the two irreps for some of the GFFs, particularly for $J_g(t)$, the $p$-values of the combined-irrep fits are comparable to those for the single-irrep fits. In future work, it will be important to repeat this calculation at different lattice spacings in order to better quantify the discretization artifacts.

\section{Remaining sources of systematic uncertainty}

This section discusses the remaining sources of systematic uncertainty that can not be fully quantified within the present calculation, including an attempt to estimate their magnitudes.

\begin{figure}[p]
    \centering
    \includegraphics[width=\textwidth]{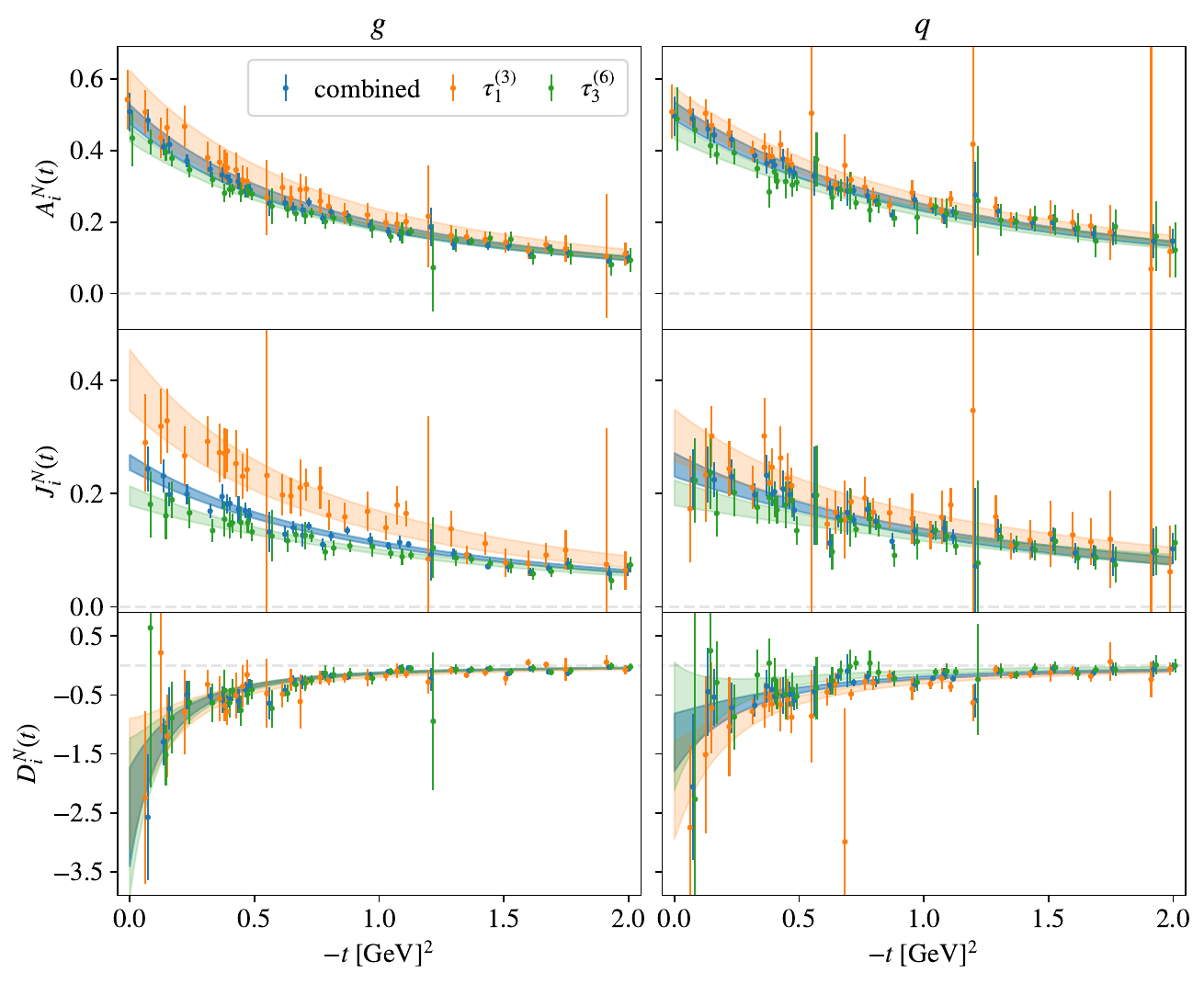}
    \caption{The renormalized isosinglet quark and gluon combined-irrep GFFs presented in the main text (blue points) shown alongside the GFFs obtained by fitting the results for irrep $\tau_1^{(3)}$ (orange points) or $\tau_3^{(6)}$ (green points) separately. The overlaid bands show dipole fits to the corresponding data.}
    \label{fig:single-irrep}
\end{figure}

\begin{itemize}
    \item Discretization effects: 

As discussed in the main text, discretization effects cannot be quantified in this first calculation using a single lattice ensemble. However, an estimate of the order of magnitude of these effects can be made by 1) comparison with previous calculations and 2) comparison of the GFFs computed in separate irreps, which will have different discretization artifacts.

No information is available from previous calculations as to the magnitude of discretization effects in the $D$-term or the $t$-dependence of the proton GFFs, as quantified, e.g., by the multipole mass. 
However, several previous studies have considered aspects of the forward-limit momentum fraction $A_i(0)$ in calculations with several lattice spacings. 
In a recent study of the purely connected isovector $u-d$ momentum fraction~\cite{Mondal:2020ela}, which employed $N_f=2+1$ ensembles with clover fermions near the physical pion mass, including the ensemble used in this calculation, a mild dependence on the lattice spacing was observed, approximately $5\%$ between $a=0.09~\text{fm}$ and the continuum limit. For the gluon momentum fraction, the study of Ref.~\cite{Fan:2022qve} using $N_f=2+1+1$ flavors of HISQ fermions (i.e., a different formulation of lattice fermions to that employed in this work) found similarly modest discretization effects. Specifically, the gluon momentum fraction was found to shift to larger values by approximately $3\%$ between $a=0.09~\text{fm}$ and the continuum limit at the physical pion mass.
Note, however, that discretization effects may be substantially larger in this calculation, both because mixing with the isosinglet quark contribution was not included in Ref.~\cite{Fan:2022qve}, and because the HISQ action is engineered specifically to reduce lattice artifacts.

Complementary information is provided by the comparison between the single-irrep and combined-irrep fits in this work, shown in Fig.~\ref{fig:single-irrep}. Taking this difference as indicative of the size of discretization artifacts, one might expect small effects on the $t$-dependence (as quantified by, e.g., dipole masses), and more significant effects as large as 10-20\% on the overall normalization and thus the forward limits. 

In summary, while discretization effects can not be quantified in this first calculation on a single lattice ensemble, it is likely that they are of order 10\% for the forward limits, and smaller for the $t$-dependence of the GFFs.

\item Finite volume corrections:

As discussed in the main text, the ensemble used in this work has $m_{\pi}L \approx 3.8$, which is slightly smaller than the typical bound of $m_\pi L > 4$ desired to limit finite-volume effects to roughly the percent level. However, the rule-of-thumb estimate of finite-volume effects gives $e^{-m_{\pi} L}\approx 2\%$, which is subdominant compared to both the statistical uncertainty of these results and the likely size of other systematic uncertainties such as discretization artifacts. 

Limited additional information on the size of finite-volume corrections can be deduced from previous lattice QCD calculations. For the isovector quark momentum fraction, the results of Ref.~\cite{Mondal:2020ela} support the conclusion that finite-volume effects are small at the parameters of this calculation, however those of Ref.~\cite{Alexandrou:2019ali} (albeit in a calculation with a significantly different setup including a different lattice action) suggests that finite-volume effects may be larger off-forward than in the forward limit for the $A_{u-d}(t)$ GFF, but not the other isovector GFFs. There is no information for the other quantities studied here.


\item Uncertainty resulting from quark masses yielding a larger-than-physical value of the pion mass:

The ensemble used in this work has quark masses tuned to yield a pion mass that is $\approx 30~\text{MeV}$ heavier than the physical value. Chiral perturbation theory predictions exist for the isovector quark, isoscalar quark, and total GFFs~\cite{Dorati:2007bk,Wein:2014wma,Alharazin:2020yjv}. However, in order to use them to extrapolate to the physical pion mass, or even to estimate the size of systematic uncertainties arising from the quark masses used in this calculation, the low energy constants would need to be determined, for example using results at several different values of $m_{\pi}$.

Again, limited information can be deduced from previous lattice QCD calculations regarding the systematic effects of quark masses corresponding to a larger-than-physical pion mass on the quantities studied in this work. The most relevant information is for the the forward-limit isovector momentum fraction $A_{u-d}(0)$, for which Ref.~\cite{Mondal:2020ela} found $O(5\%)$ effects in the extrapolation from the ensemble used here to the physical point. A similar order of effects is found in Ref.~\cite{Wang:2021vqy} for the gluon and disconnected quark contributions to $A(t)$ and $J(t)$ for a small range of $t$, albeit with a different fermion action and parameters that are not directly comparable to those studied here. There is no information available for the other quantities studied in this work.


\item Uncertainty due to the non-perturbative renormalization being computed at a different lattice spacing and pion mass to the bare GFFs:

The RI-MOM renormalization coefficients are computed on an ensemble at different physical parameters (ensemble B), as described in Ref.~\cite{Hackett:2023nkr}, for reasons of computational necessity. The corresponding results are 
\begin{equation} \label{eq:renormmatrixnumbers}
\begin{aligned}
\tau_1^{(3)}:\;\; && R_{qq}^{\overline{\text{MS}}} &= \phantom{-}1.056(27), \; & R_{qg}^{\overline{\text{MS}}} &=0.067(71) \;,\\
&& R_{gq}^{\overline{\text{MS}}} &= -0.169(22), \; & R_{gg}^{\overline{\text{MS}}} &=1.68(18) \;,\\
\tau_3^{(6)}:\;\; && R_{qq}^{\overline{\text{MS}}} &= \phantom{-}1.039(28), \; & R_{qg}^{\overline{\text{MS}}} &=0.081(19) \;,\\
&& R_{gq}^{\overline{\text{MS}}} &= -0.180(23), \; & R_{gg}^{\overline{\text{MS}}} &=1.625(48) \;,
\end{aligned}
\end{equation}
for the singlet factors and 
\begin{equation} \label{eq:renormvectornumbers}
\begin{split}
\tau_1^{(3)}\;\;:\;\; &
R_v^{\overline{\text{MS}}} = 1.067(29) \;,\\
\tau_3^{(6)}\;\;:\;\; &
R_v^{\overline{\text{MS}}} = 1.066(21) \;,
\end{split}
\end{equation}
for the nonsinglet factors.
The parameters of ensemble A, which was used for the extraction of the bare matrix elements in this work, and of ensemble B, used for the renormalization, are presented in Table~\ref{tab:ensemble}. The renormalization factor for the isovector current $u-d$ on the ensemble (ensemble A) used for the bare matrix elements in the current work was found in Ref.~\cite{Mondal:2020ela} to be within a standard deviation from the non-singlet renormalization factor that is used here. The flavor-singlet renormalization coefficient on ensemble B is also found to be consistent with the non-singlet one on ensemble B, in agreement with the perturbation theory prediction that they deviate starting at NLO~\cite{Loffler:2021afv}.
These observations suggest that the difference between ensembles A and B in the contributions to the renormalization coefficients coming from amputated three-point functions of the quark EMT currents is likely small compared to the other uncertainties in this work. For the amputated three-point functions of the gluon EMT, it can be expected that the results deviate from perturbation theory predictions that are available for the unflowed operator due to the flowing employed in the measurement, and therefore no similar argument can be made. To mitigate the effects of this limitation for the gluon EMT, the flow time used in the calculation of the renormalization was adjusted in order to match the physical scale of the gluon EMT on the bare matrix element, as noted in the beginning of Sec.~IIID in Ref.~\cite{Hackett:2023nkr}. 

\item Uncertainty due to excited-state contamination:

The stability analysis and the comparison with effective GFFs described above provide confidence that excited-state effects are well-treated in this analysis and do not appear to be a likely source of substantial unquantified systematic uncertainty.
In fact, an estimate of this uncertainty is already quantified in the analysis by the treatment of fit-range systematics with Bayesian model averaging.
However, unexpectedly large excited state effects have been observed in calculations of other hadronic structure observables~\cite{Jang:2019vkm}, highlighting the importance of employing variational methods in future calculations to better quantify the associated systematic uncertainties.

\end{itemize}

The expected size of the remaining systematic uncertainties in the lattice QCD results, as outlined above, is commensurate with the systematic uncertainties suggested by comparison with the global experimental averages of Ref.~\cite{Hou:2019efy}. Assuming these are accurate, unaffected by new physics, and without their own unquantified systematic effects, this comparison indicates a systematic uncertainty of $\approx 20\%$ in the overall normalization of the GFFs computed here.
As discussed in the main text, comparison with present experimental results, as well as the analysis of this section, suggest the $t$-dependence is less affected by these effects. 

\begin{table}
\begin{center}
\begin{tabular}{ccccccccccccccc}
\toprule
 & $L/a$   & $T/a$ & $\beta$ & $a m_l$ & $a m_s$ & $a$ [fm] & $m_{\pi}$ [MeV]  \\ \midrule
A & $48$ & $96$ & $6.3$ & $-0.2416$ &
$-0.2050$ & $0.091(1)$ & $169(1)$ &  \\ \midrule
B & $12$ & $24$ & $6.1$ & $-0.2800$ &
$-0.2450$ & $0.1167(16)$ & 
 $450(5)$ \\
\bottomrule
\end{tabular}
\end{center}
\caption{\label{tab:ensemble}Specifics of the lattice ensembles used in this work. Ensemble A, generated by the JLab/LANL/MIT/WM groups~\cite{ensembles}, is used for the calculation of the bare matrix elements presented in the main text. The calculation of the renormalization coefficients, presented in Ref.~\cite{Hackett:2023nkr}, is performed on ensemble B.}
\end{table}



\end{document}